\documentclass{amsart}

\usepackage[latin1]{inputenc}
\usepackage{dsfont}
\usepackage{graphicx}
\usepackage{amssymb}
\newcommand\R{{\ensuremath {\mathbb R} }}
\newcommand\Z{{\ensuremath {\mathbb Z} }}
\newcommand\C{{\ensuremath {\mathbb C} }}
\newcommand\1{{\ensuremath {\mathds 1} }}
\renewcommand\phi{\varphi}

\newcommand{\gS}{\mathfrak{S}}
\newcommand{\wto}{\rightharpoonup}
\renewcommand{\to}{\rightarrow}

\newcommand{\cB}{\mathcal{B}}
\newcommand{\cT}{\mathcal{T}}

\DeclareMathOperator{\tr}{{\rm Tr}}

\newcommand{\cC}{\mathcal{C}}
\newcommand{\cK}{\mathcal{K}}
\newcommand{\cX}{\mathcal{X}}
\newcommand\ii{{\ensuremath {\infty}}}
\newcommand\pscal[1]{{\ensuremath{\left\langle #1 \right\rangle}}}
\newcommand{\norm}[1]{ \left| \! \left| #1 \right| \! \right| }
\renewcommand{\epsilon}{\varepsilon}

\newtheorem{theorem}{Theorem}[section]
\newtheorem{lemma}{Lemma}[section]

\newtheorem{definition}{Definition}[section]
\newtheorem{remark}{Remark}[section]
\numberwithin{equation}{section}

\usepackage[active]{srcltx}

\begin{document}

\title{A positive density analogue of the Lieb-Thirring inequality}

\author{Rupert L. FRANK}
\address{Department of Mathematics, Princeton University, Fine Hall, Washington Road, Princeton, NJ 08544-1000, USA}
\email{rlfrank@math.princeton.edu}

\author{Mathieu LEWIN}
\address{CNRS \& Department of Mathematics (UMR 8088), University of Cergy-Pontoise, 95 000 Cergy-Pontoise,
France}
\email{mathieu.lewin@math.cnrs.fr}

\author{Elliott H. LIEB}
\address{Departments of Mathematics and Physics, Princeton University, Jadwin Hall, P.O. Box 708, Princeton, NJ
08542-0708, USA}
\email{lieb@princeton.edu}

\author{Robert SEIRINGER}
\address{Department of Mathematics and Statistics, McGill University, 805
Sherbrooke Street West, Montr\'eal, Qu\'ebec H3A 2K6,
Canada}
\email{robert.seiringer@mcgill.ca}

\date{May 15, 2012, final version to appear in the \textit{Duke Mathematical Journal}}

\begin{abstract}
The Lieb-Thirring inequalities give a bound on the
negative eigenvalues of a Schr\"odinger operator in terms
of an $L^p$ norm of the potential. These are dual to bounds
on the $H^1$-norms of a system of orthonormal functions.
Here we extend these bounds to analogous inequalities for
perturbations of the Fermi sea of non-interacting particles,
i.e., for perturbations of the continuous spectrum of the
Laplacian by local potentials.
\end{abstract}

\thanks{\copyright\, 2011 by
  the authors. This paper may be reproduced, in its entirety, for
  non-commercial purposes.}

\maketitle

\tableofcontents


\section{Introduction}

The Pauli exclusion principle for fermions in quantum mechanics has no
classical analogue. One of its primary effects is the increase in kinetic
energy that accompanies an increase in the density of such
particles. Intuitively, this increase
should be quantifiable in a manner similar to that
predicted by
the semi-classical approximation to quantum mechanics, and
it is the aim
of this paper to show that this can, indeed, be achieved in
the case of density
perturbations of an ideal Fermi gas.

We begin with some definitions \cite[Chapters 3 and 4]{LieSei-09}. The
state of a finite system of $N$ fermions of $q$ spin states each ($q=2$
for electrons) is described by a density matrix $\Gamma$, which may or not
be pure. Associated with a state is a one-body density matrix $\gamma$
(a reduction of $\Gamma$) which is an operator on $L^2(\R^d,\C^q)$. The
essential properties of $\gamma$ are that $0\leq \gamma \leq 1$ as
an operator and that $\tr\gamma = N$. It is a fact proved by Coleman in~\cite{Coleman-63} (see also
\cite[Thm. 3.2]{LieSei-09}) that any $\gamma$ with these properties arises
from some state $\Gamma$, i.e., no other restrictions on $\gamma$ are
required by quantum mechanics.

The electron density is $\rho_\gamma(x)=\tr_{\C^q} \gamma(x,x)$, in which
$\gamma(x,x)$ is a $q\times q$ matrix. The kinetic energy of the $N$ particle
system depends \textit{only}  on $\gamma$ and is given
by $\tr(-\Delta)\gamma$ in units
where $\hbar=2m=1$ and with $\Delta=\nabla^2$ denoting the Laplacian.

The semi-classical approximation for the kinetic energy is
$$
\tr(-\Delta)\gamma \approx
K_{\rm sc}(d) \int_{\R^d} \rho_\gamma(x)^{1+2/d} \,dx \,,
$$
with the constant
\begin{equation}
K_{\rm sc}(d):=\frac{d}{d+2}
\left(\frac{d(2\pi)^{d}}{q\,|S^{d-1}|}\right)^{\tfrac{2}d} \,.
\label{eq:semiclassical-ct}
\end{equation}

The Lieb-Thirring inequality \cite{LieThi-76} states that there is a constant
$0<r_d\leq 1$ such that
\begin{equation}
 \label{eq:lt}
\tr(-\Delta)\gamma \geq r_d\; K_{\rm sc}(d) \int_{\R^d} \rho_\gamma(x)^{1+2/d}
\,dx
\end{equation}
for any one-body density matrix $0\leq\gamma\leq 1$. The
original value \cite{LieThi-76} for
$r_3$ was $0.185$ but it has been improved since then to $0.672$
\cite{DolLapLos-08}, and it is current belief that it equals
one
(for $d\geq 3$). This subject continues to be actively
studied (see for instance the recent works~\cite{DolLapLos-08,BenLos-04,HunLapWei-00,LapWei-00} and the reviews~\cite{LapWei-00b,Hundertmark-07}).
Note that the inequality~\eqref{eq:lt} does not require $\tr\,\gamma$ to be
an integer; it need not even be finite.

We can turn the matter around and, instead of specifying
$\gamma$, think of specifying a
density $\rho(x)$ and
asking for the minimum kinetic energy needed to achieve this particle density.
The Lieb-Thirring inequality above gives a universal answer to this question in terms of
the semi-classical approximation. Here, we are implicitly
using the fact that for
any given function $\rho(x)\geq 0$ with $\int_{\R^d} \rho(x)\,dx = N$ there is a
fermionic $N$-particle density matrix whose one-body reduced density matrix
$\gamma$ satisfies $\rho_\gamma(x)=\rho(x)$, see \cite[Thm. 1.2]{Lieb-83b}.

It is important for many applications that the right side of
the inequality
\eqref{eq:lt} is additive in position space. If we partition $\R^3$ into disjoint subsets,
the right side is just the sum of the corresponding local energies. While this does not 
hold for the left side, it nearly does. The bound shows that
there is some truth to this approximate additivity.
This additivity, or locality, played an important role in
a proof of the stability of matter \cite{LieThi-75}.

While the inequality \eqref{eq:lt} was the object of
principal interest in \cite{LieThi-75}, the actual proof
of \eqref{eq:lt}
went via the Legendre transform of \eqref{eq:lt} with
respect to $\rho$. This is an inequality about the sum of
the negative eigenvalues $(E_j)$ of the Schr\"odinger
operator $-\Delta+V$ for an arbitrary potential $V$, namely,
\begin{equation}
 \label{eq:ltv}
 \sum_j |E_j| = \tr(-\Delta+V)_- \leq \tilde r_d\; L_{\rm
sc}(d) \int_{\R^d} V(x)_-^{1+d/2} \,dx\, ,
\end{equation}
where $X_-:=\max\{-X,\,0\}\geq 0$ denotes the negative part
of a number or a self-adjoint operator $X$,
\begin{equation}
L_{\rm sc}(d)=\frac{2q\,|S^{d-1}|}{d(d+2)\,(2\pi)^{d}} \,,
\label{eq:semiclassical-ctl}
\end{equation}
and $\tilde r_d$ is a universal constant independent of
$V$. In fact, the relation between the constants in
\eqref{eq:lt} and \eqref{eq:ltv} is given by
$$
( p \ L_{\rm sc}(d) )^{p'} ( p' \ K_{\rm sc}(d) )^{p} =1
\qquad\mathrm{and}\qquad
( \tilde r_d )^{p'} ( r_d )^{p} =1 \,,
$$
where $p=1+d/2$ and $p'= 1+2/d$; see
\cite{LieThi-75,LieSei-09}. The duality between
$\rho$ and $V$, and between kinetic energy and Schr\"odinger
eigenvalue sums is one of the important inputs in density
functional theory~\cite{Lieb-83b}.

\medskip

A question that is not only natural but of significance for condensed matter
physics is the analogue of \eqref{eq:lt} when we start, not
with the vacuum, but
with a background of fermions with some prescribed constant
density $\rho_0>0$. How
much kinetic energy does it then cost to make a local perturbation
$\delta\rho(x)$? This time $\delta\rho$ can be negative, as long as
$\rho_0+\delta\rho\geq 0$ everywhere. We would expect that the semi-classical
expression will guide us here as well and, indeed, it does
so, as we will show in
this paper.

The principal difficulty that has to be overcome is that inequality
\eqref{eq:lt} was obtained in \cite{LieThi-76} by first
proving \eqref{eq:ltv}, a route that does not seem to be
helpful now. The picture was changed by a paper of Rumin
\cite{Rumin-11} in which inequality \eqref{eq:lt} was
obtained directly, without estimates on eigenvalues. 
(The constant obtained this way is not, however, as good as
the 0.672 quoted above.) We are
able to utilize some ideas in \cite{Rumin-11} to help solve
our problem.

The first thing is to formulate a mathematically precise statement of what it
means to make a local perturbation of an ideal Fermi gas. One could think of
putting $N$ electrons in a large box of volume $v$,
computing the change in
kinetic energy, and then passing to the thermodynamic limit
$v\to\infty$ with $\rho_0 =N/v$ fixed. For this, appropriate boundary conditions have to be
imposed. To avoid this discussion
we pose the problem for an infinite sea with specified chemical potential
$\mu>0$. The chemical potential of the ideal Fermi gas is
$$
\mu = \frac{2+d}d \ K_{\rm sc}(d) \ \rho_0^{2/d} \,.
$$
It is often called the Fermi energy and can be interpreted, physically, as the
kinetic energy needed to add one more particle to the Fermi sea. 

We then look at the operator $-\Delta-\mu$ in $L^2(\R^d,\C^q)$, which, in our
context, plays the role of $-\Delta$ in inequality
\eqref{eq:lt}. The energy observable of a particle is now \emph{defined
to be} $-\Delta -\mu$, which is negative for states in the
Fermi sea and positive for states outside the sea. The
energy to create either a particle outside the Fermi
sea or a hole inside the sea is positive. The
(grand-canonical) energy of 
the unperturbed Fermi sea is 
$\tr(-\Delta-\mu)\Pi^- =-\ii$,
where $\Pi^-$ denotes the projection onto the negative
spectral subspace
of $-\Delta-\mu$. Clearly, $\Pi^-(x,x)=\rho_0/q\,\1_{\C^q}$ (we will often not write the identity matrix $\1_{\C^q}$ for simplicity). Our
interest is in the formal 
\emph{difference} in energy between the state described by a one-body density matrix $0\leq\gamma\leq1$ and the
state described by $\Pi^-$, and this is non-negative
since the minimum total energy (given $\mu$) is the
uniform, filled
Fermi sea. Our \emph{main result}  is a lower bound for
this difference 
in terms of the semi-classical expression in all dimensions $d\geq2$, namely,
\begin{multline}
\tr\bigg((-\Delta-\mu)\gamma -(-\Delta-\mu) \Pi^- \bigg)
=\tr (-\Delta-\mu) (\gamma -\Pi^-)\\
\geq r_d\ K_{\rm sc}(d) \;
\int_{\R^d}\left(\rho_\gamma(x)^{1+\tfrac{2}d}-(\rho_0)^{1+\tfrac{2}d}-\frac{2+d
}d(\rho_0)^{\tfrac{2}d}\,\big(\rho_\gamma(x)-\rho_0\big)\right)dx
\label{eq:formal-LT-rho}
\end{multline}
for some universal $r_d$ that does not depend on $\gamma$.
The trace in this expression might not exist in the usual sense, that is, $
(-\Delta-\mu) (\gamma -\Pi^-)$ might not be trace class.
This situation will be dealt with more carefully in the
sequel. 

The Legendre transform of the right side of the inequality
\eqref{eq:formal-LT-rho} will give us an inequality for the
change in energy of the Fermi sea when a one-body potential
$V$ is added to $-\Delta-\mu$. The positive density
analogue of \eqref{eq:ltv} in dimensions $d\geq2$ is
\begin{multline}
\tr\bigg((-\Delta-\mu+V)_- - (-\Delta-\mu)_-\bigg) + \rho_0
\int_{\R^d} V(x) \,dx\\
 \leq \tilde r_d \ L_{\rm sc}(d)\;
\int_{\R^d}\left((V(x)-\mu)_-^{1+\tfrac{d}2}-\mu^{1+\tfrac{d
} 2}+\frac{2+d}2\, \mu^{\tfrac{d}2}\,V(x)\right)\,dx.
\label{eq:formal-LT-V}
\end{multline}
Of course, $(-\Delta-\mu)_- = -(-\Delta-\mu)\Pi^-$. The
quantity above is not necessarily trace class but the trace
can, nevertheless, be defined; see Definition~\ref{def:total_energy} below.

The inequalities~\eqref{eq:formal-LT-rho} and~\eqref{eq:formal-LT-V} are only valid in dimensions $d\geq2$. In dimension $d=1$, a divergence related to the Peierls instability~\cite{Peierls} appears, and a Lieb-Thirring inequality of the form of~\eqref{eq:formal-LT-rho} or~\eqref{eq:formal-LT-V} cannot hold for $\mu>0$. This will be discussed in detail in this paper.

Our main inequalities \eqref{eq:formal-LT-rho} and
\eqref{eq:formal-LT-V} were announced and discussed
in~\cite{FraLewLieSei-11}. In particular, the constants $r_3\simeq
0.1279$ and $r_2\simeq 0.04493$ were given. We do not expect them to
be optimal and it is a challenge to improve them.  One interesting
case in which the sharp constant in~\eqref{eq:formal-LT-rho} can be
found is that in which $\rho_\gamma(x)$ is required to be zero for all
$x$ in some bounded domain $\Omega$. In Section~\ref{sec:Li-Yau} we
prove that if the integral on the right side of
\eqref{eq:formal-LT-rho} is taken only over $\Omega$, then $r_d =1$ in
this case, and this is obviously optimal.

Our method to prove the inequalities \eqref{eq:formal-LT-rho} and
\eqref{eq:formal-LT-V} is rather general and it can be used to treat other systems. As examples we will also discuss in this paper Lieb-Thirring inequalities in a box of size $L\gg1$ with periodic boundary conditions, the case of positive temperature, and systems with a periodic background.

\medskip

The paper is organized as follows. In the next section we introduce
some mathematical tools allowing us to give a rigorous meaning
to the Lieb-Thirring inequalities~\eqref{eq:formal-LT-rho}
and~\eqref{eq:formal-LT-V}. Our main task will be to
correctly define the traces $\tr(-\Delta-\mu)(\gamma-\Pi^-)$
and $\tr\big((-\Delta-\mu+V)_--(-\Delta-\mu)_-\big)$ in such
a way that~\eqref{eq:formal-LT-rho} and~\eqref{eq:formal-LT-V}
become dual to each other in the appropriate function spaces. In
Section~\ref{sec:perturbation_theory_1D} we consider the case of a
weak potential $tV$ with $t\ll1$, and we compute the second-order term
in $t$ of the left side of~\eqref{eq:formal-LT-V}. This will clarify
the fact that there cannot be simple Lieb-Thirring inequalities at
positive density in dimension $d=1$. The proofs of all these results
are provided in Sections~\ref{sec:proof-rho} and~\ref{sec:proof-V}. 
In Section~\ref{sec:Li-Yau} we consider the case of a density matrix $\gamma$ which vanishes on a 
given domain $\Omega$ and we derive a lower bound on the relative kinetic energy which involves 
the sharp constant $K_{\rm sc}(d)$.
In Section~\ref{sec:thermo-positive-temp} we prove Lieb-Thirring inequalities
in a box with periodic boundary conditions. This allows us to investigate
the thermodynamic limit and to extend our results to positive temperature.
Finally, in Section~\ref{sec:general} we discuss the extension of our
results to general background potentials, with an emphasis on periodic
systems.

\bigskip

\noindent\textbf{Acknowledgment.} Grants from the U.S.~NSF 
PHY-1068285 (R.F.), PHY-0965859 (E.L.), NSERC (R.S.) and from the ERC MNIQS-258023 (M.L.) are
gratefully acknowledged. R.F. would like to thank Ari Laptev for a stimulating discussion.

\section{Statement of the main results}\label{sec:results}
In this section, we provide the necessary tools to give a clear mathematical meaning to the inequalities~\eqref{eq:formal-LT-rho} and~\eqref{eq:formal-LT-V}
which we have announced in the introduction, and we state our main results.

We fix a positive number $\mu>0$ and denote by 
\begin{equation}
\Pi^-:=\1(-\Delta\leq\mu) 
\label{eq:def_Pi}
\end{equation}
the spectral projection of the Laplacian associated with the interval $(-\ii,\mu)$, describing a free Fermi gas in its ground state with
chemical potential $\mu$. As recalled before, the ambient Hilbert space is $L^2(\R^d,\C^q)$ where
$q$ is the number of spin states per particle (which is 2 for unpolarized
electrons but which will be taken arbitrary in this work). 
The gas, described by the projection $\Pi^-$,
has the constant density 
\begin{equation}
\rho_0=q\,(2\pi)^{-d}\int_{|p|^2<\mu}dp=\frac{q\,|S^{d-1}|}{d\,(2\pi)^{d}}\;\mu^
{\tfrac{d}2}.
\label{eq:relation_density_mu} 
\end{equation}
The kinetic energy per unit volume agrees with the semi-classical formula
\begin{equation}
q\,(2\pi)^{-d}\int_{|p|^2<\mu}|p|^2\,dp=(d/2)\,L_{\rm
sc}(d)\;\mu^{1+\tfrac{d}2}=K_{\rm sc}(d)\;(\rho_0)^{1+\tfrac{2}d}
\label{eq:relation_kinetic_mu} 
\end{equation}
where the semi-classical constants $K_{\rm sc}(d)$ and $L_{\rm sc}(d)$ are given by~\eqref{eq:semiclassical-ct} and~\eqref{eq:semiclassical-ctl} above.

\subsection{Lower bound on the variation of kinetic energy}\label{sec:LT_density_version}
We consider a fermionic state, with one-body density matrix $0\leq\gamma\leq1$
acting on $L^2(\R^d,\C^q)$, which we think of as a perturbation of the reference
state $\Pi^-$ defined before in~\eqref{eq:def_Pi}. We are interested in proving
a lower bound on the kinetic energy (including $\mu$) of $\gamma$, counted
relatively to that of $\Pi^-$, of the form of~\eqref{eq:formal-LT-rho}. To make
sense of this inequality, we use as main variable $Q:=\gamma-\Pi^-$ which 
satisfies the constraint
\begin{equation}
-\Pi^-\leq Q\leq 1-\Pi^-:=\Pi^+.
\label{eq:constraint_Q}
\end{equation}
Our goal is to prove a lower bound on $\tr(-\Delta-\mu)Q$. Our first task will
be to give a clear meaning to this quantity, in a rather general sense.
The constraint~\eqref{eq:constraint_Q} can also be written $(Q+\Pi^-)^2\leq Q+\Pi^-$. 
Expanding $(Q+\Pi^-)^2$ shows that~\eqref{eq:constraint_Q} is equivalent to 
\begin{equation}
Q^2\leq Q^{++}-Q^{--}
\label{eq:constraint_Q2}
\end{equation}
where we have introduced the notation
$Q^{\tau\tau'}:=\Pi^\tau Q\Pi^{\tau'}$ for $\tau,\tau'\in\{\pm\}$. 
In particular, we have $Q^{++}\geq 0$ and $Q^{--}\leq 0$. Furthermore there is
equality in \eqref{eq:constraint_Q2} if and only if $\gamma=Q+\Pi^-$ is an
orthogonal projection. 

For smooth-enough finite rank operators $Q$, the following computation is
justified:
\begin{align*}
\tr(-\Delta-\mu)Q&=\tr\Big(\Pi^+(-\Delta-\mu)Q\Pi^+ + \Pi^-(-\Delta-\mu)Q\Pi^-
\Big)\\
&=\tr|-\Delta-\mu|\big(Q^{++}-Q^{--}\big).
\end{align*}
As we have seen, we always have $Q^{++}-Q^{--}\geq0$, hence $\tr(-\Delta-\mu)Q\geq0$ (changing the density of particles inside or outside of the Fermi sea costs a positive energy once $\mu$, the energy of the Fermi level, has been subtracted).
We now use this fact to give a general meaning to $\tr(-\Delta-\mu)Q$, in the sense of quadratic forms.

\begin{definition}[Relative kinetic energy]\label{def:kinetic}
Let $Q$ be a bounded self-adjoint operator such that
$|-\Delta-\mu|^{1/2}Q^{\pm\pm}|-\Delta-\mu|^{1/2}$ are trace-class\footnote{In the whole paper we use the notation $Q^{\pm\pm}$ for the two operators $Q^{++}$ and $Q^{--}$, and the notation $Q^{\pm\mp}$ for $Q^{+-}$ and $Q^{-+}$.}. We define
\begin{equation}
\boxed{\tr_{0}\left(-\Delta-\mu\right)Q:=\tr|-\Delta-\mu|^{\tfrac12}\,\big(Q^{++
}-Q^{--}\big)\,|-\Delta-\mu|^{\tfrac12}.}
\label{eq:meaning_trace}
\end{equation}
If $Q$ is a bounded operator such that $\pm Q^{\pm\pm}\geq0$, then we extend the
previous expression by letting
$$\tr_{0}\left(-\Delta-\mu\right)Q:=+\ii$$
whenever $|-\Delta-\mu|^{1/2}Q^{++}|-\Delta-\mu|^{1/2}$ or
$|-\Delta-\mu|^{1/2}Q^{--}|-\Delta-\mu|^{1/2}$ is not trace-class.
\end{definition}

Of course we have $\tr_0(-\Delta-\mu)Q=\tr(-\Delta-\mu)Q$ (the usual trace) when
$(-\Delta-\mu)Q$ is trace-class.
The previous definition of the relative kinetic energy is inspired by similar
ideas used in the context of the Dirac equation~\cite{HaiLewSer-05a} and
of electrons in crystals~\cite{CanDelLew-08a}. Later on we will be interested in
estimating the kinetic energy of operators of the form
$Q_V=\1(-\Delta+V\leq \mu)-\Pi^-$ for a given potential $V$. In general we do not expect such
operators to be trace-class when $\mu>0$ (or even compact, see
Remark~\ref{rmk:trace-class} below). 

\begin{remark}\rm
When $\gamma=Q+\Pi^-$ is itself an orthogonal projection, $\gamma^2=\gamma$, we
have equality in \eqref{eq:constraint_Q2} and we obtain
\begin{equation}
\tr_0\left(-\Delta-\mu\right)Q=\tr|-\Delta-\mu|^{1/2}\,Q^2\,|-\Delta-\mu|^{1/2}
=\norm{Q\,|-\Delta-\mu|^{1/2}}^2_{\gS_2}
\end{equation}
where $\gS_2$ denotes the ideal of Hilbert-Schmidt operators on
$L^2(\R^d,\C^q)$.
\end{remark}

We are now ready to state our rigorous version of~\eqref{eq:formal-LT-rho}.

\begin{theorem}[Lieb-Thirring inequality, density version,
$d\geq2$]\label{thm:LT-rho}
Assume that $d\geq2$ and $\mu\geq0$. Let $Q$ be a self-adjoint operator such
that $-\Pi^-\leq Q\leq \Pi^+$ and such that 
$\big|-\Delta-\mu\big|^{1/2}Q^{\pm\pm}\big|-\Delta-\mu\big|^{1/2}$ are
trace-class. Then $Q$ is locally trace-class and the corresponding density
satisfies
\begin{equation}
\rho_Q\in L^{1+\tfrac2d}(\R^d)+L^{2}(\R^d).
\label{eq:prop_density} 
\end{equation}
Moreover, there exists a positive constant $K(d)\leq K_{\rm sc}(d)$ (depending
only on $d\geq2$) such that
\begin{multline}
\tr_0(-\Delta-\mu)Q\\
\;\geq\;
K(d)\int_{\R^d}\left(\big(\rho_0+\rho_Q(x)\big)^{1+\tfrac{2}d}-(\rho_0)^{
1+\tfrac{2}d}-\frac{2+d}d(\rho_0)^{\tfrac{2}d}\,\rho_Q(x)\right)dx
\label{eq:LT-rho}
\end{multline}
with $\rho_0$ the constant density of the Fermi gas, given
by~\eqref{eq:relation_density_mu}.
\end{theorem}

\medskip

We recall that a locally trace-class self-adjoint operator $A$ is such that
$\tr|\chi A\chi|<\ii$ for every bounded function $\chi$ of compact support. In
this case, the associated density $\rho_A$ is the unique real-valued function in
$L^1_{\rm loc}(\R^d)$ satisfying $\tr(\chi
A\chi)=\int_{\R^d}\chi(x)^2\,\rho_A(x)\,dx$.

Note that since $Q\geq-\Pi^-$ in Theorem~\ref{thm:LT-rho},
we have
$\rho_Q(x)\geq-\rho_0$ for all $x\in\R^d$. The function 
\begin{equation}
\delta \cT^{\rm
sc}_\mu(\rho):=(\rho_0+\rho)^{1+\tfrac{2}d}-(\rho_0)^{1+\tfrac{2}d}-\frac{d+2}
d(\rho_0)^{\tfrac{2}d}\,\rho 
\label{eq:def_delta_T}
\end{equation}
is non-negative and convex for $\rho\geq-\rho_0$. Hence the integrand on the
right side of \eqref{eq:LT-rho} is always non-negative. The function $\delta
\cT^{\rm sc}_\mu(\rho)$ behaves like $\rho^{1+2/d}$ for large $\rho$, and like
$\rho^2$ for small $\rho$. Moreover, it satisfies the scaling property
\begin{equation}
\delta \cT^{\rm sc}_\mu(\rho)=\mu^{d/2}\delta\cT^{\rm
sc}_1\big(\rho\mu^{-2/d}\big) 
\label{eq:scaling}
\end{equation}
and one has
$$\lim_{\mu\to0}\delta \cT^{\rm sc}_\mu(\rho)=\rho^{1+\tfrac2d}$$
uniformly on $\R^+$.
In the limit $\mu\to0$ (which is the same as $\rho_0\to0$
by~\eqref{eq:relation_density_mu}), the inequality~\eqref{eq:LT-rho} reduces to
the usual Lieb-Thirring inequality~\cite{LieThi-76,LieThi-75,LieSei-09}
\begin{equation}
\forall0\leq\gamma\leq 1,\qquad \tr(-\Delta)\gamma\geq
K(d)\int_{\R^d}\rho_\gamma(x)^{1+\tfrac2d}\,dx.
\label{eq:usual_LT-rho}
\end{equation}
The best constant in this inequality is smaller than or equal to $K_{\rm
sc}(d)$, the semi-classical constant defined above
in~\eqref{eq:semiclassical-ct}. Hence, $K(d)\leq K_{\rm
sc}(d)$ must hold. From the scaling
property~\eqref{eq:scaling}, we know that the best constant
in~\eqref{eq:LT-rho} is independent of $\mu>0$. However, the
best constant for $\mu=0$ in the Lieb-Thirring
estimate~\eqref{eq:usual_LT-rho} is not necessarily
equal to the best constant for~\eqref{eq:LT-rho}. The recent
estimates~\cite{DolLapLos-08} for the Lieb-Thirring
constant in~\eqref{eq:usual_LT-rho} do not \emph{a priori} give any information 
on the positive density analogue~\eqref{eq:LT-rho}.

The proof of Theorem~\ref{thm:LT-rho} is detailed later in
Section~\ref{sec:proof-LT-rho}. It uses the convexity of $\delta \cT^{\rm
sc}_\mu$, to estimate separately the densities corresponding to the two diagonal
terms $Q^{\pm\pm}$ and the two off-diagonal terms $Q^{\pm\mp}$. The estimate on
the diagonal terms $Q^{\pm\pm}$ is based on a new method which has recently been
introduced by Rumin~\cite{Rumin-11}. This estimate works similarly in dimension
$d=1$. The off-diagonal terms $Q^{\pm\mp}$ are studied by a direct and explicit
method which does not cover the case $d=1$.

There \emph{cannot} be an inequality like~\eqref{eq:LT-rho} in dimension $d=1$
for $\mu>0$. This surprising fact is due to a special divergence of the
off-diagonal terms $Q^{\pm\mp}$ at the Fermi points (see
Section~\ref{sec:perturbation_theory_1D} below for details). However, we can
prove the following:

\begin{theorem}[Lieb-Thirring inequality, density version,
$d=1$]\label{thm:LT-rho-1D}
Assume that $d=1$ and $\mu>0$. Let $Q$ be a self-adjoint operator such that
$-\Pi^-\leq Q\leq \Pi^+$ and such that 
$\big|-\Delta-\mu\big|^{1/2}Q^{\pm\pm}\big|-\Delta-\mu\big|^{1/2}$ are
trace-class. Then $Q$ is locally trace-class and the corresponding densities
satisfy
\begin{equation}
\rho_{Q^{\pm\pm}}\in L^{3}(\R)\cap L^{2}(\R),\quad
\int_{\R}\frac{\sqrt\mu\,|k|}{\big(\sqrt\mu+|k|\big)\,{\log\left(\frac{2\sqrt{
\mu}+|k|}{|2\sqrt{\mu}-|k||}\right)}}{\big|\widehat{\rho_{Q^{\pm\mp}}}(k)\big|^2
}\,dk<\ii,
\label{eq:prop_density-1D} 
\end{equation}
(where $\widehat{\ }$ denotes the Fourier transform).
Moreover, there exist two positive constants $K(1)\leq K_{\rm sc}(1)$ and
$K'(1)>0$ such that
\begin{align}
&\tr_0(-\Delta-\mu)Q\label{eq:LT-rho-1D}\\
&\geq
K(1)\int_{\R}\Big(\big(\rho_0+\rho_{Q^{++}}(x)+\rho_{Q^{--}}(x)\big)^{3}
-(\rho_0)^{3}
-3(\rho_0)^{2}\,\big(\rho_{Q^{++}}(x)+\rho_{Q^{--}}(x)\big)\Big)\,dx\nonumber\\
&\quad+K'(1)\;\int_{\R}\frac{\sqrt\mu\,|k|}{\big(\sqrt\mu+|k|\big)\,{
\log\left(\frac{2\sqrt{\mu}+|k|}{|2\sqrt{\mu}-|k||}\right)}}{\big|\widehat{\rho_
{Q^{+-}}}(k)+\widehat{\rho_{Q^{-+}}}(k)\big|^2}\,dk,\nonumber
\end{align} 
with $\rho_0$ the constant density of the Fermi gas, given
by~\eqref{eq:relation_density_mu}.
\end{theorem}

Note the logarithmic divergence of the function in the denominator, at $|k|=2\sqrt{\mu}$. Hence the last term is \emph{not} bounded from below by $\int_{\R}|\rho_{Q^{+-}}+\rho_{Q^{-+}}|^2$.
In Section~\ref{sec:perturbation_theory_1D} below, we will see that, up to the value of the prefactors $K(1)$ and $K'(1)$, this bound is optimal. In particular, the right side of~\eqref{eq:LT-rho-1D} cannot be replaced by a constant times 
$\int_\R\delta\cT^{\rm sc}_\mu(\rho_Q)$.
In the limit $\mu\to0$, the inequality~\eqref{eq:LT-rho-1D} nevertheless reduces to the
one-dimensional Lieb-Thirring inequality~\eqref{eq:usual_LT-rho}.

\begin{remark}\rm
 Let $\phi\in L^2(\R^d,\C^q)$ be any normalized, smooth enough function.
Applying~\eqref{eq:LT-rho} or~\eqref{eq:LT-rho-1D} to
$Q_\pm=\pm\,|\Pi^\pm\phi\rangle\langle\Pi^\pm\phi|$ and using a simple convexity
argument, we obtain the following Sobolev-like inequality:
\begin{equation}
\int_{\R^d}\big|p^2-\mu\big|\; |\widehat{\phi}(p)|^2\geq K(d)\int_{\R^d}\delta
\cT^{\rm sc}_{\mu}\big(|\phi|^2\big)
\label{eq:Sobolev_mu}
\end{equation}
for all $\phi$ with $\int_{\R^d}|\phi|^2\leq 1$, and in any dimension $d\geq1$. 
\end{remark}

\subsection{Variation of energy in presence of an external potential}\label{sec:LT_potential_version}
In this section we study the dual version of our Lieb-Thirring
inequalities~\eqref{eq:LT-rho} and~\eqref{eq:LT-rho-1D}, expressed in terms of
an external potential $V$ (the variable dual to $\rho$). We will give a rigorous
meaning to~\eqref{eq:formal-LT-V}. 

Let $V$ be a real-valued function satisfying 
\begin{equation}
V\in L^{2}(\R^d)\cap L^{1+\tfrac{d}{2}}(\R^d)\quad \text{for $d\geq2$}
\label{eq:cond_V}
\end{equation}
or
\begin{multline}
V\in L^{3/2}(\R)+L^2(\R)\\
\quad\text{with}\quad \int_{\R}\left(1+\frac{\sqrt{\mu}+|k|}{\sqrt{\mu}\,|k|}
\log\left(\frac{2\sqrt{\mu}+|k|}{|2\sqrt{\mu}-|k||}\right)\right)|\widehat{V}
(k)|^2\,dk<\ii\quad \text{for $d=1$.}
\label{eq:cond_V_1D} 
\end{multline}
Under our assumption~\eqref{eq:cond_V}, the operator $-\Delta+V$ is self-adjoint
on $H^2(\R^d)$, by the Rellich-Kato Theorem~\cite{ReeSim2}. In dimension $d=1$,
our assumption~\eqref{eq:cond_V_1D} allows to define the Friedrichs self-adjoint realization of $-\Delta+V$, by the
KLMN theorem~\cite{ReeSim2}.

We now define
\begin{equation}
\boxed{Q_V:=\Pi_V^- - \Pi^- \quad \text{where}\quad
\Pi^-_V:=\1(-\Delta+V\leq\mu),}
\end{equation}
as well as $\Pi^+_V:=1-\Pi^-_V$.

\smallskip

\begin{remark}\rm
The real number $\mu$ could \emph{a priori} be an eigenvalue of $-\Delta+V$. Then, 
Theorems~\ref{thm:LT-V} and~\ref{thm:LT-V-1D} below hold exactly the same if $\Pi^-_V$ is replaced by
$\Pi^-_V+\delta$, where $\delta$ is an orthogonal projection whose range is contained in $\ker(-\Delta+V-\mu)$.
In dimension $d\geq3$, it is indeed known~\cite{KocTat-06} that,
under our assumption~\eqref{eq:cond_V} on $V$, the self-adjoint operator
$-\Delta+V$ has no positive eigenvalue, thus $\mu$ is not is the point spectrum of $-\Delta+V$. However, $\mu$ could be an eigenvalue of $-\Delta+V$ in dimensions $d=1$ and $d=2$.
\end{remark}

\smallskip

Similarly as in Definition~\ref{def:kinetic}, we can define a relative total
energy as follows.

\begin{definition}[Relative total energy]\label{def:total_energy}
Let $R$ be a bounded self-adjoint operator such that
$|-\Delta-\mu+V|^{1/2}\Pi^\pm_V R \Pi^\pm_V|-\Delta-\mu+V|^{1/2}$ are
trace-class. We define
\begin{equation}
\boxed{\tr_V(-\Delta-\mu+V)R:=\tr|-\Delta-\mu+V|^{\tfrac12}\big( \Pi^+_VR\Pi^+_V
- \Pi^-_VR\Pi^-_V\big)|-\Delta-\mu+V|^{\tfrac12}.}
\label{eq:def_total_energy}
\end{equation}
If $R$ is a bounded operator such that $\pm \Pi_V^\pm R\Pi_V^\pm\geq0$, then we
extend the previous expression by letting
$$\tr_{V}\left(-\Delta-\mu+V\right)R:=+\ii$$
whenever $|-\Delta-\mu+V|^{1/2}\Pi_V^+ R\Pi_V^+|-\Delta-\mu|^{1/2}$ or
$|-\Delta-\mu|^{1/2}\Pi_V^- R\Pi_V^-|-\Delta-\mu|^{1/2}$ is not trace-class.
\end{definition}

Since $Q_V$ is the difference of the two orthogonal projections $\Pi^-_V$
and $\Pi^-$, we have at the same time 
$$-\Pi^-\leq Q_V \leq \Pi^+\quad\text{and}\quad -\Pi_V^-\leq -Q_V \leq
\Pi_V^+.$$
Hence both $\tr_0(-\Delta-\mu)Q_V$ and $\tr_V(-\Delta-\mu+V)Q_V$ make sense by
Definitions~\ref{def:kinetic} and~\ref{def:total_energy}. With our
definitions we have
\begin{equation}
\tr_0(-\Delta-\mu)Q_V=\norm{Q_V\,|-\Delta-\mu|^{1/2}}_{\gS_2}^2
\label{eq:kinetic_Q_V}
\end{equation}
and
\begin{equation}
\tr_V(-\Delta-\mu+V)Q_V=-\norm{Q_V\,|-\Delta-\mu+V|^{1/2}}_{\gS_2}^2.
\label{eq:energy_Q_V}
\end{equation}
In the theorem below, we show that, under suitable assumptions on $V$, the two quantities~\eqref{eq:kinetic_Q_V} and~\eqref{eq:energy_Q_V} are finite and that 
\begin{equation}
\tr_V(-\Delta-\mu+V)Q_V=\tr_0(-\Delta-\mu)Q_V+\int_{\R^d}V\rho_{Q_V},
\label{eq:relation_traces}
\end{equation}
as expected. We also derive an estimate on $\tr_V(-\Delta-\mu+V)Q_V$ which is the dual version
of~\eqref{eq:LT-rho} for $d\geq2$. 

\begin{theorem}[Lieb-Thirring inequality, potential version,
$d\geq2$]\label{thm:LT-V}
Assume that $\mu\geq0$ and $d\geq2$. Let $V$ be a real-valued function in
$L^{2}(\R^d)\cap L^{1+{d}/{2}}(\R^d)$. 

\medskip

\noindent $\bullet$ Both $Q_V\,|-\Delta-\mu|^{1/2}$ and $Q_V\, |-\Delta-\mu+V|^{1/2}$ are
Hilbert-Schmidt operators, hence~\eqref{eq:kinetic_Q_V}
and~\eqref{eq:energy_Q_V} are finite.

\medskip
 
\noindent $\bullet$ The relative total energy $\tr_V(-\Delta-\mu+V)Q_V$ can be expressed as 
\begin{multline}
\tr_V(-\Delta-\mu+V)Q_V\\
=\min_{\substack{-\Pi^-\leq Q\leq \Pi^+\\
|-\Delta-\mu|^{1/2}Q^{\pm\pm}|-\Delta-\mu|^{1/2}\in\gS_1}}
\bigg(\tr_0(-\Delta-\mu)Q+\int_{\R^d}V(x)\,\rho_Q(x)\,dx\bigg).
\label{eq:variational}
\end{multline}
The minimum in this formula is attained for $Q=Q_V$. In particular,~\eqref{eq:relation_traces} holds true.

\medskip

\noindent $\bullet$ We have the inequality
\begin{multline}
\tr_V(-\Delta-\mu+V)Q_V\\ \geq
-L(d)\int_{\R^d}\left((V(x)-\mu)_-^{1+\tfrac{d}2}-\mu^{1+\tfrac{d}2}+\frac{2+d}
2\,\mu^{\tfrac{d}2}\,V(x)\right)\,dx
\label{eq:LT-V} 
\end{multline}
with
$$L(d)=\frac2{d+2}\left(\frac{d}{(d+2)\,K(d)}\right)^{\tfrac{d}2}\geq L_{\rm
sc}(d)$$
and where $K(d)$ is the optimal constant in~\eqref{eq:LT-rho}.
\end{theorem}

We recall that the semi-classical constant $L_{\rm sc}(d)$ is defined above
in~\eqref{eq:semiclassical-ctl}.

Let us comment on our result.
We can formally write 
\begin{equation}
\tr_V(-\Delta-\mu+V)Q_V\\
\;\text{``}=\text{''}\;-\tr\Big((-\Delta-\mu+V)_- +
(-\Delta-\mu)_-\Big)-\rho_0\int_{\R^d}V
\label{eq:formal_value_total_energy}
\end{equation}
where $\rho_0$ is the constant density of the translation-invariant state
$\Pi^-$, recalled in \eqref{eq:relation_density_mu}.
The first term of the right side is the formal difference between
the total (grand-canonical) energy of the Fermi gas in the
presence of the local perturbation $V$, and its total
(grand-canonical) energy in the translation-invariant
setting without any potential. The term $\rho_0\int_{\R^d}V$,
which makes sense under the additional assumption
that $V\in L^1(\R^d)$, is also the first order term obtained
by perturbation theory when the first term is
formally expanded in powers of $V$.

The semi-classical approximation of the right side of~\eqref{eq:formal_value_total_energy} is 
$$L_{\rm sc}(d)\int_{\R^d}\left((V(x)-\mu)_-^{1+\tfrac{d}2}-\mu^{1+\tfrac{d}2}+\frac{2+d}
2\,\mu^{\tfrac{d}2}\,V(x)\right)\,dx$$
and, up to the value of the multiplicative constant $L(d)$, it is precisely the right side of our estimate~\eqref{eq:LT-V}.
Our result therefore says that the variation of energy obtained by including the potential
$V$ in the
system is $O(1)$ in the thermodynamic limit,
and~\eqref{eq:LT-V} provides a precise estimate in terms
of the size of $V$. Since the term $\rho_0\int_{\R^d}V$ is obtained via
first-order perturbation theory, the semi-classical term on the right side of~\eqref{eq:LT-V} is therefore an estimate on the validity of the first order approximation.

In Section~\ref{sec:thermo-limit}, we will render the formal equality~\eqref{eq:formal_value_total_energy} more rigorous, by means of a thermodynamic limit argument. More precisely, we show in Theorem~\ref{thm:thermo-limit} that
\begin{multline}
\tr_V(-\Delta-\mu+V)Q_{V}\\=\lim_{L\to\ii}\left(-\tr_{L^2(C_L)}(-\Delta_L-\mu+V\1_{C_L})_-+\tr_{L^2(C_L)}(-\Delta_L-\mu)_- - \rho_0\int_{C_L}V  \right)
\label{eq:thermo-limit-intro}
\end{multline}
where $-\Delta_L$ is the Laplacian on a box $C_L=[-L/2,L/2)^d$ with periodic boundary conditions. This will also justify our definition of the total free energy. A tool to prove~\eqref{eq:thermo-limit-intro} is to derive a Lieb-Thirring inequality similar to~\eqref{eq:LT-V}, for a system living in a box with periodic boundary conditions (Theorem~\ref{thm:LT-V-box}).

The estimate~\eqref{eq:LT-V} follows from the density
estimate~\eqref{eq:LT-rho} and the variational principle~\eqref{eq:variational},
by noting that 
\begin{equation*}
\tr_0(-\Delta-\mu)Q+\int_{\R^d}V\,\rho_{Q}
\geq K(d)\int_{\R^d}\delta \cT^{\rm
sc}_{\mu}\big(\rho_{Q}\big)+\int_{\R^d}V\,\rho_{Q}
\end{equation*}
for all $-\Pi^-\leq Q\leq \Pi^+$, by Theorem~\ref{thm:LT-rho}.
Optimizing the right side with respect to $\rho_Q$ (keeping in mind that
$\rho_{Q}$ is pointwise bounded from below by $-\rho_0$),
yields~\eqref{eq:LT-V}.
Similarly, if we assume that~\eqref{eq:LT-V} is known, we can
derive~\eqref{eq:LT-rho} by choosing 
$$V=-\frac{\partial (\delta \cT^{\rm sc}_\mu)}{\partial\rho}(\rho_Q).$$
Hence~\eqref{eq:LT-V} and~\eqref{eq:LT-rho} are dual to each other.

\medskip

In dimension $d=1$, using the weaker lower bound~\eqref{eq:LT-rho-1D} on
$\tr_0(-\Delta-\mu)Q$, we can prove the following result. 

\begin{theorem}[Lieb-Thirring inequality, potential version,
$d=1$]\label{thm:LT-V-1D}
Assume that $\mu>0$ and $d=1$. Let $V\in L^{3/2}(\R)+L^2(\R)$ be a real-valued function such that
\begin{equation}
\int_{\R}\frac{\sqrt{\mu}+|k|}{\sqrt{\mu}\,|k|}
\log\left(\frac{2\sqrt{\mu}+|k|}{|2\sqrt{\mu}-|k||}\right)|\widehat{V}
(k)|^2\,dk<\ii.
\end{equation}
Then all the conclusions of Theorem~\ref{thm:LT-V} remain true, except
that~\eqref{eq:LT-V} must be replaced by
\begin{multline}
\tr_V(-\Delta-\mu+V)Q_V\geq
-L(1)\int_{\R}\left((V(x)-\mu)_-^{\tfrac32}-\mu^{\tfrac32}+\frac{3}2\,\mu^{
\tfrac{1}2}\,V(x)\right)\,dx\\
-L'(1)\int_{\R} \frac{\sqrt{\mu}+|k|}{\sqrt{\mu}\,|k|}
\log\left(\frac{2\sqrt{\mu}+|k|}{|2\sqrt{\mu}-|k||}\right)|\widehat{V}(k)|^2\,dk
\label{eq:LT-V-1D} 
\end{multline}
with
$${L}(1)=\frac23\left(\frac{1}{3K(1)}\right)^{1/2}\geq L_{\rm
sc}(1)\quad\text{and}\quad L'(1)=\frac1{4K'(1)}.$$ 
\end{theorem}

We will see in Section~\ref{sec:perturbation_theory_1D} below that it is not
possible to take $L'(1)=0$.

When $\mu\to0$ the inequalities~\eqref{eq:LT-V} and~\eqref{eq:LT-V-1D} reduce
again to the usual Lieb-Thirring inequality~\cite{LieThi-75,LieThi-76} which is
the dual version of~\eqref{eq:usual_LT-rho}:
\begin{equation}
0\leq \tr(-\Delta+V)_-\leq L(d)\int_{\R^d}V(x)_-^{1+\tfrac{d}2}\,dx.
\label{eq:usual_LT-rho-V}
\end{equation}

\begin{remark}\rm\label{rmk:trace-class}
In general $Q_V$ is \emph{not} a compact operator. Indeed, it
was shown by Pushnitski~\cite{Pushnitski-08} (see also~\cite{PutYaf-09}) that
the essential spectrum of $Q_V$ is 
$$\sigma_{\rm
ess}(Q_V)=\left[-\frac{\norm{S(\mu)-1}}{2},\frac{\norm{S(\mu)-1}}{2}\right]$$
where $S(\mu)$ is the scattering matrix associated to the pair
$(-\Delta,-\Delta+V)$. Hence $Q_V$ is not compact, unless $S(\mu)= 1$. Similarly, one does not expect, in general, 
that $(-\Delta-\mu)Q_V$ and $(-\Delta-\mu+V)Q_V$ are
trace-class, rendering Definitions~\ref{def:kinetic} and~\ref{def:total_energy} necessary.
\end{remark}

\begin{remark}[Relation with the spectral shift function]\label{rmk:SSF}
\rm The spectral shift function $\zeta_V(\lambda)$ formally satisfies~\cite{Yafaev-10}
\begin{equation}
\int_{-\ii}^\mu\zeta_V(\lambda)\,d\lambda=-
\tr_V(-\Delta-\mu+V)Q_V-\rho_0\int_{
\R^d}V.
\label{eq:link_SSF}
\end{equation}
If $V$ is in $L^1(\R^d)$ and satisfies the
assumptions~\eqref{eq:cond_V} or~\eqref{eq:cond_V_1D}, it is possible to define $\zeta_V$ 
as the (distributional) derivative of the right side with respect to $\mu$. 
\end{remark}

\subsection{Second-order perturbation theory and the 1D
case}\label{sec:perturbation_theory_1D}
In this section we compute the variation of energy when a potential $tV$ is
inserted in the system, to second-order in $t$. In particular we will show that
in the one-dimensional case $d=1$, the constant $L'(1)$ in the lower
bound~\eqref{eq:LT-V-1D} cannot be taken equal to 0. 

The following result, whose proof is sketched in
Section~\ref{sec:proof-second-order-perturb}, is well known
in the physics literature \cite{Peierls}.

\begin{theorem}[Second-order perturbation
theory]\label{thm:second-order-perturb}
Assume $d\geq1$. Let $V$ be a real-valued function in $L^{1}(\R^d)\cap L^\ii(\R^d)$ and
$\mu>0$. Then, using the notation of the previous section, 
\begin{equation}
\lim_{t\to0}\frac{\tr_{t V}(-\Delta-\mu+t V)Q_{t
V}}{t^2}=-\mu^{\tfrac{d}2-1}q\int_{\R^d}\Psi_d\left(\frac{|k|}{\sqrt{\mu}}
\right)\,|\widehat{V}(k)|^2\,dk,
\label{eq:second-order-result}
\end{equation}
where
\begin{align}
\Psi_d(|k|)&=(2\pi)^{-d}\int_{\substack{|p|^2\leq1\\ |p-k|^2\geq
1}}\frac{dp}{|p-k|^2-|p|^2}\nonumber\\[0,3cm]
&=\begin{cases}
\displaystyle\frac1{4\pi|k|}\log\left(\frac{2+|k|}{|2-|k||}\right)&\text{if
$d=1$,}\\[0,4cm]
\displaystyle\frac{|S^{d-2}|}{2|k|(2\pi)^{d}}\int_{0}^1\;\log\frac{2\sqrt{1-r^2}
+|k|}{\left|2\sqrt{1-r^2}-|k|\right|}\;r^{d-2}\,dr&\text{if $d\geq2$.}
\end{cases} 
\end{align}
In particular, when $d=1$ the constant $L'(1)$ appearing in~\eqref{eq:LT-V-1D}
must satisfy $L'(1)\geq q/(12\pi)$.
\end{theorem}

Our proof is valid under much weaker assumptions on the potential $V$, but we
have not tried to optimize this. 
The divergence at $|k|= 2\sqrt{\mu}$ of $\Psi_1(\cdot/\sqrt\mu)$ in dimension $d=1$ is
well-known, and it is sometimes called the \emph{Peierls
instability}~\cite[Sec. 4.3]{Peierls}. When the interactions among the particles are turned on, the system becomes unstable because of the large number of possible electron-hole excitations between the two points $\pm 2\sqrt{\mu}$. A macroscopic deformation of the system can sometimes lead to the opening of a gap at the Fermi points~\cite{Peierls,KenLie-87,LieNac-95,LieNac-95b,LieNac-95c,Voit-95}. In higher dimensions, the second-order response function $\Psi_d$ is bounded (this also follows from our
bound~\eqref{eq:LT-V}), but it is seen to have an infinite
derivative at $|k|=2\sqrt{\mu}$, a fact sometimes referred to as \emph{Migdal-Kohn anomaly}~\cite{Migdal-58,Kohn-59}.

We note that the semi-classical approximation to the left side of~\eqref{eq:second-order-result} satisfies
\begin{equation}
\lim_{t\to0}\frac{\displaystyle -\left((t
V-\mu)_-^{1+\tfrac{d}2}-\mu^{1+\tfrac{d}2}+\frac{2+d}{2}\,\mu^{\tfrac{d}2}\,t
V\right)}{t^2}=-\frac{d(d+2)}{8}\mu^{\tfrac{d}2-1}\, V^2.
\end{equation}
This proves that for $d=1$, it is \emph{not} possible to take $L'(1)=0$
in~\eqref{eq:LT-V-1D}, since the response function diverges at the
Fermi points $k=\pm2\sqrt{\mu}$ whereas the semi-classical second-order term stays
finite. A closer inspection of the constants reveals that 
$L'(1)\geq q/(12\pi)$ must hold, as stated in Theorem~\ref{thm:second-order-perturb}.

It is possible to calculate $\Psi_2$ and $\Psi_3$ exactly:
\begin{equation}
\Psi_2(|k|)=\frac1{8\pi}-\frac1{8\pi}\sqrt{\left(1-\frac{4}{|k|^2}\right)_+},
\label{eq:formula_Psi_2}
\end{equation}
\begin{equation}
\Psi_3(|k|)=\frac{1}{16\pi^2}\left(1 +\frac{1}{|k|}\left(1-\frac{|k|^2}4\right)\log\left(\frac{2+|k|}{\big|2-|k|\big|}\right)\right).
\label{eq:formula_Psi_3}
\end{equation}
Furthermore, we have the following recursion relation
\begin{equation}
\Psi_d(|k|) = \frac{| S^{d-3}|}{(2\pi)^{d-2}} \int_0^1 \frac{ r^{d-3}}{\sqrt{1-r^2}}  \Psi_2\left(\frac{|k|}{\sqrt{1-r^2}} \right) \,dr,\qquad \text{for $d\geq3$,}
\end{equation}
which implies that $\Psi_d$ is strictly decreasing for all $d\geq3$ (whereas for $d=2$, $\Psi_2$ is constant on $[0,2]$ and strictly decreasing on $[2,\ii)$). 
We deduce that
$$\|\Psi_d\|_{L^\ii(\R^+)}=\Psi_d(0)=\frac{|S^{d-2}|}{2(2\pi)^{d}}\int_{0}
^1\frac{r^{d-2}}{\sqrt{1-r^2}}\,dr=\frac{d(d+2)}{8q}L_{\rm sc}(d),\qquad \text{for $d\geq2$.}$$
Observe that in dimensions $d\geq2$, perturbation theory predicts the same value for the constant $L(d)$ as semi-classics does.
This is not so surprising since the largest constant is obtained if $\widehat{V}$ is supported close to $0$, hence $V$ is very spread out in $x$-space, which
puts us in the semi-classical regime.  

\begin{remark}\rm
 As is detailed in Section~\ref{sec:proof-second-order-perturb} below, the
second-order perturbation of the energy arises from the first-order term in the
expansion of $Q_{t V}$. This term is purely off-diagonal (the corresponding
$(Q_{t V})^{\pm\pm}$ vanish to first order in $t$). This emphasizes the fact
that the absence of a Lieb-Thirring inequality in 1D is due to a possible
divergence of the off-diagonal densities $\rho_{Q^{\pm\mp}}$ in Fourier space at $|k| = 2\sqrt{\mu}$.

The corresponding first-order density is proportional to $\widehat\Psi_d\ast\,V$. 
For potentials $V$ whose Fourier transform does not vanish at the Fermi surface, this density decays slowly in $x$-space, due to the lack of regularity of $\Psi_d$ at $|k|=2\sqrt\mu$.
\end{remark}

\subsection{A sharp inequality}\label{sec:Li-Yau}

We state and prove in this section a lower bound on the relative kinetic energy needed to banish all the particles in a domain, $\Omega$, from
the Fermi gas. This inequality involves the \emph{sharp constant} $K_{\rm sc}(d)$ and it is the positive density analogue of a result due to Li and Yau~\cite{LiYau-83}.

\begin{theorem}[A sharp estimate for the energy shift]\label{thm:Li-Yau}
Assume $d\geq1$. Let $\Omega$ be an open subset of $\R^d$ of finite measure. Let $\mu\geq0$ and denote, as before, $\Pi^-:=\1(-\Delta\leq\mu)$.
For any fermionic density matrix such that
$$0\leq\gamma\leq \1_{\R^d\setminus\Omega},\qquad \text{i.e.,}\quad 0\leq \pscal{f,\gamma f}_{L^2(\R^d)} \leq \int_{\R^d\setminus \Omega} |f(x)|^2 dx,\quad \text{$\forall f\in L^2(\R^d)$,}$$
and such that $|-\Delta-\mu|^{1/2}Q^{\pm\pm}|-\Delta-\mu|^{1/2}$ are trace-class, with $Q:=\gamma-\Pi^-$,
we have
\begin{equation}
\tr_0(-\Delta-\mu)Q\geq \frac2d \, K_{\rm sc}(d) \rho_0^{1+\tfrac{d}2} \, |\Omega|.
\label{eq:Li-Yau}
\end{equation}
\end{theorem}

The constant in this inequality is best possible. In dimension $d\geq2$, applying Theorem~\ref{thm:LT-rho} and using that $\rho_\gamma(x)=0$ on $\Omega$, we get 
\begin{multline*}
\tr_0(-\Delta-\mu)Q\geq K(d) \, \rho_0^{1+d/2} |\Omega|\\ + K(d)\int_{\R^d\setminus \Omega} \Big( \rho_\gamma^{1+d/2} - \rho_0^{1+d/2} - \frac{2+d}{d} \rho_0^{d/2} (\rho_\gamma-\rho_0) \Big) \,dx \,. 
\end{multline*}
Here $K(d)$ is not optimal but, on the other hand, the bound also quantifies the fact that $\rho_\gamma$ cannot be equal to $\rho_0$ close to the boundary because of the Dirichlet conditions.

\begin{proof}[Proof of Theorem~\ref{thm:Li-Yau}]
We have 
$$Q^{--}=\Pi^-(\gamma-\Pi^-)\Pi^-\leq \Pi^-\1_{\R^d\setminus\Omega}\Pi^--\Pi^-=-\Pi^-\1_{\Omega}\Pi^-.$$
Using that  $Q^{++}\geq0$, we get
\begin{align*}
\tr_0(-\Delta-\mu)Q&=\tr|-\Delta-\mu|^{1/2}(Q^{++}-Q^{--})|-\Delta-\mu|^{1/2}\\
&\geq \tr(-\Delta-\mu)_-^{1/2}\1_{\Omega}(-\Delta-\mu)_-^{1/2}\\
&=(2\pi)^{-d}|\Omega|\,\int_{\R^d}(|p|^2-\mu)_-\,dp.
\end{align*}
Recalling the definition of $K_{\rm sc}(d)$, we obtain the claim.
\end{proof}

\section{Kinetic energy inequalities: Proof of Theorems~\ref{thm:LT-rho} and~\ref{thm:LT-rho-1D}}\label{sec:proof-rho}
\subsection{Preliminaries}\label{sec:preliminaries}
In this section we state and prove some preliminary results that will be useful
in the proof of our main theorems. 

Throughout the paper we denote by $\cK=\gS_\ii$
(resp. $\cB$) the algebra of compact (resp. bounded) operators on
$L^2(\R^d,\C^q)$. The usual norm of bounded operators is simply denoted by
$\norm{\,\cdot\,}$. We also denote by $\gS_p$ (for $1\leq p<\ii$) the ideal of
compact operators $A$ on $L^2(\R^d,\C^q)$ such that $\tr\,|A|^p<\ii$, endowed
with its norm $\norm{A}_{\gS_p}=(\tr\,|A|^p)^{1/p}$. 

In order to simplify the statements below, we introduce the following Banach
space
\begin{multline}
\cX:=\bigg\{ Q=Q^*\in \cB\ :\ Q|-\Delta-\mu|^{1/2}\in\gS_2,\\
|-\Delta-\mu|^{1/2}Q^{\pm\pm}|-\Delta-\mu|^{1/2}\in\gS_1\bigg\},
\label{eq:space_X}
\end{multline}
endowed with its natural norm
\begin{multline}
\norm{Q}_\cX:=\norm{Q}+\norm{Q|-\Delta-\mu|^{1/2}}_{\gS_2}+\norm{|-\Delta-\mu|^{
1/2}Q^{++}|-\Delta-\mu|^{1/2}}_{\gS_1}\\
+\norm{|-\Delta-\mu|^{1/2}Q^{--}|-\Delta-\mu|^{1/2}}_{\gS_1}.
\label{eq:norm_X}
\end{multline}
For the sake of simplicity, we do not emphasize the dependence in $\mu$ in our
notation. The space $\cX$ has a natural weak topology which is the intersection
of the ones associated with the spaces appearing in the
definition~\eqref{eq:norm_X} of $\norm{\cdot}_\cX$.  Here $Q_n\wto Q$ in $\cX$
means $Q_n\wto_\ast Q$ weakly-$\ast$ in $\cB$, $Q_n|-\Delta-\mu|^{1/2}\wto
Q|-\Delta-\mu|^{1/2}$ weakly in $\gS_2$ and
$|-\Delta-\mu|^{1/2}Q_n^{\pm\pm}|-\Delta-\mu|^{1/2}\wto
|-\Delta-\mu|^{1/2}Q^{\pm\pm}|-\Delta-\mu|^{1/2}$ weakly-$\ast$ in $\gS_1$. The
unit ball of $\cX$ is weakly compact for this topology, by the Banach-Alaoglu
theorem.
The following convex subset of $\cX$ will play an important role:
\begin{equation}
\cK:=\{Q\in\cX\ :\ -\Pi^-\leq Q\leq \Pi^+\}. 
\label{eq:convex-set}
\end{equation}

Our first result deals with the continuity of the map $Q\in\cX\mapsto \rho_Q$ in
$L^1_{\rm loc}(\R^d)$.

\begin{lemma}[Operators in $\cX$ are locally
trace-class]\label{lem:local-trace-class}
We assume that $\mu\geq0$ and $d\geq1$. Let $Q$ be a self-adjoint bounded
operator in $\cX$. Then, for every bounded function $\eta$ of compact support,
there exists a constant $C_\eta$ such that 
\begin{equation}
\norm{\eta Q\eta}_{\gS_1}\leq C_\eta \norm{Q}_\cX.
\end{equation}
Hence $Q$ is locally trace class and $\rho_Q$ is well-defined in $L^1_{\rm
loc}(\R^d)$. 

Furthermore, the map $Q\in\cX\mapsto\eta Q\eta\in\gS_1$ is weakly continuous: If
we have a sequence $\{Q_n\}$ such that $Q_n\wto Q$ weakly in $\cX$, then $\eta
Q_n\eta\to\eta Q\eta$ strongly in $\gS_1$. In particular, $\rho_{Q_n}\to\rho_Q$
strongly in $L^1_{\rm loc}(\R^d)$.
\end{lemma}

\begin{proof}
We consider the spectral projection $\Pi_1:=\1\big(-\Delta\leq\max(1,2\mu)\big)$, which localizes in a ball containing strictly the Fermi surface, and we denote by $\Pi_2=1-\Pi_1$ its complement. Then we write
$Q=\sum_{k,\ell=1,2}\Pi_kQ\Pi_\ell$ and estimate each term separately. 
We start with $\Pi_2 Q\Pi_2$ which we treat as follows
$$\eta\Pi_2 Q\Pi_2\eta=\eta\frac{\Pi_2}{|-\Delta-\mu|^{\frac12}}\;
|-\Delta-\mu|^{\frac12}Q^{++}|-\Delta-\mu|^{\frac12}\;\frac{\Pi_2}{
|-\Delta-\mu|^{\frac12}}\eta$$
where we have used that $\Pi_2=\Pi_2\Pi^+$. Since $\eta$ and
$\Pi_2|-\Delta-\mu|^{-1/2}$ are bounded, it is clear that the previous operator
is trace-class. Furthermore, we know that if $T_n\wto T$ weakly-$\ast$ in
$\gS_1$ and $K$ is compact, then $KT_nK\to KTK$ strongly in $\gS_1$. Hence the
weak continuity follows from the fact that $\eta\Pi_2|-\Delta-\mu|^{-1/2}$ is
compact.
For $\Pi_1 Q\Pi_2$, we write similarly
$$\eta\Pi_1
Q\Pi_2\eta=\eta\Pi_1\;Q|-\Delta-\mu|^{\frac12}\;\frac{\Pi_2}{|-\Delta-\mu|^{
\frac12}}\eta$$
and use that $\eta\Pi_1\in\gS_2$, $Q|-\Delta-\mu|^{1/2}\in\gS_2$ and
$\Pi_2|-\Delta-\mu|^{-1/2}\eta\in\cK$. The argument is then similar as before. 
Finally, for $\Pi_1 Q\Pi_1$, we simply use that $\eta\Pi_1\in\gS_2$ and that $Q$
is bounded. The rest follows.
\end{proof}

\begin{remark}\rm 
The previous proof does not use the fact that
$|-\Delta-\mu|^{1/2}Q^{--}|-\Delta-\mu|^{1/2}$ is trace-class. 
\end{remark}

The following says that finite rank operators are dense in $\cX$ in the
appropriate sense.

\begin{lemma}[Density of finite rank operators]\label{lem:density}
For every $Q\in\cX$, there exists a sequence $Q_n\in\cX$ of finite rank
operators, such that $(-\Delta)Q_n\in\cB$ and
\begin{itemize}
\item $Q_n\to Q$ strongly (that is, $Q_n f\to Qf$ strongly in $L^2(\R^d,\C^q)$
for every fixed $f\in L^2(\R^d,\C^q)$);

\smallskip

\item $\displaystyle\lim_{n\to\ii}\norm{(Q_n-Q)|-\Delta-\mu|^{1/2}}_{\gS_2}=0$;

\smallskip

\item
$\displaystyle\lim_{n\to\ii}\norm{|-\Delta-\mu|^{1/2}(Q_n-Q)^{\pm\pm}
|-\Delta-\mu|^{1/2}}_{\gS_1}=0$;

\smallskip

\item $\rho_{Q_n}\to\rho_Q$ strongly in $L^1_{\rm loc}(\R^d)$.
\end{itemize}
Furthermore, if $Q$ belongs to the convex set $\cK$ defined
in~\eqref{eq:convex-set}, then $Q_n$ can be chosen in $\cK$ for all $n$.
\end{lemma}

Note that operators $Q\in\cX$ are not all compact, hence in general
$\norm{Q_n-Q}\not\rightarrow0$.

\begin{proof}
We start by approximating $Q$ by a sequence of Hilbert-Schmidt
operators $Q_n$, with $(Q_n)^{\pm\pm}\in\gS_1$. Let us define the orthogonal projection
$P_n:=\1\big(1/n\leq|-\Delta-\mu|\leq n\big)$, which localizes in momentum space
away from the Fermi surface and from infinity. We now define $Q_n:=P_nQP_n$. It
is easy to verify that $Q_n$ is a Hilbert-Schmidt operator by choice of $P_n$
and, similarly that $(Q_n)^{\pm\pm}$ are trace-class. We have $P_n\to1$ strongly
in $L^2(\R^d)$. Since $Q$ is bounded, we obtain that $Q_n\to Q$ strongly. Also,
it is well-known that when $A\in\gS_p$ for some $1\leq p\leq\ii$, then
$P_nAP_n\to A$ strongly in $\gS_p$. In particular, we have that
$P_nQ|-\Delta-\mu|^{1/2}P_n=Q_n|-\Delta-\mu|^{1/2}\to Q|-\Delta-\mu|^{1/2}$
strongly in $\gS_2$, using that $P_n$ commutes with $|-\Delta-\mu|^{1/2}$. The
convergence of the trace-class terms is similar, and the strong convergence of
$\rho_{Q_n}$ in $L^1_{\rm loc}(\R^d)$ follows from
Lemma~\ref{lem:local-trace-class}.
Finally, we note that, since $P_n$ commutes with $\Pi^-$, $Q_n$ belongs to $\cK$
for all $n$, whenever $Q$ is itself in $\cK$. 

For a proof that $Q_n$ can itself
be approximated by smooth finite rank operators in $\cK$, see~\cite[Theorem
6]{HaiLewSer-09}.
\end{proof}

\subsection{Proof of Theorem~\ref{thm:LT-rho}: kinetic Lieb-Thirring inequality for $d\geq2$}\label{sec:proof-LT-rho}
In this section we prove Theorem~\ref{thm:LT-rho} for $\mu>0$ (the case $\mu=0$
is well-known~\cite{LieThi-75,LieThi-76,LieSei-09}).
Replacing $Q$ by $U_\mu QU_\mu^*$ where $(U_\mu f)(x):=\mu^{d/4}f(\sqrt{\mu}x)$,
it is easy to verify that~\eqref{eq:LT-rho} follows from the case $\mu=1$, which
we will assume throughout the proof. Also we assume for simplicity that the
number of spin states is $q=1$ but the proof for the general case is identical. Finally, since the semi-classical energy
difference $\delta \cT^{\rm sc}_1$ (defined in~\eqref{eq:def_delta_T}) is
non-negative, the right side of our Lieb-Thirring inequality \eqref{eq:LT-rho} is
lower semi-continuous with respect to $\rho_Q$. This shows that, by
Lemma~\ref{lem:density}, we can prove~\eqref{eq:LT-rho} assuming that $Q$ is a
smooth-enough finite rank operator, and deduce the general case by density.

Recall our notation $Q^{--}=\Pi^- Q\Pi^-$, $Q^{++}=\Pi^+ Q\Pi^+$ and so on. We
will estimate the density arising from each term separately. The constraint
$-\Pi^-\leq Q\leq \Pi^+$ is equivalent to $Q^2\leq Q^{++}-Q^{--}$.

\subsubsection*{Step 1. Estimate on $Q^{\pm\pm}$}
In order to bound the density arising from the diagonal terms, we will use the
following generalization of the Lieb-Thirring inequality.

\begin{lemma}[Lieb-Thirring inequality with positive Fermi
level]\label{lem:generalized_LT}
Assume $d\geq1$. Let $0\leq\gamma\leq1$ be a self-adjoint operator on
$L^2(\R^d)$ such that $|\Delta+1|^{1/2}\gamma|\Delta+1|^{1/2}$ is trace-class.
Then $\gamma$ is locally trace-class and its density satisfies
\begin{equation}
\tr|\Delta+1|^{1/2}\gamma|\Delta+1|^{1/2}\geq \hat{K}(d)\int_{\R^d} \delta
\cT^{\rm sc}_1\big(\rho_\gamma(x)\big)\,dx
\label{eq:generalized_LT}
\end{equation}
where 
$$\delta \cT^{\rm
sc}_1(\rho)=(\rho_0+\rho)^{1+\tfrac2d}-(\rho_0)^{1+\tfrac2d}-\frac{2+d}{d}
(\rho_0)^{\tfrac2d}\rho$$ 
with $\rho_0=|S^{d-1}|\,(2\pi)^{d}/d$,
and where $\hat{K}(d)$ is a positive constant depending only on $d$.
\end{lemma}

The proof of Lemma~\ref{lem:generalized_LT} follows ideas of
Rumin~\cite{Rumin-11}. Note that Lemma~\ref{lem:generalized_LT} is also valid in
dimension $d=1$.

\begin{proof}
We follow a recent method of Rumin~\cite{Rumin-11}.
We introduce the spectral projection $P_e:=\1\big(|\Delta+1|\geq e\big)$ in such a way
that we have the layer cake representation
$$|\Delta+1|=\int_{0}^\ii P_e\, de.$$
Let now $0\leq\gamma\leq1$ be a smooth-enough finite rank operator. We have
\begin{equation}
\tr|\Delta+1|\gamma=\int_0^\ii de\; \tr(P_e\gamma P_e)=\int_0^\ii
de\;\int_{\R^d} \rho_e(x)\,dx
\label{eq:estim_trace_Rumin} 
\end{equation}
where $\rho_e$ is the density of the finite-rank operator $P_e\gamma P_e$. We
now consider a bounded set $A\subset \R^d$  and estimate
\begin{align}
\int_A \rho_e(x)\,dx=\tr(\1_A P_e\gamma P_e)&=\norm{\1_A
P_e\gamma^{1/2}}_{\gS_2}^2\nonumber\\
&\geq \left(\norm{\1_A \gamma^{1/2}}_{\gS_2}-\norm{\1_A P_e^\perp
\gamma^{1/2}}_{\gS_2}\right)^2_+\nonumber\\
&= \left(\left(\int_A\rho\right)^{1/2}-\norm{\1_A P_e^\perp
\gamma^{1/2}}_{\gS_2}\right)^2_+.\label{eq:estim_rho_Rumin}
\end{align}
Note that, since $\norm{\gamma}\leq1$,
\begin{equation*}
\norm{\1_A P_e^\perp \gamma^{1/2}}_{\gS_2}^2=\tr(\1_A P_e^\perp \gamma
P_e^\perp\1_A)
\leq \norm{\1_A P_e^\perp}_{\gS_2}^2\norm{\gamma}\leq |A|\; f(e)
\end{equation*}
where
\begin{equation*}
f(e):= (2\pi)^{-d}\;\Big|\big\{\big|p^2-1\big|\leq
e\big\}\Big|=\frac{|S^{d-1}|}{d\,(2\pi)^{d}}\;\left(
(1+e)^{d/2}-(1-e)_+^{d/2}\right).
\end{equation*}
Taking $A$ to be a ball of radius $\epsilon\to0$ centered at $x$, we obtain
from~\eqref{eq:estim_rho_Rumin} the pointwise estimate
$$\rho_e(x)\geq \left(\sqrt{\rho(x)}-\sqrt{f(e)}\right)^2_+.$$
We may now insert this in~\eqref{eq:estim_trace_Rumin} and obtain 
\begin{equation*}
\tr|\Delta+1|\gamma\geq  \int_{\R^d}\,dx\;\int_0^\ii \,de
\left(\sqrt{\rho(x)}-\sqrt{f(e)}\right)^2_+=\int_{\R^d}R_d\big(\rho(x)\big)\,dx
\end{equation*}
with
\begin{equation}
R_d(\rho):=\int_0^\ii \left(\sqrt{\rho}-\sqrt{f(e)}\right)^2_+\, de.
\label{eq:def_Rumin}
\end{equation}
At zero we have
$$R_d(\rho)\underset{\rho\to0}{\sim} \frac{(2\pi)^d}{6|S^{d-1}|}\rho^2.$$
At infinity, one can compute that
$$R_d(\rho)\underset{\rho\to\ii}{\sim}\frac{d}{d+4}\;K_{\rm sc}(d)\;\rho^{1+2/d}.$$
Hence there is a constant $\hat{K}(d)$ such that~\eqref{eq:generalized_LT}
holds.

We have written the proof for a smooth enough finite-rank operator. The
general case follows from an approximation argument based on
Lemma~\ref{lem:density}. This completes the proof of
Lemma~\ref{lem:generalized_LT}.
\end{proof}

Since $|\Delta+1|^{1/2}Q^{\pm\pm}|\Delta+1|^{1/2}$ is trace-class by assumption
and $0\leq Q^{++}\leq \Pi^+\leq1$, $-1\leq-\Pi^-\leq Q^{--}\leq 0$, we
immediately obtain from Lemma~\ref{lem:generalized_LT} that
\begin{equation}
\pm\tr|\Delta+1|^{1/2}Q^{\pm\pm}|\Delta+1|^{1/2}\geq \hat{K}(d)\int_{\R^d} \delta
\cT^{\rm sc}_1\big(\rho_{Q^{\pm\pm}}(x)\big)\,dx.
\label{eq:final_estim_diag}
\end{equation}
It therefore remains to estimate the density arising from the off-diagonal terms
$Q^{+-}$ and $Q^{-+}$.

\medskip

\subsubsection*{Step 2. Estimate on $Q^{\pm\mp}$}
It is enough to consider $Q^{-+}=\Pi^-\, Q\Pi^+$, since $\rho_{Q^{+-}}+\rho_{Q^{-+}}=2\Re\,\rho_{Q^{-+}}$.
In order to estimate the density $\rho_{Q^{-+}}$ in the whole space $\R^d$, we
argue by duality and write
\begin{align}
\int_{\R^d}V\rho_{Q^{-+}}&=\tr(V\Pi^- Q\Pi^+)\nonumber\\
&=\tr\left( \frac{\Pi^+}{|\Delta+1|^{1/4}}V\frac{\Pi^-}{|\Delta+1|^{1/4}}\;
|\Delta+1|^{1/4}Q|\Delta+1|^{1/4}\right).\label{eq:form_tr_off_diagonal} 
\end{align}
This calculation is valid if $V$ is bounded and compactly supported,
since $Q$ is a smooth-enough finite-rank operator. Using Schwarz's
inequality and that $Q^2\leq Q^{++}-Q^{--}$, we have
\begin{align*}
\norm{|\Delta+1|^{1/4}Q|\Delta+1|^{1/4}}_{\gS_2}&\leq
\norm{|\Delta+1|^{1/2}Q}_{\gS_2}\\
&=\left(\tr|\Delta+1|Q^2\right)^{1/2}\leq \left(\tr_0(-\Delta-1)Q\right)^{1/2}. 
\end{align*}
Returning to~\eqref{eq:form_tr_off_diagonal}, we obtain 
$$\left|\int_{\R^d}V\rho_{Q^{-+}}\right| \leq
\norm{\frac{\Pi^+}{|\Delta+1|^{1/4}}V\frac{\Pi^-}{|\Delta+1|^{1/4}}}_{\gS_2}
\left(\tr_0(-\Delta-1)Q\right)^{1/2}.$$
We now compute
\begin{align}
 \norm{\frac{\Pi^+}{|\Delta+1|^{1/4}}V\frac{\Pi^-}{|\Delta+1|^{1/4}}}_{\gS_2}
^2&=(2\pi)^{-d}\int_{\substack{|p|^2\leq 1\\
|q|^2\geq1}}\frac{|\widehat{V}(p-q)|^2}{\big(1-|p|^2\big)^{1/2}
\;\big(|q|^2-1\big)^{1/2}}\,dp\,dq\nonumber\\
&=(2\pi)^{-d}\int_{\R^d}|\widehat{V}(k)|^2\, \Phi_d(|k|)\,dk
\label{eq:HS_off_diagonal_Phi_d}
\end{align}
where
\begin{equation}
\Phi_d(|k|):=\int_{\substack{|p|\leq 1\\
|p-k|\geq1}}\frac{dp}{\big(1-|p|^2\big)^{1/2}\;\big(|p-k|^2-1\big)^{1/2}}.
\label{eq:def_Phi_d} 
\end{equation}

We will use the following fundamental result.

\begin{lemma}\label{lem:Phi_bounded}
For $d\geq2$, the function $\Phi_d$ is bounded on $\R^d$. The function $\Phi_1$
is \emph{not} bounded in a neighborhood of $k=2$.
\end{lemma}

For clarity the proof of Lemma~\ref{lem:Phi_bounded} is postponed until the end
of the proof of Theorem~\ref{thm:LT-rho}. We deduce that
$$\norm{\frac{\Pi^+}{|\Delta+1|^{1/4}}V\frac{\Pi^-}{|\Delta+1|^{1/4}}}_{\gS_2}
\leq (2\pi)^{-d/2}\norm{\Phi_d}_{L^\ii(\R^+)}^{1/2}\norm{V}_{L^2(\R^d)},$$
which leads to the estimate
\begin{equation}
\int_{\R^d}|\rho_{Q^{-+}}|^2\leq
(2\pi)^{-d}\norm{\Phi_d}_{L^\ii(\R^+)}\;\tr_0(-\Delta-1)Q.
\label{eq:final_estim_off_diag_1} 
\end{equation}

We can extend $\delta\cT^{\rm sc}_1$ for $\rho\leq-\rho_0$ linearly
as follows
$$\delta\cT^{\rm
sc}_1(\rho)=(\rho_0+\rho)_+^{1+\tfrac2d}-(\rho_0)^{1+\tfrac2d}-\frac{2+d}{d}
(\rho_0)^{\tfrac2d}\rho.$$
The function is now convex on the whole line $\R$.
Note that for $d\geq2$, we have $|\rho|^2\geq c\,\delta\cT^{\rm sc}_1(\rho)$ for
all $\rho$, hence we have also shown that 
\begin{equation}
c\int_{\R^d}\delta\cT^{\rm sc}_1\big(\rho_{Q^{-+}}\big)\leq \tr_0(-\Delta-1)Q
\label{eq:final_estim_off_diag_2} 
\end{equation}
for a small enough constant $c>0$.

\begin{remark}\rm
Modifying the previous proof by using
$\Pi^+|\Delta+1|^{-\alpha}\,V\,\Pi^-|\Delta+1|^{-\alpha}$ with an appropriate
power $\alpha$ and $\gS_p$ norms, one can show that 
\begin{equation}
\int_{\R^d}|\rho_{Q^{-+}}|^p+\int_{\R^d}|\rho_{Q^{+-}}|^p\leq C(d,p)\;
\tr_0(-\Delta-1)Q
\label{eq:estim_all_p_all_d}
\end{equation}
holds for all $2\leq p<\ii$ and all $d\geq2$.
\end{remark}

\smallskip

\subsubsection*{Conclusion}
Putting~\eqref{eq:final_estim_diag} and~\eqref{eq:final_estim_off_diag_2}
together, we deduce by convexity of $\delta\cT^{\rm sc}_1$ that
\begin{align*}
\tr_0(-\Delta-1)Q&\geq c\int_{\R^d} \delta\cT^{\rm
sc}_1\big(\rho_{Q^{++}}\big)+\delta\cT^{\rm
sc}_1\big(\rho_{Q^{--}}\big)+\delta\cT^{\rm
sc}_1\big(\rho_{Q^{-+}}\big)+\delta\cT^{\rm sc}_1\big(\rho_{Q^{+-}}\big)\\
&\geq 4c\int_{\R^d} \delta\cT^{\rm
sc}_1\left(\frac{\rho_{Q^{++}}+\rho_{Q^{--}}+\rho_{Q^{-+}}+\rho_{Q^{+-}}}
4\right)\\
&\geq K(d)\int_{\R^d} \delta\cT^{\rm sc}_1\left(\rho_{Q}\right)
\end{align*}
for a small enough constant $K(d)>0$. This completes the proof of
Theorem~\ref{thm:LT-rho}.\qed

\medskip

\begin{remark}\rm 
Our method yields explicit values for the constant $K(d)$ appearing in the statement of
Theorem~\ref{thm:LT-rho}, see~\cite{FraLewLieSei-11}. 
\end{remark}

\medskip

It remains to prove Lemma~\ref{lem:Phi_bounded}. 

\begin{proof}[Proof of Lemma~\ref{lem:Phi_bounded}]
To study $\Phi_d(k)$ for $d\geq2$, we make the decomposition
$p=(s,p_\perp)$ with $s=p\cdot \vec{k}$ and find
\begin{align}
\Phi_d(k)&=\int_{\substack{s^2+|p_\perp|^2\leq 1\\
(s-k)^2+|p_\perp|^2\geq1}}\frac{ds\,dp_\perp}{\big(1-s^2-|p_\perp|^2\big)^{1/2}
\;\big((s-k)^2+|p_\perp|^2-1\big)^{1/2}}\nonumber\\
&=|S^{d-2}|\int_{\substack{s^2+r^2\leq 1\\ (s-k)^2+r^2\geq1\\
r\geq0}}\frac{ds\;r^{d-2}dr}{\big(1-s^2-r^2\big)^{1/2}\;\big((s-k)^2+r^2-1\big)^
{1/2}}.\label{eq:1st_formula_Phi}
\end{align}
For $k\geq2$ and $|s|\leq1$, it is clear that $(s-k)^2+r^2\geq1$. The
integration domain is therefore independent of $k$ when $k\geq2$. It is then
easy to verify that $\Phi_d$ is decreasing and continuous on $(2,\ii)$. Hence we
only have to prove that it is bounded in a neighborhood of $[0,2]$. 
Next we note that
\begin{equation}
\Phi_d(k)
=|S^{d-2}|\int_0^1\frac{r^{d-2}}{\sqrt{1-r^2}}\Phi_1\left(\frac{k}{\sqrt{1-r^2}}
\right)\,dr\label{eq:formula_Phi} 
\end{equation}
where we recall that
\begin{equation*}
\Phi_1(x)=\int_{\substack{v^2\leq 1\\
(v-x)^2\geq1}}\frac{dv}{\sqrt{1-v^2}\sqrt{(v-x)^2-1}}=\int_{-1}^{\min(1,-1+x)}
\frac{dv}{\sqrt{1-v^2}\sqrt{(v-x)^2-1}}.
\end{equation*}
It is an exercise to verify that $\Phi_1$ is a continuous function on
$\R^+\setminus\{2\}$ (in particular it has a finite limit at $x=0$), and that
$$\Phi_1(x)\underset{x\to2}{\sim}-\frac12\log|x-2|,\qquad
\Phi_1(x)\underset{x\to\ii}{\sim}\frac\pi x\,.$$
Using that, for instance,
$$\Phi_1(x)\leq \frac{C}{|x-2|^{\tfrac14}},$$
and letting $u=\sqrt{1-r^2}$, we obtain
\begin{equation*}
\Phi_d(k)\leq C\int_0^1\frac{du}{\sqrt{1-u^2}}\frac{u}{|k-2u|^{\frac14}}\leq
C\left(\int_0^1\frac{du}{(1-u)^{\frac34}}\right)^{\frac23}\left(\int_0^1\frac{du
}{|k-2u|^{\frac34}}\right)^{\frac13}
\end{equation*}
by H\"older's inequality.
The right side is bounded with respect to $k$, hence $\Phi_d$ is uniformly
bounded for $d\geq2$. By a similar proof one can verify that $\Phi_d$ is also a
continuous function on $\R^+$. This completes the proof of
Lemma~\ref{lem:Phi_bounded}.
\end{proof}

\begin{remark}\rm 
It is possible to calculate the exact maximum value of $\Phi_d$, which might be
interesting for physical applications~\cite{FraLewLieSei-11}. 
Starting from~\eqref{eq:1st_formula_Phi} and letting $t=r^2$, we obtain
\begin{align*}
\Phi_3(k)&=\pi\int_{\substack{s^2+t\leq 1\\ (s-k)^2+t\geq1\\
r\geq0}}\frac{ds\;dt}{\big(1-s^2-t\big)^{1/2}\;\big((s-k)^2+t-1\big)^{1/2}}\\
&=\pi\int_{\min(1,-1+k)}^{\min(1,k/2)} ds \int_{1-(s-k)^2}^{1-s^2}
\frac{dt}{\big(1-s^2-t\big)^{1/2}\big((s-k)^2+t-1\big)} \\
& \qquad\qquad  + \pi\int_{-1}^{\min(1,-1+k)} ds \int^{1-s^2}_0
\frac{dt}{\big(1-s^2-t\big)^{1/2}\big((s-k)^2+t-1)^{1/2}}.
\end{align*}
In order to compute these integrals we use the fact that
\begin{equation}
 \label{eq:primitive}
\int \frac{dt}{\sqrt{(a-t)(t-b)}} = -2 \arcsin\sqrt\frac{a-t}{a-b}
\end{equation}
whenever $a>b$. We find for $0\leq k\leq 2$, with $s=-1+ku$, 
\begin{equation}
\Phi_3(k) =\pi^2  + 2\,\pi\,k
\left(\int_{0}^{1}\arcsin\sqrt{u\frac{2-ku}{2+k(1-2u)}}\,du-\frac\pi4\right).
\label{eq:change_variable} 
\end{equation}
We have 
$$\frac{d}{dk}\left(\frac{2-ku}{2+k(1-2u)}\right)=\frac{2(u-1)}{(2+k(1-2u))^2}
\leq0$$
hence the function 
$$f(k):=\int_{0}^{1}\arcsin\sqrt{u\frac{2-ku}{2+k(1-2u)}}\,du-\frac\pi4$$
appearing in the parenthesis in \eqref{eq:change_variable} is decreasing with
respect to $k$, by monotonicity of $t\mapsto \arcsin(\sqrt{t})$. Its value at
$k=0$ is
$$f(0)=\int_{0}^{1}\arcsin\sqrt{u}\,du-\frac\pi4=0.$$
Therefore, $f(k)\leq0$ for $0<k\leq2$. Now
$\Phi_3'(k)=2\pi(f(k)+k\,f'(k))\leq0$, hence $\Phi_3$ is decreasing on $[0,2]$.
Since we know already that $\Phi_3$ also decreases on $[2,\infty)$, we conclude
that
$$\max_{\R^+}\Phi_3=\Phi_3(0)=\pi^2.$$

Similarly as in~\eqref{eq:formula_Phi}, we can express $\Phi_d$ in terms of $\Phi_3$ for $d\geq4$ by assuming, for instance, $\vec{k}=k(1,0,...,0)$ and writing $p=(q,p_\perp)$ with $q\in\R^3$ and $p_\perp\in \R^{d-3}$. We obtain the recursion relation
\begin{equation}
\Phi_d(k)=|S^{d-4}|\int_0^1\sqrt{1-r^2}\;\Phi_3\left(\frac{k}{\sqrt{1-r^2}}\right)r^{d-4}\,dr, \qquad \text{for $d\geq4$.}
\label{eq:relation_Phi_d_Phi_3}
\end{equation}
As we have shown above that $\Phi_3$ is strictly decreasing, this proves that $\Phi_d$ is also strictly decreasing, hence that
$$\max_{\R^+}\Phi_d=\Phi_d(0)=\pi^2|S^{d-4}|\int_0^1\sqrt{1-r^2}\,r^{d-4}\,dr, \qquad \text{for $d\geq4$.}$$
\end{remark}

\subsection{Proof of Theorem~\ref{thm:LT-rho-1D}: kinetic Lieb-Thirring inequality for $d=1$}\label{sec:proof-LT-rho-1D}
In the one-dimensional case $d=1$, the same proof as that of
Theorem~\ref{thm:LT-rho} leads to a bound of the form
\begin{multline}
\tr_0(-\Delta-1)Q\geq K(1)\int_{\R}\delta\cT^{\rm
sc}\big(\rho_{Q^{++}}+\rho_{Q^{--}}\big)\\+K''(1)\int_{\R}\frac{\big|\widehat{
\rho_{Q^{+-}}}(k)+\widehat{\rho_{Q^{-+}}}(k)\big|^2}{\Phi_1(|k|)}\,dk.
\label{eq:LT-rho-1D-proof}
\end{multline}
Using the known behavior of $\Phi_1$ at $|k|=2$ and when $|k|\to\ii$, one can
state this bound as in~\eqref{eq:LT-rho-1D}.\qed

\section{Potential inequalities: Proof of Theorems~\ref{thm:LT-V},
\ref{thm:LT-V-1D} and~\ref{thm:second-order-perturb}}\label{sec:proof-V}
For the standard Lieb-Thirring inequalities~\cite{LieThi-75,LieThi-76} (the case where $\mu=0$), there is a duality between the kinetic energy and the potential versions of the inequality, and this duality is based on a variational principle for sums of eigenvalues. A similar variational principle is also valid inside the continuous spectrum and can be used to deduce Theorems~\ref{thm:LT-V} and~\ref{thm:LT-V-1D} from Theorems~\ref{thm:LT-rho} and~\ref{thm:LT-rho-1D}.

\begin{theorem}[Variational principle]\label{thm:variational_principle}
Let $\mu\geq0$ and $V$ be a real-valued function. Assume that 
$V\in L^{2}(\R^d)\cap L^{1+{d}/{2}}(\R^d)$ when $d\geq2$, and that
$V\in L^{3/2}(\R)+L^2(\R)$ with
\begin{equation}
\int_{\R}\frac{\sqrt{\mu}+|k|}{\sqrt{\mu}\,|k|}
\log\left(\frac{2\sqrt{\mu}+|k|}{|2\sqrt{\mu}-|k||}\right)|\widehat{V}
(k)|^2\,dk<\ii
\end{equation}
when $d=1$.
Then both $Q_V\,|-\Delta-\mu|^{1/2}$ and $Q_V\, |-\Delta-\mu+V|^{1/2}$ are
Hilbert-Schmidt operators and
\begin{equation}
\tr_V(-\Delta-\mu+V)Q_V
=\inf_{Q\in\cK}
\bigg(\tr_0(-\Delta-\mu)Q+\int_{\R^d}V(x)\,\rho_Q(x)\,dx\bigg),
\label{eq:variational-proof}
\end{equation}
where $\cK$ was defined in~\eqref{eq:convex-set}. The infimum in~\eqref{eq:variational-proof} is attained for $Q=Q_V$.
\end{theorem}

To motivate this theorem, we explain its analogue for self-adjoint finite-dimensional matrices $A$ and $B$. The starting point is the well-known formula for the sum of eigenvalues~\cite[Thm. 12.1]{LieLos-01}
$$-\tr(A+B)_-=\inf_{0\leq \gamma\leq 1}\tr(A+B)\gamma.$$
Introducing the spectral projection $\Pi^-=\1(A\leq0)$ onto the negative spectral subspace of $A$ and changing variables, $\gamma=Q+\Pi^-$, we obtain
$$-\tr(A+B)_-=\inf_{-\Pi^-\leq Q\leq 1-\Pi^-}\tr(A+B)Q+\tr(A+B)\Pi^-,$$
that is, with the notation $\Pi^-_B=\1(A+B\leq0)$,
$$\tr(A+B)(\Pi^-_B-\Pi^-)=\inf_{-\Pi^-\leq Q\leq 1-\Pi^-}\tr(A+B)Q.$$
The right-side is obviously the analogue of the corresponding term in~\eqref{eq:variational-proof}, with $A=-\Delta-\mu$ and $B=V$. The left-side is negative, which can be seen by taking $Q=0$ on the right, or by noticing that
$$\tr(A+B)(\Pi^-_B-\Pi^-)=-\tr|A+B|(\Pi^-_B-\Pi^-)^2.$$
This, clearly, is the analogue of $\tr_V(-\Delta+V-\mu)Q_V$, see Definition~\ref{def:total_energy}.

\subsection{Proof of the Lieb-Thirring inequalities in a potential $V$}\label{sec:proof-LT-V}

Here we explain how to prove the Lieb-Thirring inequalities~\eqref{eq:LT-V} and~\eqref{eq:LT-V-1D}, assuming Theorem~\ref{thm:variational_principle}. 
As in the proof of Theorem \ref{thm:LT-rho}, we assume $\mu=1$, the general case
being obtained by a simple scaling argument. 
By Theorems~\ref{thm:variational_principle} and~\ref{thm:LT-rho}, we have
for $d\geq2$
\begin{align*}
0&\geq \tr_V(-\Delta+V-1)Q_V\\
&\geq \inf_{\rho\geq-\rho_0}\left(K(d)\int_{\R^d}\delta\cT^{\rm
sc}(\rho)+\int_{\R^d}V\,\rho\right)\\
&=-L(d)\int_{\R^d}\left((V(x)-1)_-^{1+\tfrac{d}2}-1+\frac{2+d}2\,V(x)\right)\,
dx.
\end{align*}
The second equality follows from a simple optimization argument.
When $d=1$, we argue similarly. We decompose $\rho=\rho_{Q^{++}}+\rho_{Q^{--}}$ and $\rho'=
\rho_{Q^{-+}}+\rho_{Q^{-+}}$ and use~\eqref{eq:LT-rho-1D-proof} to obtain
\begin{align*}
0&\geq \tr_V(-\Delta+V-1)Q_V\\
&\geq \inf_{\rho\geq-\rho_0}\left(K(1)\!\int_{\R}\delta\cT^{\rm
sc}(\rho)+\int_{\R}V\,\rho\right)+\inf_{\rho'}\left(K'(1)\!\int_{\R}\frac{
|\widehat{\rho'}(k)|^2}{F_1(|k|)}\,dk+\int_{\R}V\,\rho'\right)\\
&=-{L}(1)\int_{\R}\left((V(x)-1)_-^{\tfrac32}-1+\frac{3}2\,V(x)\right)\,
dx-L'(1)\int_{\R}F_1(|k|)|\widehat{V}(k)|^2\,dk
\end{align*}
with 
$$F_1(|k|)=\frac{1+|k|}{|k|} \log\left(\frac{2+|k|}{|2-|k||}\right)$$
and
$${L}(1)=\frac23\left(\frac{1}{3K(1)}\right)^{1/2},\qquad
L'(1)=\frac1{4K'(1))}.$$ 
This concludes the proof of~\eqref{eq:LT-V} and~\eqref{eq:LT-V-1D}.\qed

\subsection{Proof of Theorem~\ref{thm:variational_principle}: the variational principle}
As before we assume $\mu=1$.
Let us denote by $I(V)$ the infimum appearing in~\eqref{eq:variational-proof}:
$$I(V):=\inf_{Q\in\cK}
\bigg(\tr_0(-\Delta-1)Q+\int_{\R^d}V(x)\,\rho_Q(x)\,dx\bigg).$$
Note that by Lemma~\ref{lem:density} we can restrict the infimum to finite-rank states
$Q\in\cK$. 

We split the proof of the theorem into two parts. First we show that
\begin{equation}
0\geq \tr_V(-\Delta-1+V)Q_V=-\norm{Q_V|-\Delta-1+V|^{1/2}}_{\gS_2}^2\geq I(V)
\label{eq:1st_step}
\end{equation}
This will show that $Q_V|-\Delta-1+V|^{1/2}$ is Hilbert-Schmidt. We will also find that
$Q_V|-\Delta-1|^{1/2}$ is Hilbert-Schmidt. To prove \eqref{eq:1st_step}, we
approximate $Q_V$ by a well-chosen sequence $Q_V^\epsilon$ of smooth operators
in $\cX$ satisfying the constraint $-\Pi^-\leq Q_V^\epsilon\leq\Pi^+$.

In a second step we prove the converse inequality
\begin{equation}
\tr_V(-\Delta-1+V)Q_V\leq I(V),
\label{eq:2nd_step}
\end{equation}
using the information that $Q_V|-\Delta-1+V|^{1/2}\in\gS_2$ and the density of
finite-rank operators in $\cK$, as stated in Lemma~\ref{lem:density}.

\subsubsection*{Step 1. Proof of the lower bound~\eqref{eq:1st_step}}
We introduce the following function
$$h(x):=-|x|\,\1(|x|\leq 1)+(2-|x|)\,\1(1\leq |x|\leq 2)$$
and replace $-\Delta$ by $K_\epsilon:=-\Delta+\epsilon\, h(-i\nabla)$ for a
small $\epsilon>0$. The gain is that $-\Delta+\epsilon\,h(-i\nabla)$ now has a
gap $(1-\epsilon,1+\epsilon)$ in its spectrum. Note also that we have 
$\Pi^-=\1(K_\epsilon\leq1)$
for all $\epsilon>0$, hence the free Fermi sea is not changed.

Let us introduce the corresponding regularized operator
$$Q_V^\epsilon:=\1\big(-\Delta+\epsilon\,h(-i\nabla)+V\leq1\big)-\Pi^-.$$
Note that $-\Delta+\epsilon\,h(-i\nabla)+V\to -\Delta+V$ in the norm resolvent
sense. 
When $d\geq3$, our assumption that $V\in L^{2}(\R^d)\cap L^{1+{d}/{2}}(\R^d)$
implies $V\in L^{(d+1)/2}(\R^d)$ hence it follows from a result of Koch and
Tataru~\cite{KocTat-06} that $-\Delta+V$ has no positive eigenvalue.
This in turn implies that $Q^\epsilon_V\to Q_V$ strongly by, e.g., \cite[Thm.
VIII.24]{ReeSim1}.

In dimensions $d=1$ and $d=2$, it was shown by von
Neumann-Wigner~\cite{NeuWig-29} and Ionescu-Jerison~\cite{IonJer-03} that there
exist potentials $V$ satisfying our assumptions for which
$\ker(-\Delta+V-1)\neq\{0\}$. For this reason, when $d=1,2$, we assume first
that $V$ has a compact support and is bounded (then $-\Delta+V$ has no positive
eigenvalue and $Q_V^\epsilon\to Q_V$ strongly), and only remove this assumption
at the very end of the proof.

What we have gained is that the operator $Q_V^\epsilon$ is now Hilbert-Schmidt,
whereas $Q_V$ is not even compact in general (Remark~\ref{rmk:trace-class}).

\begin{lemma}\label{lem:Q_V_epsilon}
Under our assumptions on $V$, 
\begin{equation}
Q_V^\epsilon\in\gS_2,\quad (-\Delta)\,Q_V^\epsilon\in\gS_2\quad\text{and}\quad
(K_\epsilon+V)\,Q_V^\epsilon\in\gS_2
\label{eq:reg_Q_V} 
\end{equation}
for all $\epsilon>0$. 
\end{lemma}

For clarity we postpone the proof of Lemma~\ref{lem:Q_V_epsilon}. 

The facts that $Q_V^\epsilon\in\gS_2$ and $(-\Delta)Q_V^\epsilon\in\gS_2$ imply
in particular $(Q_V^\epsilon)^{\pm\pm}\in\gS_1$,
$|1+\Delta|^{1/2}Q_V^\epsilon\in\gS_2$ and
$|1+\Delta|^{1/2}(Q_V^\epsilon)^{\pm\pm}|1+\Delta|^{1/2}\in\gS_1$. In particular
$Q_V^\epsilon\in\cX$, the Banach space introduced before in~\eqref{eq:space_X}.
By Theorem~\ref{thm:LT-rho} in dimensions $d\geq2$, we deduce that
$\rho_{Q^\epsilon_V}$ is a well-defined function such that $\delta\cT^{\rm
sc}_1(\rho_{Q^\epsilon_V})\in L^1(\R^d)$ and that
$$\tr_0(-\Delta-1)Q^\epsilon_V=\norm{Q_V^\epsilon|-\Delta-1|^{1/2}}^2_{\gS_2}
\geq K(d)\int_{\R^d}\delta\cT^{\rm sc}_1(\rho_{Q^\epsilon_V}).$$
In dimension $d=1$ we at least know that $\rho_{Q_V^\epsilon}\in L^1_{\rm
loc}(\R)$ by Lemma~\ref{lem:local-trace-class}.
			  
\begin{lemma}
We have the following equality:
\begin{multline}
\tr_0(-\Delta-1)Q^\epsilon_V+
\epsilon\,
\tr|h(-i\nabla)|\Big((Q^\epsilon_V)^{++}-(Q^\epsilon_V)^{--}\Big)+\int_{\R^d}
V\rho_{Q^\epsilon_V}\\
=-\tr|K_\epsilon+V-1|^{1/2}(Q^\epsilon_V)^2|K_\epsilon+V-1|^{1/2}\leq0.
\label{eq:equality_alpha}
\end{multline}
\end{lemma}

\begin{proof}
It is possible to approximate $Q_V^\epsilon$ by a sequence $\{R_n\}$ of smooth
finite rank operators such that $-\Pi^-\leq R_n\leq \Pi^+$, $(-\Delta+i)R_n\to
(-\Delta+i)Q_V^\epsilon$ strongly in $\gS_2$ and
$(-\Delta+i)(R_n)^{\pm\pm}(-\Delta+i)\to
(-\Delta+i)(Q_V^\epsilon)^{\pm\pm}(-\Delta+i)$ strongly in $\gS_1$. See, e.g.,
\cite[Prop. 2 \& App. B]{HaiLewSer-09}. We then write
\begin{align*}
\tr(-\Delta-1+\epsilon
h(-i\nabla)+V)R_n&=\tr(-\Delta-1)R_n+\epsilon\,\tr\,h(-i\nabla) R_n
+\int_{\R^d}V\rho_{R_n}\\
&=\tr|K_\epsilon-1+V|\big(\Pi^+_{V,\epsilon}R_n\Pi^+_{V,\epsilon}-\Pi^-_{V,
\epsilon}R_n\Pi^-_{V,\epsilon}\big) 
\end{align*}
where $\Pi^-_{V,\epsilon}=\1(K_\epsilon+V\leq1)$ and pass to the limit
$n\to\ii$.
\end{proof}

Since $Q_V^\epsilon\in\cX$, we deduce from~\eqref{eq:equality_alpha} that
\begin{equation}
-\tr|K_\epsilon+V-1|^{1/2}(Q^\epsilon_V)^2|K_\epsilon+V-1|^{1/2} \geq I(V)
\label{eq:lower_bd_epsilon_HS} 
\end{equation}
for all $\epsilon\geq0$. In particular $Q^\epsilon_V|K_\epsilon+V-1|^{1/2}$ is
uniformly bounded in the Hilbert-Schmidt class $\gS_2$. Note that the weak
limit of $Q^\epsilon_V|K_\epsilon+V-1|^{1/2}$ in $\gS_2$ can only be
$Q_V|-\Delta+V-1|^{1/2}$, since $Q_V^\epsilon\to Q_V$ strongly and
$|K_\epsilon+V-1|^{1/2}y\to |-\Delta+V-1|^{1/2}y$ for every $y\in H^2(\R^d)$.
This latter statement can be seen by writing\footnote{In dimension $d=1$, the
domain of $-\Delta+V$ (hence of $K_\epsilon+V$) contains $H^2(\R)$ by choice of
the Friedrichs extension via the KLMN Theorem.}
\begin{align*}
|K_\epsilon+V-1|^{1/2}y&=\frac{|K_\epsilon+V-1|^{1/2}}{K_\epsilon+V+i}
(-\Delta+V+i)y+\epsilon \frac{|K_\epsilon+V-1|^{1/2}}{K_\epsilon+V+i}h(-i\nabla)\,y\\
&\underset{\epsilon\to0}{\rightarrow}\frac{|-\Delta+V-1|^{1/2}}{-\Delta+V+i}
(-\Delta+V+i)y=|-\Delta+V-1|^{1/2}y
\end{align*}
and using that $f(x)=|x-1|^{1/2}(x+i)^{-1}$ is a continuous function tending to
zero at infinity, thus $\norm{f(K_\epsilon+V-1)- f(-\Delta-1+V)}\to0$ by
\cite[Thm. VIII.20]{ReeSim1}. Hence we have 
$$|K_\epsilon+V-1|^{1/2}Q^\epsilon_V\wto |-\Delta+V-1|^{1/2}Q_V\quad\text{weakly
in $\gS_2$}$$
and, passing to the weak limit in~\eqref{eq:lower_bd_epsilon_HS}, we obtain the
claimed inequality~\eqref{eq:1st_step}.

From~\eqref{eq:equality_alpha}, we also have the following bound
\begin{equation}
\norm{Q^\epsilon_V|\Delta+1|^{1/2}}_{\gS_2}^2+\int
V\rho_{Q_V^\epsilon}=\tr_0(-\Delta-1)Q_V^\epsilon +\int V\rho_{Q_V^\epsilon}\leq
0
\label{eq:lower_bd_negative_Q_V_epsilon}
\end{equation}
for all $\epsilon>0$. We deduce for instance that
\begin{equation}
\norm{Q^\epsilon_V|\Delta+1|^{1/2}}_{\gS_2}^2\leq
-\left(\tr_0(-\Delta-1)Q_V^\epsilon +2\int V\rho_{Q_V^\epsilon}\right)\leq
-I(2V).
\label{eq:unif_bd_kinetic_Q_V_epsilon} 
\end{equation}
This uniform bound proves that $Q_V|\Delta+1|^{1/2}\in\gS_2$ and that
$$Q^\epsilon_V|\Delta+1|^{1/2}\wto Q_V|\Delta+1|^{1/2}\quad\text{weakly in
$\gS_2$}.$$

In dimensions $d=1,2$, we have only written the proof for $V$ a bounded function
of compact support. If $V$ is an arbitrary function satisfying our
assumptions~\eqref{eq:cond_V} and~\eqref{eq:cond_V_1D}, we apply the result to
$V_R(x):=V(x)\1(|x|\leq R)\1(|V(x)|\leq R)$ and, from~\eqref{eq:1st_step} and
\eqref{eq:unif_bd_kinetic_Q_V_epsilon}, we obtain uniform estimates of the form
\begin{equation*}
\tr|-\Delta+V_R-1|^{1/2}(Q_{V_R})^2|-\Delta+V_R-1|^{1/2}\leq -I(V_R)
\end{equation*}
and
$$\tr|\Delta+1|^{1/2}\big(Q_{V_R}^{++}-Q_{V_R}^{--}\big)|\Delta+1|^{1/2}\leq -
I(2V_R).$$
Extracting subsequences we now have at best that $Q_{V_R}\wto Q_{V}+\delta$
weakly as $R\to\ii$, where $0\leq\delta\leq\1(-\Delta+V=1)$. Passing to weak limits as
before, we therefore obtain that
\begin{equation*}
\tr|-\Delta+V-1|^{1/2}(Q_{V})^2|-\Delta+V-1|^{1/2}\leq -I(V)
\end{equation*}
as was claimed, and that
\begin{align*}
\tr|\Delta+1|^{1/2}(Q_V^{++}-Q_V^{--})|\Delta+1|^{1/2}&\leq
\tr|\Delta+1|^{1/2}(Q_V^{++}-Q_V^{--}+\delta)|\Delta+1|^{1/2}\\
&\leq -I(2V).
\end{align*}

\medskip

It remains to provide the

\begin{proof}[Proof of Lemma~\ref{lem:Q_V_epsilon}]
Our claim~\eqref{eq:reg_Q_V} follows from Cauchy's formula and the resolvent
expansion:
\begin{multline}
Q_V^\epsilon=-\frac{1}{2i\pi}\sum_{k=1}^J(-1)^k\oint_\cC\frac{1}{K_\epsilon-z}
\left(V\frac{1}{K_\epsilon-z}\right)^{k}\,dz\\
+\frac{(-1)^{J}}{2i\pi}\oint_\cC\frac{1}{K_\epsilon-z}\left(V\frac{1}{
K_\epsilon-z}\right)^{J+1}(K_\epsilon-z)\frac{1}{K_\epsilon+V-z}\,dz. 
\label{eq:Cauchy_Q}
\end{multline}
Under our assumptions the function $V$ is $K_\epsilon$--compact, hence
$K_\epsilon+V$ has the gap $(1-\epsilon,1+\epsilon)$ in its essential spectrum
and it is bounded from below. In~\eqref{eq:Cauchy_Q}, we choose for $\cC$ a
smooth curve enclosing the spectra of $K_\epsilon$ and $K_\epsilon+V$ below $1$,
without intersecting them. We will explain below how to choose $J$.

In order to show that $(1-\Delta)Q_V^\epsilon$ is a Hilbert-Schmidt operator for
all $\epsilon>0$, we estimate each term in~\eqref{eq:Cauchy_Q}. Our bounds will
depend on $\epsilon$. 
We start by noticing that there is a uniform bound of the form
\begin{equation}
\forall z\in\cC,\qquad
\norm{\frac{1-\Delta}{K_\epsilon-z}}+\norm{(K_\epsilon-z)\frac{1}{K_\epsilon+V-z
}}\leq C.
\label{eq:bound_curve} 
\end{equation}
The constant $C$ diverges when $\epsilon\to0$ but we do not emphasize this in
our notation.
To estimate the last term of~\eqref{eq:Cauchy_Q}, we use (for $d\geq2$) that
\begin{equation}
\norm{V\frac{1}{K_\epsilon-z}}_{\gS_{1+d/2}}\leq C\norm{V}_{L^{1+d/2}(\R^d)} 
\label{eq:apply_KSS_Q_V}
\end{equation}
by the Kato-Seiler-Simon inequality \cite[Thm 4.1]{Simon-79},
\begin{equation}
\forall p \geq 2,\qquad  \| f(- i \nabla) g(x) \|_{\gS_p} \leq \frac{1}{(2
\pi)^\frac{d}{p}}
\| g \|_{L^p(\R^d)} \| f \|_{L^p(\R^d)}.
\label{eq:KSS}
\end{equation}
The constant $C$ in~\eqref{eq:apply_KSS_Q_V} also depends on $\epsilon$.
Choosing $J\geq 1/2+d/4$ in \eqref{eq:Cauchy_Q}, we obtain by
H\"older's inequality and~\eqref{eq:bound_curve}
\begin{equation*}
\norm{(1-\Delta)\oint_\cC\frac{1}{K_\epsilon-z}\left(V\frac{1}{K_\epsilon-z}
\right)^{J+1}\!\!(K_\epsilon-z)\frac{1}{K_\epsilon+V-z}\,dz}_{\gS_2}
\leq C\norm{V}_{L^{1+d/2}(\R^d)}^{J+1}. 
\end{equation*}
We now treat the term corresponding to $k=1$ in the first sum
of~\eqref{eq:Cauchy_Q} and start by noticing that
$$\oint_\cC\frac{\Pi^-}{K_\epsilon-z}V\frac{\Pi^-}{K_\epsilon-z}\,
dz=\oint_\cC\frac{\Pi^+}{K_\epsilon-z}V\frac{\Pi^+}{K_\epsilon-z}\,dz=0.$$
For the other terms, we simply write, for instance,
$$\norm{(1-\Delta)\oint_\cC\frac{\Pi^-}{K_\epsilon-z}V\frac{\Pi^+}{K_\epsilon-z}
\,dz}_{\gS_2}\leq C\norm{\Pi^- V}_{\gS_2}\leq C\norm{V}_{L^2(\R^d)}$$
since $\Pi^-=\1(|\nabla|\leq1)$ has a compact support in Fourier space.

The argument is the same for the other terms of the first sum
in~\eqref{eq:Cauchy_Q}: We write 
$$\frac{1}{K_\epsilon-z}=\frac{\Pi^-}{K_\epsilon-z}+\frac{\Pi^+}{K_\epsilon-z}$$
and note first that the term which has only $\Pi^+$ vanishes after integrating
over $z\in\cC$, by the residue formula (the same holds for the term which has
only $\Pi^-$). The other terms contain at least one $\Pi^-$ and can be estimated
similarly as before.

We deduce, as was claimed in~\eqref{eq:reg_Q_V}, that
$(1-\Delta)Q_V^\epsilon\in\gS_2$ for every $\epsilon>0$. Since
$(Q^\epsilon_V)^2=(Q^\epsilon_V)^{++}-(Q^\epsilon_V)^{--}$, this implies that
$(1-\Delta)(Q^\epsilon_V)^{\pm\pm}(1-\Delta)\in\gS_1$.
Finally, $(K_\epsilon+V)(1-\Delta)^{-1}$ being bounded, we have that
$(K_\epsilon+V)Q_V^\epsilon\in\gS_2$. 

We have written the proof for $d\geq2$. The case $d=1$ is similar and left to
the reader (see also the proof of Theorem~\ref{thm:second-order-perturb} below).
\end{proof}

\subsubsection*{Step 2. Proof of the upper bound~\eqref{eq:2nd_step}}
To finish the proof, it remains to show the inequality~\eqref{eq:2nd_step}, that
is $\tr_V(-\Delta-1+V)Q_V\leq I(V)$.

We pick a smooth finite rank operator $Q$ such that $-\Pi^-\leq Q\leq \Pi^+$ and
$Q(-\Delta)$ is bounded, and note that
\begin{align*}
&\tr_0(-\Delta-1)Q+\int_{\R^d}V\,\rho_Q\\
&\qquad\qquad\quad=\tr(-\Delta-1+V)Q\\
&\qquad\qquad\quad=\tr|-\Delta-1+V|^{1/2}\big(\Pi_V^+Q\Pi_V^+ -
\Pi_V^-Q\Pi_V^-\big)|-\Delta-1+V|^{1/2}.
\end{align*}
We now use that 
$$|-\Delta-1+V|^{1/2}\Pi_V^\pm Q_V\Pi_V^\pm|-\Delta-1+V|^{1/2}\in\gS_1$$
as we have shown in Step 1. Writing $Q=(Q-Q_V)+Q_V$ we obtain
\begin{align*}
\tr_0(-\Delta-1)Q+\int_{\R^d}V\,\rho_Q&
=\tr_V(-\Delta-1+V)(Q-Q_V)+\tr_V(-\Delta-1+V)Q_V\\
&\geq \tr_V(-\Delta-1+V)Q_V.
\end{align*}
In the second line we have used that
$$\tr_V(-\Delta-1+V)(Q-Q_V)\geq0$$
since $-\Pi_V^-\leq Q-Q_V\leq \Pi_V^+$.
By the density of finite rank operators in $\cX$ (see Lemma~\ref{lem:density}),
we deduce that 
\begin{equation*}
\tr_V(-\Delta+V-1)Q_V\leq I(V),
\end{equation*}
which finishes the proof of Theorem~\ref{thm:variational_principle}.\qed

\begin{remark}\rm 
Our proof also yields the limit
\begin{equation}
\lim_{\epsilon\to0}\norm{|K_\epsilon+V-1|^{1/2}Q^\epsilon_V -
|-\Delta+V-1|^{1/2}Q_V}_{\gS_2}=0.
\label{eq:approx_epsilon}
\end{equation}
Indeed, from~\eqref{eq:lower_bd_epsilon_HS}, we know that 
$$\limsup_{\epsilon\to0}\norm{|K_\epsilon+V-1|^{1/2}Q^\epsilon_V}_{\gS_2}^2\leq
-I(V)=\norm{|-\Delta+V-1|^{1/2}Q_V}_{\gS_2}^2$$
where the last equality follows from Theorems~\ref{thm:LT-V}
and~\ref{thm:LT-V-1D}. Since we also have proved that 
$|K_\epsilon+V-1|^{1/2}Q^\epsilon_V \wto |-\Delta+V-1|^{1/2}Q_V$ weakly in
$\gS_2$, the statement~\eqref{eq:approx_epsilon} follows.
\end{remark}

\subsection{Proof of 
Theorem~\ref{thm:second-order-perturb}: second-order perturbation theory}\label{sec:proof-second-order-perturb}
In this section we sketch the proof of Theorem~\ref{thm:second-order-perturb}.
We detail first the one-dimensional case $d=1$ and mention the
necessary modifications in higher dimensions afterwards.

We could embark upon expanding $Q_{tV}$ in powers of $t$ by directly using the
resolvent formula. Since we want to avoid a tedious justification of this
expansion, we instead work with the approximate state 
$$Q^\epsilon_{tV}:=\1(K_\epsilon+tV\leq1)-\Pi^-$$ 
which we have already introduced in the proof of Theorems~\ref{thm:LT-V}
and~\ref{thm:LT-V-1D}. We will prove bounds in $t$ which are uniform in
$\epsilon$, and pass to the limit $\epsilon\to0$ in the end,
using~\eqref{eq:approx_epsilon}. The same method of proof can be used to justify
an expansion of $Q_{tV}$ to any order in $t$.

We come back to the resolvent expansion~\eqref{eq:Cauchy_Q} for
$Q^\epsilon_{tV}$ which we have already mentioned in the proof of
Lemma~\ref{lem:Q_V_epsilon} above. In dimension $d=1$, we write 
\begin{equation}
Q_{tV}^\epsilon=t\, Q_1^\epsilon+t^2\, Q_2^\epsilon+t^3\, Q_3^\epsilon(t)
\label{eq:decomp_Q}
\end{equation}
where
$$Q_1^\epsilon=\frac1{2i\pi}\oint_{\cC}\frac1{K_\epsilon-z}V\frac1{K_\epsilon-z
}\,dz,\qquad
Q_2^\epsilon=-\frac1{2i\pi}\oint_{\cC}\left(\frac1{K_\epsilon-z}V\right)^2\frac1{
K_\epsilon-z}\,dz$$
and
$$Q_3^\epsilon(t)=\frac1{2i\pi}\oint_{\cC}\left(\frac1{K_\epsilon-z}
V\right)^3\frac1{K_\epsilon+t V-z}\,dz.$$
In the above formulas, we choose for $\cC$ a curve in the complex plane 
enclosing the interval $[-R,1]\subset\R$, where $-R<\inf\sigma(K_\epsilon+tV)$
for all $0<\epsilon<1$ and all $|t|<1$. 
To simplify certain estimates below, we
also assume that $|\Im z|\leq 1/2$ for all $z\in\cC$ (in such a way that
$\log|\Im z|^{-1}\geq0$). For convenience  we will make the assumption that
$1\notin\sigma(K_\epsilon+tV)$ for all $t$ small enough. If $1$ is an eigenvalue of $K_\epsilon+tV$, one
has to let the curve $\cC$ depend on $\epsilon$, and modify it a bit in a
neighborhood of $z=1$. It can then be verified that our estimates below still
hold true. These details are left to the reader for brevity.

Note that $Q_1^\epsilon$ is purely off-diagonal, i.e.
$(Q_1^\epsilon)^{\pm\pm}=0$.
Using that $|K_\epsilon-z|>0$ for all $z\in\cC$ by definition of $K_\epsilon$,
one can prove (similarly as in the proof of Lemma~\ref{lem:Q_V_epsilon}), that
$Q_1^\epsilon$, $Q_2^\epsilon$ and $Q_3^\epsilon(t)$ are trace-class, and that
\begin{multline}
\tr_{tV}(K_\epsilon-1+tV)Q_{tV}^\epsilon=t^2\big\{\tr(Q_1^\epsilon V)+\tr
Q_2^\epsilon(K_\epsilon-1)\big\}\\
+t^3\big\{\tr (Q_2^\epsilon V) +\tr Q^\epsilon_3(t)(K_\epsilon-1+tV)\big\}.
\end{multline}
Each of the terms of the right side makes sense and can be bounded uniformly in
$t$ and $\epsilon$, as we now explain.
First, we have
$$\norm{Q_2^\epsilon V}_{\gS_1}\leq (2\pi)^{-1}\oint_\cC
\norm{\left(\frac1{K_\epsilon-z} V\right)^3}_{\gS_1}|dz|,$$
and, similarly,
$$\norm{Q^\epsilon_3(t)(K_\epsilon-1+tV)}_{\gS_1}\leq C\,\oint_\cC
\norm{\left(\frac1{K_\epsilon-z} V\right)^3}_{\gS_1}|dz|,$$
since $(K_\epsilon+tV-1)(K_\epsilon+t V-z)^{-1}$ is uniformly bounded for
$z\in\cC$, by choice of the curve $\cC$ in the complex plane. We now use that 
$$\norm{\frac{|-\Delta-1|}{K_\epsilon-z}}\leq C$$
for a constant $C$ independent of $z\in\cC$ and $\epsilon$, to deduce that 
\begin{equation*}
\norm{\left(\frac1{K_\epsilon-z} V\right)^3}_{\gS_1}\leq C\,|\Im
z|^{-1/2}\norm{\frac1{|-\Delta-z|^{1/2}}
V\frac1{|-\Delta-z|^{1/2}}}_{\gS_1}^2\norm{\frac1{|-\Delta-z|^{1/2}} V}_{\gS_2}.
\end{equation*}
We have the bound
\begin{align}
\norm{\frac1{|-\Delta-z|^{1/2}} V\frac1{|-\Delta-z|^{1/2}}}_{\gS_1}&\leq
\norm{\frac1{|-\Delta-z|^{1/2}} \sqrt{|V|}}_{\gS_2}^2\nonumber\\
&= (2\pi)^{-1}\norm{V}_{L^1(\R)}\int_{\R}\frac{dp}{\sqrt{(p^2-\Re z)^2+(\Im
z)^2}}\nonumber\\
&\leq C\norm{V}_{L^1(\R)}\log|\Im z|^{-1},\label{eq:HS_bound_AD}
\end{align}
and, in a similar fashion,
$$\norm{\frac1{|-\Delta-z|^{1/2}} V}_{\gS_2}\leq C\norm{V}_{L^2(\R)}\log|\Im
z|^{-1}.$$
Using these two bounds we deduce that
$$\norm{\left(\frac1{K_\epsilon-z} V\right)^3}_{\gS_1}\leq
C\norm{V}_{L^1(\R)}^2\norm{V}_{L^2(\R)}|\Im z|^{-1/2}\big(\log|\Im z|^{-1})^3.$$
Integrating over $z\in\cC$, this eventually shows that
$$\norm{Q_2^\epsilon V}_{\gS_1}
+\norm{Q^\epsilon_3(t)(K_\epsilon-1+tV)}_{\gS_1}\leq
C\norm{V}_{L^1(\R)}^2\norm{V}_{L^2(\R)},$$
hence that
\begin{multline}
\Big|\tr_{tV}(K_\epsilon-1+tV)Q_{tV}^\epsilon-t^2\big\{\tr(Q_1^\epsilon V)+\tr
Q_2^\epsilon(K_\epsilon-1)\big\}\Big|\\
\leq Ct^3\norm{V}_{L^1(\R)}^2\norm{V}_{L^2(\R)}
\label{eq:estim_unif_epsilon}
\end{multline}
with a constant $C$ independent of $\epsilon$ and $t$. 

Using the residue formula we find
$$\tr(Q_1^\epsilon V)=-2(2\pi)^{-1}\iint_{\substack{|p|^2\leq1\\ |q|^2\geq
1}}\frac{|\widehat{V}(p-q)|^2}{|q|^2-|p|^2+\epsilon(h(q)-h(p))}\,dp\,dq$$
and
$$\tr Q_2^\epsilon (K_\epsilon-1)=(2\pi)^{-1}\iint_{\substack{|p|^2\leq1\\
|q|^2\geq
1}}\frac{|\widehat{V}(p-q)|^2}{|q|^2-|p|^2+\epsilon(h(q)-h(p))}\,dp\,dq.$$
The result in the case $d=1$ now follows from taking first the limit
$\epsilon\to0$ in~\eqref{eq:estim_unif_epsilon},
using~\eqref{eq:approx_epsilon}, and then $t\to0$.

When $d\geq2$, the proof is similar but a bit more tedious. We start again with
the resolvent expansion~\eqref{eq:Cauchy_Q}, to an order $J$ such that the last
term becomes trace-class (when multiplied by $K_\epsilon+tV-1$). This means we
write
\begin{align}
\tr_{tV}Q^\epsilon_{tV}(K_\epsilon+tV-1)=&-t^2(2\pi)^{-d}\iint_{\substack{
|p|^2\leq1\\ |q|^2\geq
1}}\frac{|\widehat{V}(p-q)|^2}{|q|^2-|p|^2+\epsilon(h(q)-h(p))}\,dp\,
dq\nonumber\\
&+\sum_{j=3}^{J}(-t)^j\;\tr\oint_{\cC}dz\,\left(\frac{1}{K_\epsilon-z}
V\right)^j\frac{K_\epsilon-1}{K_\epsilon-z}\nonumber\\
&-\sum_{j=3}^{J+1}(-t)^j\;\tr\oint_{\cC}dz\,\left(\frac{1}{K_\epsilon-z}
V\right)^j\nonumber\\
&+(-t)^{J+1}\;\tr
\oint_{\cC}dz\,\left(\frac{1}{K_\epsilon-z}V\right)^{J+1}\frac{K_\epsilon+tV-1}{
K_\epsilon+tV-z}.
\label{eq:decomp_energy_Q}
\end{align}
We fix a $J\geq 1+d/2$ and deduce, similarly as before, that
\begin{align*}
&\norm{\left(\frac{1}{K_\epsilon-z}V\right)^{J+1}}_{\gS_1}\\
&\qquad \leq C\,|\Im z|^{-1/2}\norm{\frac1{|-\Delta-z|^{1/2}}
V\frac1{|-\Delta-z|^{1/2}}}_{\gS_{1+d/2}}^J\norm{\frac1{|-\Delta-z|^{1/2}} V}\\
&\qquad \leq C \,|\Im z|^{-1/2}\left(\norm{V}_{L^1(\R^d)}\log|\Im
z|^{-1}+\norm{V}_{L^{1+d/2}(\R^ d)} \right)^J\times\\
&\qquad\qquad\qquad\qquad\qquad \qquad \times\left(\norm{V}_{L^2(\R^d)}\log|\Im
z|^{-1}+\norm{V}_{L^\ii(\R^d)}\right)
\end{align*}
with a constant $C$ that is independent of $\epsilon$.
For the other terms in~\eqref{eq:decomp_energy_Q}, we have to work a bit more.
As an illustration, we only consider the term
$$\tr_0\oint_{\cC}dz\,\left(\frac{1}{K_\epsilon-z}V\right)^3,$$
the other terms are treated by the same argument. We decompose 
$$\frac{1}{K_\epsilon-z}=\frac{\Pi^-}{K_\epsilon-z}+\frac{\Pi^+}{K_\epsilon-z}$$
and expand $((K_\epsilon-z)^{-1}V)^3$ accordingly. The terms which have only
$\Pi^+$ or only $\Pi^-$ vanish after the integration over the curve $\cC$, by
the residue formula. For the other terms, $\Pi^-V\Pi^+$ (or its adjoint) must
appear at least twice in the trace to be estimated. For instance, we look at the
term
\begin{equation}
\tr\oint_\cC\,dz
\frac{\Pi^+}{K_\epsilon-z}V\frac{\Pi^+}{K_\epsilon-z}V\frac{\Pi^-}{K_\epsilon-z}
V=\tr\oint_\cC\,dz
\frac{\Pi^+}{K_\epsilon-z}V\frac{\Pi^+}{K_\epsilon-z}V\frac{\Pi^-}{K_\epsilon-z}
V\;\Pi^+.
\label{eq:3rd_term_example} 
\end{equation}
By cyclicity of the trace, this term can be estimated by
\begin{equation}
|\eqref{eq:3rd_term_example}|\leq
\oint_\cC\,|dz|\,\norm{\frac{\Pi^+}{|-\Delta-z|^{1/2}}V\frac{\Pi^+}{|-\Delta-z|^{
1/2}}}\;\norm{\frac{\Pi^+}{|-\Delta-z|^{1/2}}V\frac{\Pi^-}{|-\Delta-z|^{1/2}}}_{
\gS_2}^2.
\label{eq:estim_term_example} 
\end{equation}
Decomposing
$|-\Delta-z|^{1/2}=|-\Delta-z|^{1/2}\1(|\Delta+1|\geq1)+|-\Delta-z|^{1/2}\1(
|\Delta+1|\leq1)$ and using that $V\in L^1(\R^d)\cap L^{1+d/2}(\R^d)$, we find
\begin{align*}
&\norm{\frac{\Pi^+}{|-\Delta-z|^{1/2}}V\frac{\Pi^+}{|-\Delta-z|^{1/2}}}\\
&\qquad\qquad\leq \norm{\frac{1}{|-\Delta-z|^{1/2}}\sqrt{|V|}}^2\\
&\qquad\qquad\leq
\left(\norm{\frac{\1(|\Delta+1|\leq1)}{|-\Delta-z|^{1/2}}\sqrt{|V|}}_{\gS_2}
+\norm{\frac{\1(|\Delta+1|\geq1)}{|-\Delta-z|^{1/2}}\sqrt{|V|}}_{\gS_{2+d}}
\right)^2 \\
&\qquad\qquad\leq C\left(\norm{V}_{L^1(\R^d)}\log|\Im
z|^{-1}+\norm{V}_{L^{1+d/2}(\R^d)}\right).
\end{align*}
For the second term in the right side of~\eqref{eq:estim_term_example}, we use
that
\begin{multline*}
\norm{\frac{\Pi^+}{|-\Delta-z|^{1/2}}V\frac{\Pi^-}{|-\Delta-z|^{1/2}}}_{\gS_2}\\
\leq
\norm{\frac{\Pi^+\1(|\Delta+1|\leq1)}{|-\Delta-z|^{1/2}}V\frac{\Pi^-}{
|-\Delta-z|^{1/2}}}_{\gS_2}+\norm{\frac{\Pi^+\1(|\Delta+1|\geq1)}{|-\Delta-z|^{
1/2}}V\frac{\Pi^-}{|-\Delta-z|^{1/2}}}_{\gS_2}.
\end{multline*}
The first term on the right side is estimated as before. For the second one, we
use that
\begin{multline*}
\norm{\frac{\Pi^+\1(|\Delta+1|\geq1)}{|-\Delta-z|^{1/2}}V\frac{\Pi^-}{
|-\Delta-z|^{1/2}}}_{\gS_2}\\
\leq C|\Im z|^{-1/4}
\norm{\frac{\Pi^+}{|-\Delta-1|^{1/4}}V\frac{\Pi^-}{|-\Delta-1|^{1/4}}}_{\gS_2}.
\end{multline*}
This term is now exactly the one which we have calculated before
in~\eqref{eq:HS_off_diagonal_Phi_d} and it is finite under our assumptions on
$V$. Summarizing, we have proved that the term~\eqref{eq:3rd_term_example} is
bounded uniformly in $\epsilon$. 

The same argument can be applied to all the terms in~\eqref{eq:decomp_energy_Q},
showing that they are bounded uniformly in $\epsilon$. This concludes our sketch of the proof of
Theorem~\ref{thm:second-order-perturb}. \qed

\section{Thermodynamic limit and positive temperature}\label{sec:thermo-positive-temp}
\subsection{Lieb-Thirring inequalities in a box}
In this section, we extend our inequalities~\eqref{eq:LT-rho} and~\eqref{eq:LT-V} to the case of a system living in a box of size $L$, with constants independent of $L$. For simplicity we restrict ourselves to periodic boundary conditions and dimensions $d\geq2$. 

We denote by $-\Delta_L$ the Laplacian on $C_L=[-L/2,L/2)^d$, with periodic boundary conditions, and, for any chosen $\mu>0$, we introduce $\Pi^-_{L,\mu}:=\1(-\Delta_L\leq\mu)$. Note that since the spectrum of $-\Delta_L$ is discrete in $\R^+$, $\Pi^-_{L,\mu}$ has finite rank for every finite $L>0$ and $\mu\geq0$.
The following is a generalization of the density inequality~\eqref{eq:LT-rho}.

\begin{theorem}[Lieb-Thirring inequality in a box, density version,
$d\geq2$]\label{thm:LT-rho-box}
We assume that $d\geq2$, $\mu\geq0$ and $L>0$. Let $Q$ be a self-adjoint operator of finite rank such
that $-\Pi^-_{L,\mu}\leq Q\leq 1-\Pi^-_{L,\mu}$. Then there exists positive constants $\tilde{K}(d)$ and $C$ (depending
only on $d\geq2$) such that
\begin{multline}
\tr_{L^2(C_L)}(-\Delta_L-\mu)Q\\
\geq \tilde{K}(d)\begin{cases}
\displaystyle\int_{C_L}\delta\cT_{\mu}^{\rm sc}\left(\left(|\rho_Q(x)|-CL^{-1}\mu^{\tfrac{d-1}{2}}\right)_+\right)\,dx&\text{for $\mu>1/L^2$,}\\[0,4cm]
\displaystyle\int_{C_L}\delta\cT_{0}^{\rm sc}\left(\left(|\rho_Q(x)|-CL^{-d}\right)_+\right)\,dx&\text{for $\mu\leq 1/L^2$,}
\end{cases}
\label{eq:LT-rho-box} 
\end{multline}
where we recall that
$$\delta \cT^{\rm
sc}_\mu(\rho):=(\rho_0+\rho)^{1+\tfrac{2}d}-(\rho_0)^{1+\tfrac{2}d}-\frac{d+2}
d(\rho_0)^{\tfrac{2}d}\,\rho$$
with $\rho_0=\mu^
{\tfrac{d}2}\;q\,(2\pi)^{-d}|S^{d-1}|/d$.
\end{theorem}

The function appearing in the integrand of~\eqref{eq:LT-rho-box} vanishes for $\rho\leq CL^{-1}\mu^{\frac{d-1}{2}}$ (in the case $\mu>1/L^2$) or for $\rho\leq CL^{-d}$ (in the case $\mu\leq1/L^2$), and it converges to $\delta\cT_\mu(|\rho_Q|)$ in the limit $L\to\ii$. Note the absolute value which we have used to simplify our statement. Of course, $\delta\cT_\mu^{\rm sc}(|\rho_Q|)$ is comparable to $\delta\cT_\mu^{\rm sc}(\rho_Q)$.

Using Theorem~\ref{thm:LT-rho-box}, we can now deduce the (dual) potential version in the box. Again, note that for $V\in L^{1+d/2}(C_L)$, the spectrum of $-\Delta_L+V$ is discrete and bounded from below, hence there is only a finite number of eigenvalues below each chosen Fermi level $\mu$.

\begin{theorem}[Lieb-Thirring inequality in a box, potential version,
$d\geq2$]\label{thm:LT-V-box}
Assume that $\mu\geq0$, $d\geq2$ and $L>0$. Let $V$ be a real-valued function in
$L^{1+{d}/{2}}(C_L)$. Then we have
\begin{multline}
0\geq -\tr(-\Delta_L+V-\mu)_-+\tr(-\Delta_L-\mu)_- - \rho_0\int_{C_L}V\\
\geq -\tilde{L}(d)\int_{C_L}\left((V(x)-\mu)_-^{1+\tfrac{d}2}-\mu^{1+\tfrac{d}2}+\frac{2+d}
2\,\mu^{\tfrac{d}2}\,V(x)+\frac{\mu^{\tfrac{d-1}2}}{L}|V(x)|\right)\,dx
\label{eq:LT-V-box}
\end{multline}
when $\mu>1/L^2$, and
\begin{multline}
0\geq -\tr(-\Delta_L+V-\mu)_-+\tr(-\Delta_L-\mu)_- - \rho_0\int_{C_L}V\\
\geq -\tilde{L}(d)\int_{C_L}\left(V(x)_-^{1+\tfrac{d}2}+\frac{1}{L^d}|V(x)|\right)\,dx
\label{eq:LT-V-box-small-mu}
\end{multline}
when $\mu\leq1/L^2$. The constant $\tilde{L}(d)$ only depends on $d$.
\end{theorem}

Since all operators are finite-rank, the proof simply reduces to computing the Legendre transform of
$\rho\mapsto \big(1+(|\rho|-\epsilon)_+\big)^\alpha-1-\alpha(|\rho|-\epsilon)_+$. We skip the details and only provide the proof of Theorem~\ref{thm:LT-rho-box}.

\bigskip

\noindent\textit{Proof of Theorem~\ref{thm:LT-rho-box}.}
The proof follows the same two steps as that of Theorem~\ref{thm:LT-rho}, but it is slightly more tedious.

\subsubsection*{Step 1. Estimate on $Q^{\pm\pm}$}
We start by estimating the diagonal densities $\rho_{Q^{\pm\pm}}$. 
Following the strategy of the proof of Lemma~\ref{lem:generalized_LT}, we get, with $\gamma=Q^{\pm\pm}$,
$$\tr_{L^2(C_L)}|-\Delta_L-\mu|\gamma\geq \int_{C_L}R_{d,\mu,L}\big(\rho_\gamma(x)\big)\,dx$$
where
\begin{equation}
R_{d,\mu,L}(\rho)=\int_0^\ii\left(\sqrt{\rho}-\sqrt{f_{d,\mu,L}(e)}\right)_+^2\,de 
\label{eq:Rumin_L}
\end{equation}
and
\begin{align*}
f_{d,\mu,L}(e)&=\frac{1}{L^d}\;\#\left\{p\in(2\pi\Z/L)^d\ :\ \big|p^2-\mu\big|\leq e\right\}\\
&=\frac{1}{L^d}\left\{\;\#\Z^d\cap B\left(\frac{L\sqrt{\mu}}{2\pi}\left(1+\frac{e}\mu\right)^{1/2}\right)-\#\Z^d\cap B\left(\frac{L\sqrt{\mu}}{2\pi}\left(1-\frac{e}\mu\right)_+^{1/2}\right)\right\}. 
\end{align*}

The following gives an estimate on the function $f_{d,\mu,L}$. 

\begin{lemma}[Estimates on $f_{d,\mu,L}$]\label{lem:estim_f_L}
When $\mu> 1/L^2$, we have
\begin{equation}
f_{d,\mu,L}(e)\leq C\left(\frac{\mu^{\frac{d-1}{2}}}{L}+\mu^{\frac{d}2-1}e\,\1(e\leq\mu)+e^{d/2}\,\1(e\geq\mu)\right)
\label{eq:estim_f_L_1}
\end{equation}
whereas when $0\leq\mu\leq 1/L^2$, we have 
\begin{equation}
f_{d,\mu,L}(e)\leq C\left(\frac{1}{L^d}+e^{d/2}\right),
\label{eq:estim_f_L_2}
\end{equation}
for all $e>0$.
\end{lemma}

Note that the estimate~\eqref{eq:estim_f_L_2} on $f_{d,\mu,L}$ in the case $\mu\leq1/L^2$ is a bit weaker than the one~\eqref{eq:estim_f_L_1} for $\mu>1/L^2$.

\begin{proof}[Proof of Lemma~\ref{lem:estim_f_L}]
First, we recall the following well-known property
\begin{equation}
\left|\#\Z^d\cap B(R) - \frac{|S^{d-1}|}{d}R^d\right|\leq C\max\big(1,R^{d-1}\big),
\label{eq:nb_points_volume}
\end{equation}
which says that the number of points of the lattice $\Z^d$ inside a ball of radius $R$, behaves like the volume of the ball $B(R)$ in the limit of large $R$, whereas it is just bounded for small $R$. The error term can even be replaced by $o(R^{d-1})$ but we do not need this here. 
Note that the bound~\eqref{eq:nb_points_volume} implies $\#\Z^d\cap B(R)\leq C(1+R^d)$.

The proof of~\eqref{eq:estim_f_L_2} is now straightforward: Assuming $\mu\leq 1/L^2$, we simply write
$$f_{d,\mu,L}(e)\leq \frac{1}{L^d}\#\Z^d\cap B\left(\frac{L}{2\pi}\sqrt{\mu+e}\right)
\leq \frac{C}{L^d}\left(1+L^d\big(\mu^{d/2}+e^{d/2}\big)\right)
\leq \frac{2C}{L^d}+C\,e^{d/2}.$$

In order to prove~\eqref{eq:estim_f_L_1} we need another estimate. 
Let $M>0$ and $0< x\leq x_0$ for some fixed $x_0>0$. Using~\eqref{eq:nb_points_volume} we obtain
\begin{eqnarray}
&&\#\Z^d\cap B\left(M(1+x)^{1/2}\right) - \#\Z^d\cap B\left(M(1-x)_+^{1/2}\right)\nonumber\\
&&\qquad\leq \frac{|S^{d-1}|\,M^d}{d}\left((1+x)^{1/2}-(1-x)_+^{1/2}\right)+2C\max\left(1,M^{d-1}\right)(1+x_0)^{\frac{d-1}{2}}\nonumber\\
&&\qquad\leq C\frac{|S^{d-1}|\,M^d}{d}x+2C\max\left(1,M^{d-1}\right)(1+x_0)^{\frac{d-1}{2}}\nonumber\\
&&\qquad\leq C\left(M^d\,x+\max\left(1,M^{d-1}\right)\right).\label{eq:estim_diff_nb_ball}
\end{eqnarray}
We have used that $(1+x)^{1/2}-(1-x)_+^{1/2}\leq Cx$ for all $0\leq x\leq x_0$, where $C$ only depends on $x_0$. 
We can use~\eqref{eq:estim_diff_nb_ball} to prove~\eqref{eq:estim_f_L_1}, assuming now $\mu>1/L^2$. For $e\leq3\mu/2$, we use~\eqref{eq:estim_diff_nb_ball} with $M=L\sqrt{\mu}/2\pi\geq 1/(2\pi)$ and $x=e/\mu\leq 3/2$. We obtain 
$$f_{d,\mu,L}(e)\leq \frac{C}{L^d}\left(\frac{L^d\mu^{\frac{d}2}}{(2\pi)^d}\frac{e}{\mu}+L^{d-1}\mu^{\frac{d-1}2}\right)\leq C\left(\mu^{\frac{d}2-1}e+\frac{\mu^{\frac{d-1}2}}L\right).$$
Finally, for $e\geq 3\mu/2$ we have
\begin{multline*}
f_{d,\mu,L}(e)=\frac{1}{L^d}\#\Z^d\cap B\left(\frac{L\sqrt{\mu}}{2\pi}\left(1+\frac{e}\mu\right)^{1/2}\right)\\
\leq C\left(\mu^{d/2}+e^{d/2}+\frac{1}{L^d}\right)\leq C\left(\frac{\mu^{\frac{d-1}2}}{L}+e^{\frac{d}2}\right) 
\end{multline*}
where in the last estimate we have used both $L^{-1}\leq \mu^{1/2}$ and $e\geq 3\mu/2$. This finishes the proof of Lemma~\ref{lem:estim_f_L}.
\end{proof}

Using the bounds~\eqref{eq:estim_f_L_1} and~\eqref{eq:estim_f_L_2} on $f_{d,\mu,L}$, we can now deduce an estimate on $R_{d,\mu,L}$ appearing in~\eqref{eq:Rumin_L}. To simplify our argument, we introduce
\begin{equation}
g_{d,\mu,L}(e)=\begin{cases}
\mu^{\frac{d}2-1}e\,\1(e\leq\mu)+e^{d/2}\,\1(e\geq\mu)&\text{for $\mu>1/L^2$,}\\
e^{d/2}&\text{for $\mu\leq1/L^2$.}
\end{cases}
\label{def_g_d_mu} 
\end{equation}
such that~\eqref{eq:estim_f_L_1} and~\eqref{eq:estim_f_L_2} can be rewritten as
$$f_{d,\mu,L}(e)\leq \epsilon_{d,\mu,L}+Cg_{d,\mu,L}(e)$$
with 
$$\epsilon_{d,\mu,L}=C\begin{cases}
L^{-1}\mu^{\frac{d-1}{2}}&\text{for $\mu>1/L^2$,}\\[0,2cm]
L^{-d} & \text{for $\mu\leq 1/L^2$.}
\end{cases}$$
We then have in all cases
\begin{align*}
\tilde{R}_{d,\mu,L}(\rho)&=\int_0^\ii\left(\sqrt{\rho}-\sqrt{f_{d,\mu,L}(e)}\right)_+^2\,de\\
& \geq \int_0^\ii\left(\sqrt{\rho}-\sqrt{\epsilon_{d,\mu,L}}-\sqrt{g_{d,\mu,L}(e)}\right)_+^2\,de={S}_{d,\mu,L}\left(\big(\sqrt{\rho}-\sqrt{\epsilon_{d,\mu,L}}\big)^2_+\right), 
\end{align*}
with 
\begin{align*}
{S}_{d,\mu,L}(\rho)&=\int_0^\ii\left(\rho-\sqrt{g_{d,\mu,L}(e)}\right)_+^2\,de\\
&\geq C\begin{cases}
\mu^{\frac{d-2}d}\rho^{2}\1(\rho\leq \mu^{2/d})+\rho^{1+2/d}\1(\rho\geq \mu^{2/d})&\text{for $\mu>1/L^2$,}\\[0,2cm]
\rho^{1+2/d}&\text{for $\mu\leq1/L^2$.}
\end{cases} 
\end{align*}
To conclude, it suffices to note that
$$\big(\sqrt{\rho}-\sqrt{\epsilon_{d,\mu,L}}\big)^2_+\geq \alpha_\theta\big(\rho-\theta\epsilon_{d,\mu,L}\big)_+$$
for any $\theta$ bounded away from 0 and $\alpha_\theta$ small enough, and that $S_{d,\mu,L}(\alpha_\theta \rho)\geq \beta_\theta\, S_{d,\mu,L}(\rho)$. 

\subsubsection*{Step 2. Estimate on $Q^{\pm\mp}$}
We again separate the cases $\mu>1/L^2$ and $\mu\leq1/L^2$. 

We start with the case $\mu>1/L^2$ and decompose $Q^{+-}$ as 
$$Q^{+-}=\Pi^+Q\Pi^-=(\Pi^+_0+\Pi^+_1)Q(\Pi^-_0+\Pi^-_1)=Q^{+-}_{00}+Q^{+-}_{01}+Q^{+-}_{10}+Q^{+-}_{11}$$
where
$$\Pi^+_0=\1\big(\mu\leq p^2\leq \mu+\sqrt{\mu}/L\big)\quad\text{and}\quad \Pi^-_0=\1\big(\mu-\sqrt{\mu}/L\leq p^2\leq \mu\big)$$
(we remove the index on $\Pi_{\mu,L}^\pm$ for simplicity).
We have, with $e_k:=L^{-d/2}e^{ik\cdot x}$,
$$\rho_{00}^{+-}=\frac{1}{L^d}\sum_{\substack{k,\ell\in(2\pi\Z/L)^d\\ \mu\leq k^2\leq \mu+\sqrt{\mu}/L\\ \mu-\sqrt{\mu}/L\leq \ell^2\leq \mu}}\pscal{e_k,Qe_\ell}e^{ix(k-\ell)}.$$
The matrix $\pscal{e_k,Qe_\ell}$ has a norm $\leq1$, hence we deduce by Schwarz's inequality that
\begin{equation}
|\rho_{00}^{+-}|\leq \frac{1}{L^d}\sqrt{\#\{\mu\leq k^2\leq \mu+\sqrt{\mu}/L\}}\sqrt{\#\{\mu-\sqrt{\mu}/L\leq \ell^2\leq \mu\}}\leq C\frac{\mu^{\frac{d-1}2}}{L}.
\label{eq:Q00} 
\end{equation}
In the last bound we have used \eqref{eq:estim_diff_nb_ball} and the assumption that $\mu>1/L^2$. For $\rho_{10}^{+-}$, we write, this time,
\begin{align*}
|\tr(VQ^{+-}_{10})|&=\left|\tr\left(\Pi^{-}_0V\frac{\Pi^{+}_1}{|-\Delta_L-\mu|^{1/2}}|-\Delta_L-\mu|^{1/2}Q\right)\right|\\
&\leq\sqrt{\tr(-\Delta_L-\mu)Q}\ \norm{\Pi^{-}_0V\frac{\Pi^{+}_1}{|-\Delta_L-\mu|^{1/2}}}_{\gS_2}. 
\end{align*}
We now have
\begin{align*}
\norm{\Pi^{-}_0V\frac{\Pi^{+}_1}{|-\Delta_L-\mu|^{1/2}}}_{\gS_2}^2&=\frac{1}{L^{2d}}\sum_{\substack{\mu-\sqrt{\mu}/L\leq p^2\leq\mu\\  q^2>\mu+\sqrt{\mu}/L}}\frac{|\widehat{V}(p-q)|^2}{|q^2-\mu|}\\
&\leq \frac{1}{L^{2d}}\frac{L}{\sqrt{\mu}}\sum_{\substack{\mu-\sqrt{\mu}/L\leq p^2\leq\mu\\ |p-k|^2\geq\mu+\sqrt{\mu}/L}}|\widehat{V}(k)|^2\leq C\mu^{\frac{d-2}2}\int_{C_L}|V|^2
\end{align*}
where, in the last estimate we have again used that 
$$L^{-d}\;\#\left\{\mu-\sqrt{\mu}/L\leq p^2\leq \mu\right\}\leq C\frac{\mu^{\frac{d-1}{2}}}{L}.$$
From these bounds we deduce that
\begin{equation}
\int_{C_L}|\rho_{10}^{+-}|^2\leq C\mu^{\frac{d-2}{2}} \tr(-\Delta_L-\mu)Q.
\label{eq:Q10} 
\end{equation}
The term $\rho^{+-}_{01}$ is treated similarly.
We conclude this paragraph with an estimate on $\rho^{+-}_{11}$, which we derive by the same method as for~\eqref{eq:final_estim_off_diag_1}, in the proof of Theorem~\ref{thm:LT-rho}:
\begin{equation}
 \int_{C_L}|\rho_{Q^{+-}_{11}}|^2\leq
(2\pi)^{-d}\norm{\Phi_{d,\mu,L}}_{L^\ii((2\pi\Z/L)^d)}\;\tr_{L^2(C_L)}(-\Delta_L-1)Q,
\label{eq:Q11} 
\end{equation}
with
\begin{equation}
\Phi_{d,\mu,L}(k):=\frac{1}{L^d}\sum_{\substack{p\in(2\pi\Z/L)^d\\ |p|^2\leq \mu-\sqrt{\mu}/L\\
|p-k|^2\geq\mu+\sqrt{\mu}/L}}\frac{1}{\big(\mu-|p|^2\big)^{1/2}\;\big(|p-k|^2-\mu\big)^{1/2}}.
\label{eq:def_Phi_d_L} 
\end{equation}
The function $\Phi_{d,\mu,L}$ is a Riemann approximation of $\mu^{(d-2)/2}\Phi_d(\cdot/\sqrt{\mu})$.
In order to prove that $\Phi_{d,\mu,L}$ is uniformly bounded on $(2\pi\Z/L)^d$ by $C\mu^{(d-2)/2}$, independently of $L$, we compare it with its limit. For every $p$ in the sum above, we introduce the ball $B_p$ of radius $\eta/L$, centered at $p$. We will fix the value of $\eta$ later, but as a first constraint we impose that
\begin{equation}
\sqrt{\mu-\frac{\sqrt{\mu}}L}+\frac{\eta}{L}\leq \sqrt{\mu-\frac{\sqrt{\mu}}{2L}}\quad \text{ and }\quad  \sqrt{\mu+\frac{\sqrt{\mu}}L}-\frac{\eta}{L}\geq \sqrt{\mu+\frac{\sqrt{\mu}}{2L}},
\label{eq:constraint-balls} 
\end{equation}
for all $\mu>1/L^2$ and $L\geq1$. It is easy to verify that the previous condition is satisfied when, for instance, $\eta\leq 1/8$. 
The constraints~\eqref{eq:constraint-balls} imply that
\begin{equation}
\forall p'\in B_p,\qquad |p'|^2\leq \mu-\frac{\sqrt{\mu}}{2L},\qquad |p'-k|^2\geq\mu+\frac{\sqrt{\mu}}{2L}.
\label{eq:prop_p_ball} 
\end{equation}
Next we compute the gradient 
\begin{multline*}
\nabla_p\frac{1}{\big(\mu-|p|^2\big)^{1/2}\;\big(|p-k|^2-\mu\big)^{1/2}}\\=\left(\frac{p}{\mu-|p|^2}-\frac{p-k}{|p-k|^2-\mu}\right)\frac{1}{\big(\mu-|p|^2\big)^{1/2}\;\big(|p-k|^2-\mu\big)^{1/2}}.
\end{multline*}
For $p'$ satisfying~\eqref{eq:prop_p_ball}, we have
$$\frac{|p'|}{\mu-|p'|^2}\leq \frac{\sqrt{\mu-\frac{\sqrt{\mu}}{2L}}}{\frac{\sqrt{\mu}}{2L}}=2L\sqrt{1-\frac{1}{2L\sqrt\mu}}\leq2\, L$$
and
$$\frac{|p'-k|}{|p'-k|^2-\mu}\leq \frac{\sqrt{\mu+\frac{\sqrt{\mu}}{2L}}}{\frac{\sqrt{\mu}}{2L}}=2L\sqrt{1+\frac{1}{2L\sqrt\mu}}\leq\sqrt6\, L.$$
We therefore deduce by Taylor's formula, that for every $p'\in B_p$
\begin{multline*}
\left|\frac{1}{\big(\mu-|p|^2\big)^{1/2}\;\big(|p-k|^2-\mu\big)^{1/2}}-\frac{1}{\big(\mu-|p'|^2\big)^{1/2}\;\big(|p'-k|^2-\mu\big)^{1/2}} \right|\\
\leq (2+\sqrt6)\;(2\eta)\;\sup_{q\in B_p}\frac{1}{\big(\mu-|q|^2\big)^{1/2}\;\big(|q-k|^2-\mu\big)^{1/2}}.
\end{multline*}
Choosing $\eta$ small enough, we can therefore make sure that
$$\sup_{q\in B_p}\frac{1}{\big(\mu-|q|^2\big)^{1/2}\;\big(|q-k|^2-\mu\big)^{1/2}}\leq 2\frac{1}{\big(\mu-|p|^2\big)^{1/2}\;\big(|p-k|^2-\mu\big)^{1/2}}$$
and then that
$$\frac{1}{\big(\mu-|p|^2\big)^{1/2}\;\big(|p-k|^2-\mu\big)^{1/2}}\leq 4\inf_{q\in B_p}\frac{1}{\big(\mu-|q|^2\big)^{1/2}\;\big(|q-k|^2-\mu\big)^{1/2}}.$$
Using that the balls $B_p$ are disjoint for $\eta$ small enough, we finally obtain 
\begin{align*}
\Phi_{d,\mu,L}(k)&\leq \frac{4}{L^d}\sum_{\substack{p\in(2\pi\Z/L)^d\\ |p|^2\leq \mu-\sqrt{\mu}/L\\
|p-k|^2\geq\mu+\sqrt{\mu}/L}}\frac{1}{|B_p|}\int_{B_p}\frac{1}{\big(\mu-|p'|^2\big)^{1/2}\;\big(|p'-k|^2-\mu\big)^{1/2}}dp'\\
&\leq \frac{4}{|S^{d-1}|\eta^d}\,\mu^{\frac{d-2}2}\Phi_d(|k|/\sqrt{\mu})\leq C\mu^{\frac{d-2}2},
\end{align*}
since $\Phi_d$ is bounded by Lemma~\ref{lem:Phi_bounded}.
Summarizing all our estimates, we have proved that 
$$\mu^{1-\frac{2}d}\int_{C_L}|\rho^{+-}-\rho_{00}^{+-}|^2\leq C \tr_{L^2(C_L)}(-\Delta_L-\mu)Q.$$
Using now both that $|x-\epsilon|\geq (|x|-\epsilon)_+$ and $\rho_{00}^{+-}\leq C\mu^{(d-1)/2}/L$, we deduce that
$$\mu^{1-\frac{2}d} \int_{C_L}\left(|\rho^{+-}|-C\mu^{(d-1)/2}/L\right)_+^2\leq C \tr_{L^2(C_L)}(-\Delta_L-\mu)Q$$
with a constant $C$ that does not depend on $L$, for $\mu>1/L^2$.
This completes the proof of~\eqref{eq:LT-rho-box} when $\mu>1/L^2$.

The case $\mu\leq1/L^2$ is similar, except that we only decompose 
$$\Pi^+=\1(\mu\leq p^2\leq \mu+1/L^2)+\1(p^2>\mu+1/L^2),$$ 
and retain $\Pi^-$.
We get two terms $Q^{+-}_0$ and $Q^{+-}_1$. We estimate $\rho^{+-}_0$ in $L^\ii$ as in~\eqref{eq:Q00}, getting
$$|\rho^{+-}_0|\leq\frac{1}{L^d}\sqrt{\#\{\mu\leq k^2\leq \mu+1/L^2\}}\sqrt{\#\{\ell^2\leq \mu\}}\leq \frac{C}{L^d},
$$
since $\mu\leq 1/L^2$ (each of the two sets above contains a finite number of points which does not increase with $L$). Finally, we estimate $\rho^{+-}_1$ as in~\eqref{eq:Q10} and obtain 
$$|\tr_{L^2(C_L)}(VQ^{+-}_1)|\leq \norm{\Pi^{-}V\frac{\Pi^{+}_1}{|-\Delta_L-\mu|^{1/2}}}_{\gS_2}\;\sqrt{\tr_{L^2(C_L)}(-\Delta_L-\mu)Q}.$$
This time we have 
\begin{align*}
\norm{\Pi^{-}V\frac{\Pi^{+}_1}{|-\Delta_L-\mu|^{1/2}}}_{\gS_2}^2&=\frac{1}{L^{2d}}\sum_{\substack{p^2\leq\mu\\  q^2>\mu+1/L^2}}\frac{|\widehat{V}(p-q)|^2}{q^2-\mu}\\
&\leq \frac{1}{L^{2d}}{L^2}\sum_{\substack{p^2\leq\mu\\  q^2>\mu+1/L^2}}|\widehat{V}(k)|^2\leq C\,L^{2-d}\int_{C_L}|V|^2.
\end{align*}
Since $d\geq2$, this completes the proof of Theorem~\ref{thm:LT-rho-box}.\qed

\subsection{Thermodynamic limit}\label{sec:thermo-limit}

With the Lieb-Thirring inequality~\eqref{eq:LT-V-box} at hand, we can now relate the well-defined total relative energy in a large box to the one we have defined in Section~\ref{sec:LT_potential_version}. The following can therefore serve as an \emph{a posteriori} justification of our definition of $\tr_V(-\Delta+V-\mu)Q_V$.

\begin{theorem}[Thermodynamic Limit, $d\geq2$]\label{thm:thermo-limit}
We assume that $d\geq2$ and $\mu\geq0$. Let $V$ be a real-valued function in $L^1(\R^d)\cap L^{1+d/2}(\R^d)$. Then we have
\begin{multline}
\tr_V(-\Delta+V-\mu)Q_V\\
=\lim_{L\to\ii}\left(-\tr_{L^2(C_L)}(-\Delta_L+V\1_{C_L}-\mu)_-+\tr_{L^2(C_L)}(-\Delta_L-\mu)_- - \mu\int_{C_L}V\right),
\label{eq:thermo-limit}
\end{multline}
where the left side is defined in Definition~\ref{def:total_energy}, and $-\Delta_L$ is the Laplacian with periodic boundary conditions on $C_L=[-L/2,L/2)^d$.
\end{theorem}

\medskip

\begin{proof}[Sketch of the proof]
We quickly explain the main steps of the proof, which proceeds by showing an upper and a lower bound.

Let us fix a smooth finite-rank operator $Q\in\cK$ which we write in the form
$$Q=\sum_{i,j=1}^K \bigg(q_{i,j}^{++}|u_i\rangle\langle u_j|+q_{i,j}^{+-}|u_i\rangle\langle v_j|+q_{i,j}^{-+}|v_i\rangle\langle u_j|+q_{i,j}^{--}|v_i\rangle\langle v_j|\bigg)$$
where $(u_i)_{i=1}^K$ and $(v_i)_{i=1}^K$ are orthonormal systems of the kernel and the range of $\Pi^-$, respectively. The constraint that $-\Pi^-\leq Q=Q^*\leq1-\Pi^-$ is then only reflected in the coefficients $q_{ij}^{\pm/\pm}$ (see~\cite[App. B]{HaiLewSer-09} for an explicit representation of $q_{i,j}^{\pm/\pm}$). By assumption, the functions $u_i$ and $v_i$ are all smooth. Now we build from $Q$ a test state $Q_L$ in the box $C_L$, by simply replacing the $u_i$'s and $v_i$'s by orthonormal sequences $(u_{i,L})_{i=1}^K$ and $(v_{i,L})_{i=1}^K$ in, respectively, the kernel and the range of $\Pi^-_{\mu,L}$. We can do this in such a way that $\1_{C_L}u_{i,L}\to u_i$, $\1_{C_L}v_{i,L}\to v_i$, $\1_{C_L}\nabla u_{i,L}\to\nabla u_i$ and $\1_{C_L}\nabla v_{i,L}\to\nabla v_i$ in $L^2(\R^d)$, as $L\to\ii$. One simple way to realize that is to periodize the functions as 
$$\tilde{u}_{i,L}(x)=L^{-d}\sum_{k\in(2\pi\Z/L)^d}\widehat{u_i}(k)\,e^{-ik\cdot x},$$
and then to orthonormalize the so-obtained system. Similar arguments have already been used and detailed in~\cite{CanDelLew-08a}. The test state is then defined as
$$Q_L:=\sum_{i,j=1}^K \bigg(q_{i,j}^{++}|u_{i,L}\rangle\langle u_{j,L}|+q_{i,j}^{+-}|u_{i,L}\rangle\langle v_{j,L}|+q_{i,j}^{-+}|v_{i,L}\rangle\langle u_{j,L}|+q_{i,j}^{--}|v_{i,L}\rangle\langle v_{j,L}|\bigg)$$
and it satisfies the constraint $-\Pi^-_{L,\mu}\leq Q_L\leq 1-\Pi^-_{L,\mu}$ by construction. 
We also have 
$$\lim_{L\to\ii}\left(\tr(-\Delta_L-\mu)Q_L+\int_{C_L}V\rho_{Q_L}\right)=\tr(-\Delta-\mu)Q+\int_{\R^d}V\rho_Q.$$
Because we obviously have a variational principle in the box,
\begin{multline*}
-\tr_{L^2(C_L)}(-\Delta_L+V\1_{C_L}-\mu)_-+\tr_{L^2(C_L)}(-\Delta_L-\mu)_- - \mu\int_{C_L}V\\
=\inf_{-\Pi^-_{L,\mu}\leq Q\leq 1-\Pi^-_{L,\mu}}\tr(-\Delta_L-\mu)Q+\int_{C_L}\rho_{Q}V, 
\end{multline*}
we deduce the upper bound
\begin{multline*}
\limsup_{L\to\ii}\left(-\tr_{L^2(C_L)}(-\Delta_L+V\1_{C_L}-\mu)_-+\tr_{L^2(C_L)}(-\Delta_L-\mu)_- - \mu\int_{C_L}V\right)\\
\leq  \tr(-\Delta-\mu)Q+\int_{\R^d}V\rho_Q.
\end{multline*}
From the variational principle~\eqref{eq:variational} in the whole space and the density of smooth finite-rank operators in $\cK$, as stated in Lemma~\ref{lem:density}, we conclude that
\begin{multline*}
\limsup_{L\to\ii}\left(-\tr_{L^2(C_L)}(-\Delta_L+V\1_{C_L}-\mu)_-+\tr_{L^2(C_L)}(-\Delta_L-\mu)_- - \mu\int_{C_L}V\right)\\
\leq  \tr_V(-\Delta+V-\mu)Q_V.
\end{multline*}

In a second step we prove the reverse inequality, with the $\limsup$ replaced by a $\liminf$. We consider a sequence $L_n\to\ii$ realizing this $\liminf$. Denoting by 
$$Q_n=\1(-\Delta_{L_n}+V\1_{C_{L_n}}\leq\mu)-\1(-\Delta_{L_n}\leq\mu)$$ 
the corresponding state, we know from our estimates that 
$$\norm{|-\Delta_{L_n}-\mu|^{1/2}Q_n^{\pm\pm}|-\Delta_{L_n}-\mu|^{1/2}}_{\gS_1(L^2(C_{L_n}))}\leq C,$$
$$\norm{Q_n|-\Delta_{L_n}-\mu|^{1/2}}_{\gS_2(L^2(C_{L_n}))}\leq C,$$
$$\norm{Q_n}\leq 1,$$
and that $\1_{C_{L_n}}\rho_{Q_n}=\rho^1_n+\rho^2_n$ where $\norm{\rho^2_n}_{L^\ii}\leq C\mu^{(d-1)/2}(L_n)^{-1}$ and $\rho^1_n$ is bounded in $L^2(\R^d)+L^{1+2/d}(\R^d)$. Passing to weak limits, using $V\in L^1(\R^d)\cap L^{1+d/2}(\R^d)$, we deduce that
\begin{multline*}
 \liminf_{L\to\ii}\left(-\tr_{L^2(C_L)}(-\Delta_L+V\1_{C_L}-\mu)_-+\tr_{L^2(C_L)}(-\Delta_L-\mu)_- - \mu\int_{C_L}V\right)\\
\geq\tr(R^{++}+R^{--})+\int_{\R^d}V\rho
\end{multline*}
where $\1_{C_{L_n}}\rho_{Q_n}\wto\rho$ and $\1_{C_{L_n}}|-\Delta_{L_n}-\mu|^{1/2}Q_n^{\pm\pm}|-\Delta_{L_n}-\mu|^{1/2}\1_{C_{L_n}}\wto R^{\pm\pm}$. Similarly, we have $\1_{C_{L_n}}Q_n|-\Delta_{L_n}-\mu|^{1/2}\1_{C_{L_n}}\wto S$ weakly-$\ast$ in $\gS_2(L^2(\R^d))$ and $\1_{C_{L_n}}Q_n\1_{C_{L_n}}\wto Q$ weakly-$\ast$ in $\cB$. We now claim that $Q\in\cK$, $\rho_Q=\rho$, $S=Q|-\Delta-\mu|^{1/2}$, and $R^{\pm\pm}=|-\Delta-\mu|^{1/2}Q^{\pm\pm}|-\Delta-\mu|^{1/2}$. All this can be seen by testing against smooth functions of compact support, and we skip the details. We conclude that
\begin{multline*}
 \liminf_{L\to\ii}\left(-\tr_{L^2(C_L)}(-\Delta_L+V\1_{C_L}-\mu)_-+\tr_{L^2(C_L)}(-\Delta_L-\mu)_- - \mu\int_{C_L}V\right)\\
\geq\tr_V(-\Delta+V-\mu)Q_V
\end{multline*}
by the variational principle~\eqref{eq:variational}. This completes our sketch of the proof of Theorem~\ref{thm:thermo-limit}.
\end{proof}

\subsection{Extension to positive temperature}
In this section we extend our results to smooth partition functions, following~\cite{FraLewLieSei-11}. This means we consider a smooth function $f:\R\to\R$ tending to zero at infinity, and we look for a lower bound on the formal expression
\begin{equation}
\tr\bigg(f(-\Delta+V) -f(-\Delta)-f'(-\Delta)V\bigg). 
\label{eq:goal-positive-temp}
\end{equation}
Our results above dealt with the function $f_{0,\mu}(x)=-(x-\mu)_-$. Here we typically think of the free energy for a Fermi-Dirac distribution at positive temperature $T$ and chemical potential $\mu$, corresponding to
\begin{equation}
f_{T,\mu}(x)=-T\log\left(1+e^{-(x-\mu)/T}\right),
\label{eq:def_free_energy_positive-temp} 
\end{equation}
which converges to $f_{0,\mu}$ in the limit $T\to0$. We will, however, be able to treat general functions $f$, provided they are concave and decay fast enough at infinity. The trick is to write $f$ as an average of the reference functions $f_{0,\mu}$ as 
\begin{equation}
f(x)=\int_{\R}(x-\lambda)_-\, f''(\lambda)\,d\lambda,
\label{eq:integral-formula-f} 
\end{equation}
leading to the formal expression
\begin{equation}
\tr\bigg(f(-\Delta+V) -f(-\Delta)-f'(-\Delta)V\bigg)=-\int_\R \tr_V(-\Delta+V-\lambda)Q_{\lambda,V} \; f''(\lambda)\,d\lambda
\label{eq:def_positive_temp_abstract} 
\end{equation}
where
$$Q_{\lambda,V}:=\1(-\Delta+V\leq\lambda)-\1(-\Delta\leq\lambda).$$
When $f$ is concave, the integrand in the right side of~\eqref{eq:def_positive_temp_abstract} is $\geq0$ since $\tr_V(-\Delta+V-\lambda)Q_{\lambda,V}\leq0$, hence the integral always makes sense in $\R^+\cup\{+\ii\}$. We may thus use this as a definition for the left side. In the following result we justify this formal calculation by a thermodynamic limit, and we state the corresponding Lieb-Thirring inequality.

\begin{theorem}[Lieb-Thirring inequality for smooth partition functions, $d\geq2$]\label{thm:thermo-limit-positive-temp}
Let $f:\R\to\R$ be a concave function such that $f''\in L^\ii_{\rm loc}(\R)$ and
\begin{equation}
\int_0^{\ii}\lambda^{1+\frac{d}{2}}|f''(\lambda)|\,d\lambda<\ii
\label{eq:assumption_f_positive_temp} 
\end{equation}
for some $d\geq2$. Then, for $V\in L^1(\R^d)\cap L^{1+d/2}(\R^d)$, we have
\begin{multline}
\lim_{L\to\ii}\left\{\tr_{L^2(C_L)}f\big(-\Delta_L+V\1_{C_L}\big) -\tr_{L^2(C_L)}f(-\Delta_L)-\int_{C_L}\rho_{f'(-\Delta_L)}V\right\}\\
= -\int_\R \tr_V(-\Delta+V-\lambda)Q_{\lambda,V} \; f''(\lambda)\,d\lambda,
\label{eq:thermo-limit-positive-temp}
\end{multline}
where, as before, $-\Delta_L$ is the Laplacian with periodic boundary conditions on $C_L=[-L/2,L/2)^d$.
Moreover, we have the following inequality 
\begin{multline}
\int_\R \tr_V(-\Delta+V-\lambda)Q_{\lambda,V} \; f''(\lambda)\,d\lambda\\
\leq L(d)\int_\R d\lambda\,f''(\lambda) \int_{\R^d}dx\;\left((V(x)-\lambda)_-^{1+\tfrac{d}2}-\lambda_+^{1+\tfrac{d}2}+\frac{2+d}
2\,\lambda_+^{\tfrac{d}2}\,V(x)\right).
\label{eq:LT-positive-temp}
\end{multline}
\end{theorem}

\bigskip

The result holds the same under a weaker assumption on $f$ than~\eqref{eq:assumption_f_positive_temp}, provided that the right side of~\eqref{eq:thermo-limit-positive-temp} is interpreted in a suitable manner. As such, Theorem~\ref{thm:thermo-limit-positive-temp} already applies to the Fermi-Dirac free energy $f_{T,\mu}$ as given in~\eqref{eq:def_free_energy_positive-temp}, since we have
$$f''_{T,\mu}(\lambda)=- \frac{e^{(\lambda-\mu)/T}}{T\left(1+e^{(\lambda-\mu)/T}\right)^2}$$
in this case.

\medskip

\begin{proof}
The Lieb-Thirring inequality~\eqref{eq:LT-positive-temp} is an immediate consequence of~\eqref{eq:LT-V} and we only explain the thermodynamic limit~\eqref{eq:thermo-limit-positive-temp}. First, it follows from the integral formula~\eqref{eq:integral-formula-f} and our assumption~\eqref{eq:assumption_f_positive_temp}, that $f(-\Delta_L)$ and $f(-\Delta_L+V\1_{C_L})$ are both trace-class. Using~\eqref{eq:integral-formula-f}, we obtain the identity
\begin{multline*}
\tr_{L^2(C_L)}f\big(-\Delta_L+V\1_{C_L}\big) -\tr_{L^2(C_L)}f(-\Delta_L)-\int_{C_L}\rho_{f'(-\Delta_L)}V\\
=-\int_\R \tr_{L^2(C_L)}(-\Delta_L+V\1_{C_L}-\lambda)Q_{\lambda,V,L} \; f''(\lambda)\,d\lambda
\end{multline*}
with $Q_{\lambda,V,L}=\1(-\Delta_L+V\1_{C_L}\leq\lambda)-\1(-\Delta_L\leq\lambda)$. By the Lieb-Thirring inequality~\eqref{eq:LT-V-box} in the box, we have for $\lambda\geq1/L^2$
\begin{align*}
&-\tr_{L^2(C_L)}(-\Delta_L+V\1_{C_L}-\lambda)Q_{\lambda,V,L}\\
&\qquad \leq\tilde{L}(d)\int_{C_L}\left((V(x)-\lambda)_-^{1+\tfrac{d}2}-\lambda^{1+\tfrac{d}2}+\frac{2+d}
2\,\lambda^{\tfrac{d}2}\,V(x)+\frac{\lambda^{\tfrac{d-1}2}}{L}|V(x)|\right)\,dx\\
&\qquad \leq C\left(1+\lambda^{\frac{d-2}{2}}+\frac{\lambda^{\frac{d-1}{2}}}L\right).
\end{align*}
The last estimate is obtained by first replacing the domain of integration $C_L$ by $\R^3$ (the integrand being $\geq0$), and then using that
$$\int_{\R^3}\left((V(x)-\lambda)_-^{1+\tfrac{d}2}-\lambda^{1+\tfrac{d}2}+\frac{2+d}
2\,\lambda^{\tfrac{d}2}\,V(x)\right)\,dx\underset{\lambda\to\ii}{\sim} \frac{d(d+2)}{8}\lambda^{\frac{d-2}{2}}\,\int_{\R^3}|V|^2.$$
For $0\leq\lambda\leq1/L^2$, we use~\eqref{eq:LT-V-box-small-mu} instead and obtain
$$-\tr_{L^2(C_L)}(-\Delta_L+V\1_{C_L}-\lambda)Q_{\lambda,V,L}\leq C\left(1+L^{-d}\right).$$
Finally, for $\lambda<0$, we simply note that
\begin{equation*}
-\tr_{L^2(C_L)}(-\Delta_L+V\1_{C_L}-\lambda)Q_{\lambda,V,L}=\tr(-\Delta_L+V\1_{C_L}-\lambda)_-.
\end{equation*}
This last term vanishes when $\lambda\leq \inf\sigma(-\Delta_L+V\1_{C_L})$ and it is bounded by $\tr(-\Delta_L+V\1_{C_L})_-\leq C(1+L^{-d})$ otherwise.
As a conclusion, for $L$ large enough we have a uniform bound 
\begin{equation}
 -\tr_{L^2(C_L)}(-\Delta_L+V\1_{C_L}-\lambda)Q_{\lambda,V,L}\leq C\left(1+\lambda^{\frac{d-1}{2}}\right)\1(\lambda\geq-M)
\end{equation}
with $M<\liminf_{L\to\ii}\inf\sigma(-\Delta_L+V\1_{C_L})$. On the other hand we know by Theorem~\ref{thm:thermo-limit} that
$$\lim_{L\to\ii}\tr_{L^2(C_L)}(-\Delta_L+V\1_{C_L}-\lambda)Q_{\lambda,V,L}=\tr_V(-\Delta+V-\lambda)Q_{\lambda,V}$$
for every fixed $\lambda$. Now~\eqref{eq:thermo-limit-positive-temp} simply follows from Lebesgue's dominated convergence theorem.
\end{proof}

\section{Extension to more general background operators}\label{sec:general}

In the previous sections, we have considered perturbations of a constant density $\rho_0$. Our approach is, in fact, more general and we explain now how to handle other background densities. We typically think of a periodic background but, since we actually need very few assumptions, we state below an abstract theorem. We comment on the assumptions in the periodic case in Section~\ref{sec:periodic}.

\subsection{An abstract Lieb-Thirring inequality with positive background}
We consider a bounded-below self-adjoint operator $H$ in $L^2(\R^d,\C^q)$, with $d\geq2$, and we fix a real number $\mu\in\R$. We assume that there is a constant $C$ and an $\epsilon>0$ such that

\smallskip

\begin{enumerate}
\item[(A1)] $\displaystyle\rho_{\1(|H-\mu|\leq E)}(x)\leq C\left(E+E^{{d}/{2}}\right)$ \quad for all $E\geq0$ and a.e. $x\in\R^d$;

\medskip

\item[(A2)] $\displaystyle \norm{\frac{\1(\mu\leq H\leq \mu+\epsilon)}{|H-\mu|^{1/4}}\;V\;\frac{\1(\mu-\epsilon\leq H\leq \mu)}{|H-\mu|^{1/4}}}_{\gS_2}\leq C\norm{V}_{L^2}$\ for all $V\in L^2(\R^d)$;

\medskip

\item[(A3)] $\displaystyle \rho_0(x):=\rho_{\1(H\leq\mu)}(x)\geq\epsilon$ for a.e. $x\in\R^d$.
\end{enumerate}

\smallskip

\noindent We define 
$$\Pi^-:=\1_{(-\ii,\mu)}(H)\qquad\text{and}\qquad \Pi^+=1-\Pi^-.$$
We emphasize that (A1) implies that $\rho_{\1(H=\mu)}\equiv0$, hence that $\mu$ is not an eigenvalue of $H$. With respect to the projections $\Pi^-$ and $\Pi^+$ we can decompose any bounded operator $Q=Q^{++}+Q^{-+}+Q^{+-}+Q^{--}$.
Similarly as in Definition~\ref{def:kinetic}, we define the relative kinetic energy by
$$\tr_0(H-\mu)Q=\tr|H-\mu|^{1/2}\big(Q^{++}-Q^{--}\big)|H-\mu|^{1/2}$$
for any bounded self-adjoint operator $Q$ such that $|H-\mu|^{1/2}Q^{\pm\pm}|H-\mu|^{1/2}$ are trace-class.

\begin{theorem}[Abstract Lieb-Thirring inequality, density version, $d\geq2$]\label{thm:abstract}
We assume that the bounded-below self-adjoint operator $H$ satisfies (A1)--(A3).
Let $Q$ be a self-adjoint operator such
that $-\Pi^-\leq Q\leq \Pi^+$ and such that 
$\big|H-\mu\big|^{1/2}Q^{\pm\pm}\big|H-\mu\big|^{1/2}$ are
trace-class. Then $Q$ is locally trace-class and the corresponding density
satisfies
\begin{equation}
\rho_Q\in L^{1+\tfrac2d}(\R^d)+L^{2}(\R^d).
\label{eq:prop_density_abstract} 
\end{equation}
Moreover, there exists a positive constant $K$ (depending
only on $d$, $q$, $\mu$, $C$ and $\epsilon$) such that
\begin{multline}
\tr_0(H-\mu)Q\\
\;\geq\;
K\int_{\R^d}\left(\big(\rho_0(x)+\rho_Q(x)\big)^{1+\tfrac{2}d}-\rho_0(x)^{
1+\tfrac{2}d}-\frac{2+d}d \rho_0(x)^{\tfrac{2}d}\,\rho_Q(x)\right)dx
\label{eq:LT-rho_abstract}
\end{multline}
with $\rho_0(x)$ the background density of $\Pi^-$, defined above in (A3).
\end{theorem}

\begin{remark}\rm
For simplicity we restrict our attention to $d\geq2$ but, with appropriate modifications, a similar result holds for $d=1$. If in Assumption (A1) the exponent $d/2$ is replaced by $\delta\geq1$, our method still applies but the resulting lower bound is of course different. For instance, if relativistic effects are taken into account, $d/2$ should be $d$ in (A1), in which case the exponent $1+2/d$ in~\eqref{eq:LT-rho_abstract} becomes $1+1/d$.
\end{remark}

\begin{remark}\rm 
Similarly to Theorem~\ref{thm:LT-V}, it is possible to deduce from \eqref{eq:LT-rho_abstract} a dual estimate on $\tr_V(H+V-\mu)Q_V$, where $Q_V=\1(H+V\leq \mu)-\Pi^-$, for any potential $V\in L^2(\R^d)\cap L^{1+d/2}(\R^d)$. For brevity we will not discuss this here. 
\end{remark}

In applications, we typically think of $H=-\Delta+W(x)$ where $W$ is a sufficiently regular function, and of $\mu$ strictly above the infimum of the essential spectrum of $H$. In Assumption (A1), the $E^{d/2}$ behavior of the density for large $E$ is a rather general fact which we discuss below. On the other hand, the small $E$ behavior in (A1) as well as (A2) are assumptions on $H$ close to the Fermi surface. Vaguely speaking, (A1) is a (rather weak) assumption on the regularity of the spectral projections uniformly in $x$-space, whereas (A2) controls the interactions between particles inside and outside the Fermi sea.

Next we show how to verify the large $E$ behavior in (A1), under the assumption that $H=-\Delta+W(x)$ with $W$ bounded from below. 

\begin{lemma}\label{lem:Feynman-Kac}
Let $W\in L^1_{\rm loc}(\R^d)$ such that $W_-\in L^\ii(\R^d)$, and consider the Friedrichs self-adjoint extension of $-\Delta+W$ on $C^\ii_c(\R^d)$. Then $\rho_{\1(-\Delta+W\leq E)}$ is uniformly bounded on $\R^d$ for every $E\in\R$ and 
$$\rho\left[\1(-\Delta+W\leq E)\right]\leq \left(\frac{e}{2\pi d}\right)^{d/2}\left(\norm{W_-}_{L^\ii}+E\right)^{d/2},$$
holds almost everywhere.
\end{lemma}

\begin{proof}
Since $W$ is bounded from below, we have, by the Feynman-Kac formula, 
$$\rho\left[e^{-t\left(-\Delta+W\right)}\right]\leq \frac{e^{t\norm{W_-}_{L^\ii}}}{(4\pi t)^{d/2}}$$
where $\rho[A](x)=A(x,x)$ denotes the density of an operator $A$. 
Using the inequality $\1(x\leq E)\leq e^{-t(x-E)}$ we deduce that 
\begin{equation*}
\rho\left[\1(-\Delta+W\leq E)\right]\leq e^{tE}\rho\left[e^{-t\left(-\Delta+W\right)}\right]\leq \frac{e^{t(\norm{W_-}_{L^\ii}+E)}}{(4\pi t)^{d/2}}.
\end{equation*}
Optimizing this bound with respect to $t$ gives the result.
\end{proof}

\subsection{Application to periodic backgrounds}\label{sec:periodic}
In this section we restrict ourselves to periodic systems, that is, we take 
$$H=-\Delta+W(x)$$
where $W$ is a $\Z^d$-periodic function which we assume to be sufficiently regular.
Of course, we could as well consider other lattices than $\Z^d$.
It is well known, see, e.g.,~\cite[Sec. XIII.16]{ReeSim4}, that the spectrum of $H$ is the union of bands
$$\sigma(H)=\sigma_{\rm ess}(H)=\bigcup_{n\geq1}\left\{\lambda_n(\xi),\ \xi\in[-\pi,\pi]^d\right\},$$
where $\lambda_n(\xi)$ denotes the sequence of Bloch-Floquet eigenvalues of $H$ with corresponding eigenvectors $u_n(\xi,x)$. Each $\lambda_n$ is a periodic Lipschitz function of $\xi$, but the map $\xi\mapsto u_n(\xi)\in L^2((0,1)^d)$ is only piecewise smooth because of possible degeneracies. Writing for instance $H=-\Delta/2+(-\Delta/2+W)\geq -\Delta/2-C$ and comparing the $\lambda_n(\xi)$ with the eigenvalues of the periodic Laplacian in each Bloch sector, it can be seen that
$$\lambda_n(\xi)\geq a\, n^{2/d}-b$$
for some constants $a,b>0$ independent of $\xi$. Hence for every fixed $\mu\in\R$, there is only a finite number of $n$'s such that $\lambda_n(\xi)=\mu$ for some $\xi\in[-\pi,\pi)^d$. 

Let us fix $\mu>\inf\sigma_{\rm ess}(-\Delta+W)$. Then we have
$$\rho_0(x)=(2\pi)^{-d}\sum_{n\geq1}\int_{[-\pi,\pi)^d}d\xi\;\1(\lambda_n(\xi)\leq\mu)\,|u_n(\xi,x)|^2. $$
Since $u_1(0,x)$ is strictly positive, we easily conclude, by continuity in $\xi$, that $\rho_0(x)\geq\epsilon>0$, and hence that (A3) is verified.

\begin{figure}[h]
\centering
\includegraphics{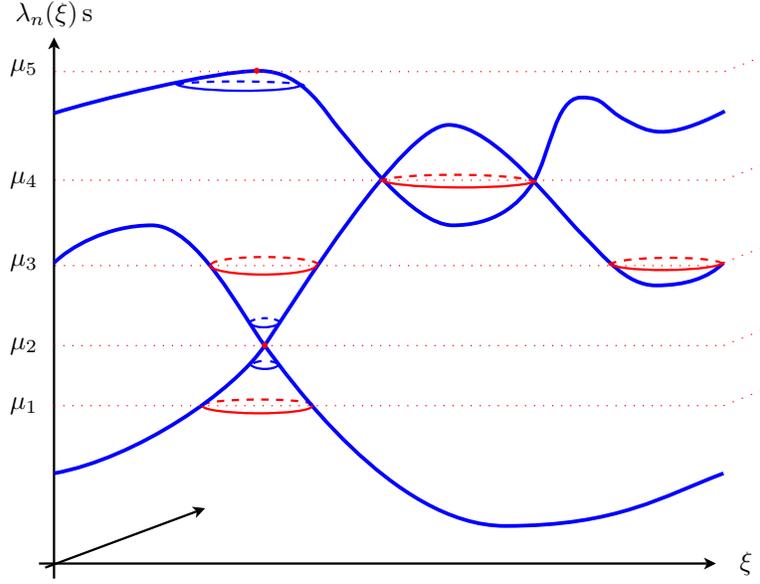}
\caption{Typical Fermi surfaces for a two-dimensional periodic Schrödinger operator.\label{fig:bands}}
\end{figure}

Now we give some ideas on how one can verify Assumptions (A1) and (A2) in practice. First, we have
$$\rho_{\1(|-\Delta+W-\mu|\leq E)}(x)=(2\pi)^{-d}\sum_{n\geq1}\int_{[-\pi,\pi)^d}d\xi \;\1(|\lambda_n(\xi)-\mu|\leq E)\,|u_n(\xi,x)|^2. $$
Under suitable assumptions on $W$, $u_n(\cdot,\xi)$ is bounded in $L^\ii(\R^d)$, uniformly with respect to $\xi$, for each fixed $n\geq1$. In this case, Assumption (A2) follows if the eigenvalues satisfy the following property:
\begin{equation}
\left|\left\{\xi\in[-\pi,\pi]^d,\ |\lambda_n(\xi)-\mu|\leq E\right\}\right|\leq CE.
\label{eq:size_set_periodic}
\end{equation}
This is generically true: If there is a unique $n$ such that the graph of $\lambda_n$ crosses $\mu$, and if $\nabla_\xi\lambda_n(\xi)\neq0$ for all $\xi$ with $\lambda_n(\xi)=\mu$, one can easily verify that~\eqref{eq:size_set_periodic} is satisfied ($\mu_1$ in Figure~\ref{fig:bands}). At a point $\xi$ such that $\nabla_\xi\lambda_n(\xi)=0$, the validity of~\eqref{eq:size_set_periodic} depends on the order of vanishing at this point. If, for instance, the second derivative is invertible, then~\eqref{eq:size_set_periodic} holds in any dimension $d\geq2$ ($\mu_5$ in Fig.~\ref{fig:bands}). Only the $\xi$'s which have a high (depending on the dimension $d$) order of vanishing can make~\eqref{eq:size_set_periodic} fail. When the Fermi surface is disconnected, each component being as before, the result is the same ($\mu_3$ in Fig.~\ref{fig:bands}). Finally, if $\lambda_n(\xi)=\lambda_m(\xi)=\mu$ for $n\neq m$, the analysis is similar. For instance, transversal crossing of surfaces ($\mu_4$ in Fig.~\ref{fig:bands}) as well as Dirac-type cone singularities ($\mu_2$ in Fig.~\ref{fig:bands}) are allowed.

Verifying (A2) is much more subtle and requires a detailed analysis of the bands close to the Fermi surface. An exception is when $\mu$ lies in or at the edge of a gap, in which case (A2) is trivially satisfied (the estimate on $\rho_{Q^{\pm\mp}}$ in $L^2$ was already obtained in this case in~\cite{CanDelLew-08a}). In the case where $\mu$ lies in the interior of the essential spectrum, we expect (A2) to be true, as soon as the Fermi surface is sufficiently regular. To make this intuition precise, a possible line of attack could be as follows. We assume again, for simplicity, that there is a unique $n$ such that the graph of $\lambda_n$ crosses $\mu$. Then we have to prove that the operator whose kernel is 
\begin{equation}
\int_{\mu-\epsilon\leq\lambda_n(\xi)\leq\mu}d\xi\, \int_{\mu\leq\lambda_n(\xi')\leq\mu+\epsilon}d\xi'\; \frac{u_n(\xi,x)\,\overline{u_n(\xi',x)}\,\overline{u_n(\xi,x')}\,u_n(\xi',x')}{\sqrt{\mu-\lambda_n(\xi)}\sqrt{\lambda_n(\xi')-\mu}},
\end{equation}
is bounded on $L^2(\R^d)$. The main idea is now that, for the question of boundedness, each Bloch function $u_n(\xi,x)$ can be replaced by the corresponding plane wave $\exp(ix\cdot\xi)$. Arguments of this sort have been carried out in a similar context  in~\cite{Sobolev-91,BirYaf-94,BirSlo-10,FraSim-11}, for instance. When $\nabla_\xi\lambda_n(\xi)\neq0$ for all $\xi$ such that $\lambda_n(\xi)=\mu$, the Fermi surface is smooth and can be locally replaced by a sphere. This reduces the computation to what we have done in Lemma~\ref{lem:Phi_bounded}, in the translation-invariant case.

This concludes our intuitive description of how to verify Assumptions (A1) and (A2) for periodic backgrounds. Rendering all this rigorous is beyond the scope of this paper, however.

\subsection{Sketch of the proof of Theorem~\ref{thm:abstract}}
The proof follows again the same two steps as that of Theorem~\ref{thm:LT-rho}.

\subsubsection*{Step 1. Estimate on $Q^{\pm\pm}$}
We start by estimating the diagonal densities $\rho_{Q^{\pm\pm}}$. 
Following the proof of Lemma~\ref{lem:generalized_LT}, Assumption (A1) implies that 
$$\tr|H-\mu|\gamma\geq \int_{\R^d}\tilde{R}\big(\rho_\gamma(x)\big)\,dx$$
for every $0\leq\gamma\leq1$, where
$$\tilde{R}(\rho)=\int_0^\ii \left(\sqrt{\rho}-\sqrt{Cq(e+e^{d/2})}\right)^2_+\, de.$$
One has
$$\tilde{R}(\rho)\underset{0}{\sim}\frac{\rho^2}{6Cq}\quad\text{and}\quad \tilde{R}(\rho)\underset{\ii}{\sim}\frac{d^2}{(d+4)(d+2)(Cq)^{2/d}}\;\rho^{1+2/d}.$$
Since $\epsilon<\rho_0\leq M$ by Assumptions (A1) and (A3), we therefore also have a lower bound
\begin{multline*}
\int_{\R^d}\tilde{R}\big(\rho_\gamma(x)\big)\,dx\\
\geq c\int_{\R^d} \left(\big(\rho_0(x)+\rho_\gamma(x)\big)^{1+\tfrac{2}d}-\rho_0(x)^{
1+\tfrac{2}d}-\frac{2+d}d \rho_0(x)^{\tfrac{2}d}\,\rho_\gamma(x)\right)dx
\end{multline*}
with a constant $c$ only depending on $d$, $\epsilon$, $M$ and $Cq$.
Applying these estimates to $\gamma=\pm Q^{\pm\pm}$, we obtain the estimate analogous to~\eqref{eq:final_estim_diag}.

\medskip

\subsubsection*{Step 2. Estimate on $Q^{\pm\mp}$}
Following the corresponding step in the proof of Theorem~\ref{thm:LT-rho}, a bound of the form
$$\int_{\R^d}|\rho_{Q^{\pm\mp}}|^2\leq c\tr_0(H-\mu)Q$$
can be derived from the estimate
\begin{enumerate}
\item[(A2')] $\displaystyle \norm{\frac{\Pi^+}{|H-\mu|^{1/4}}\;V\;\frac{\Pi^-}{|H-\mu|^{1/4}}}_{\gS_2}\leq C\norm{V}_{L^2}$\quad for all $V\in L^2(\R^d)$.
\end{enumerate}
We now explain how to derive this bound from (A1) and (A2).

To this end, we split
$$\Pi^-=\Pi^-_<+\Pi^-_>=\1(\mu-\epsilon\leq H\leq \mu)+\1(H< \mu-\epsilon),$$
$$\Pi^+=\Pi^+_<+\Pi^+_>=\1(\mu\leq H\leq \mu+\epsilon)+\1(H> \mu+\epsilon)$$
and estimate each term separately. The bound on $\Pi^+_< V \Pi^-_<$ is exactly our assumption (A2). For $\Pi^+_> V \Pi^-$, we write that
\begin{equation}
\norm{\frac{\Pi^+_>}{|H-\mu|^{1/4}} V \frac{\Pi^-}{|H-\mu|^{1/4}}}_{\gS_2}^2
\leq\frac1{\sqrt{\epsilon}}\norm{V \frac{\Pi^-}{|H-\mu|^{1/4}}}_{\gS_2}^2
=\frac1{\sqrt{\epsilon}}\int_{\R^d}|V|^2\,\rho\left[\frac{\Pi^-}{|H-\mu|^{1/2}}\right].\label{eq:estim-far-Fermi}
\end{equation}
The density appearing on the right side is uniformly bounded on $\R^d$. Indeed, one has, more generally,
\begin{equation}
\rho\left[\frac{\1(|H-\mu|\leq a)}{|H-\mu|^{1/2}}\right]\in L^\ii(\R^d)
\label{eq:bound_rho_unif}
\end{equation}
for every fixed $a>0$. To see this, we write
$$\frac{\1(|x|\leq a)}{|x|^{1/2}}=\frac12 \int_0^\ii \1\big(|x|\leq \min(y,a)\big)\frac{dy}{y^{3/2}}$$
which, by (A1), implies the uniform bound
$$\rho\left[\frac{\1(|H-\mu|\leq a)}{|H-\mu|^{1/2}}\right]\leq \frac{C}2 \int_0^\ii \left(\min(y,a)+\min(y,a)^{d/2}\right)\frac{dy}{y^{3/2}}.$$
Inserting~\eqref{eq:bound_rho_unif} in~\eqref{eq:estim-far-Fermi} and using the fact that $H$ is bounded from below, we obtain the desired estimate for $\Pi^+_< V \Pi^-$. The term corresponding to $\Pi^+_<V\Pi^-_>$ is estimated similarly, using
the fact that $\Pi^-_>\,|H-\mu|^{-1/4}\leq \epsilon^{-1/4}$.

This completes our sketch of the proof of Theorem~\ref{thm:abstract}.\qed

\bibliographystyle{siam}
\bibliography{biblio.bib}

\begin{thebibliography}{10}

\bibitem{BenLos-04}
{\sc R.~D. Benguria and M.~Loss}, {\em Connection between the {L}ieb-{T}hirring
  conjecture for {S}chr{\"o}dinger operators and an isoperimetric problem for
  ovals on the plane}, in Partial differential equations and inverse problems,
  vol.~362 of Contemp. Math., Amer. Math. Soc., Providence, RI, 2004,
  pp.~53--61.

\bibitem{BirSlo-10}
{\sc M.~S. Birman and V.~A. Sloushch}, {\em Discrete spectrum of the periodic
  {S}chr{\"o}dinger operator with a variable metric perturbed by a nonnegative
  potential}, Math. Model. Nat. Phenom., 5 (2010), pp.~32--53.

\bibitem{BirYaf-94}
{\sc M.~S. Birman and D.~R. Yafaev}, {\em The scattering matrix for a
  perturbation of a periodic {S}chr{\"o}dinger operator by decreasing
  potential}, Algebra i Analiz, 6 (1994), pp.~17--39.

\bibitem{CanDelLew-08a}
{\sc {\'E}.~Canc{\`e}s, A.~Deleurence, and M.~Lewin}, {\em A new approach to
  the modelling of local defects in crystals: the reduced {H}artree-{F}ock
  case}, Commun. Math. Phys., 281 (2008), pp.~129--177.

\bibitem{Coleman-63}
{\sc A.~Coleman}, {\em Structure of fermion density matrices}, Rev. Modern
  Phys., 35 (1963), pp.~668--689.

\bibitem{DolLapLos-08}
{\sc J.~Dolbeault, A.~Laptev, and M.~Loss}, {\em Lieb-{T}hirring inequalities
  with improved constants}, J. Eur. Math. Soc. (JEMS), 10 (2008),
  pp.~1121--1126.

\bibitem{FraLewLieSei-11}
{\sc R.~L. Frank, M.~Lewin, E.~H. Lieb, and R.~Seiringer}, {\em Energy {C}ost
  to {M}ake a {H}ole in the {F}ermi {S}ea}, Phys. Rev. Lett., 106 (2011),
  p.~150402.

\bibitem{FraSim-11}
{\sc R.~L. Frank and B.~Simon}, {\em Critical {L}ieb-{T}hirring bounds in gaps
  and the generalized {N}evai conjecture for finite gap {J}acobi matrices},
  Duke Math. J., 157 (2011), pp.~461--493.

\bibitem{HaiLewSer-05a}
{\sc C.~Hainzl, M.~Lewin, and {\'E}.~S{\'e}r{\'e}}, {\em Existence of a stable
  polarized vacuum in the {B}ogoliubov-{D}irac-{F}ock approximation}, Commun.
  Math. Phys., 257 (2005), pp.~515--562.

\bibitem{HaiLewSer-09}
\leavevmode\vrule height 2pt depth -1.6pt width 23pt, {\em Existence of atoms
  and molecules in the mean-field approximation of no-photon quantum
  electrodynamics}, Arch. Ration. Mech. Anal., 192 (2009), pp.~453--499.

\bibitem{Hundertmark-07}
{\sc D.~Hundertmark}, {\em Some bound state problems in quantum mechanics}, in
  Spectral theory and mathematical physics: a {F}estschrift in honor of {B}arry
  {S}imon's 60th birthday, vol.~76 of Proc. Sympos. Pure Math., Amer. Math.
  Soc., Providence, RI, 2007, pp.~463--496.

\bibitem{HunLapWei-00}
{\sc D.~Hundertmark, A.~Laptev, and T.~Weidl}, {\em New bounds on the
  {L}ieb-{T}hirring constants}, Invent. Math., 140 (2000), pp.~693--704.

\bibitem{IonJer-03}
{\sc A.~D. Ionescu and D.~Jerison}, {\em On the absence of positive eigenvalues
  of {S}chr{\"o}dinger operators with rough potentials}, Geom. Funct. Anal., 13
  (2003), pp.~1029--1081.

\bibitem{KenLie-87}
{\sc T.~Kennedy and E.~H. Lieb}, {\em Proof of the {P}eierls instability in one
  dimension}, Phys. Rev. Lett., 59 (1987), pp.~1309--1312.

\bibitem{KocTat-06}
{\sc H.~Koch and D.~Tataru}, {\em Carleman estimates and absence of embedded
  eigenvalues}, Comm. Math. Phys., 267 (2006), pp.~419--449.

\bibitem{Kohn-59}
{\sc W.~Kohn}, {\em Image of the {F}ermi surface in the vibration spectrum of a
  metal}, Phys. Rev. Lett., 2 (1959), p.~393.

\bibitem{LapWei-00b}
{\sc A.~Laptev and T.~Weidl}, {\em Recent results on {L}ieb-{T}hirring
  inequalities}, in Journ{\'e}es ``{\'E}quations aux {D}{\'e}riv{\'e}es
  {P}artielles'' ({L}a {C}hapelle sur {E}rdre, 2000), Univ. Nantes, Nantes,
  2000, pp.~Exp.\ No.\ XX, 14.

\bibitem{LapWei-00}
\leavevmode\vrule height 2pt depth -1.6pt width 23pt, {\em Sharp
  {L}ieb-{T}hirring inequalities in high dimensions}, Acta Math., 184 (2000),
  pp.~87--111.

\bibitem{LiYau-83}
{\sc P.~Li and S.~T. Yau}, {\em On the {S}chr\"odinger equation and the
  eigenvalue problem}, Comm. Math. Phys., 88 (1983), pp.~309--318.

\bibitem{Lieb-83b}
{\sc E.~H. Lieb}, {\em Density functionals for {C}oulomb systems}, Int. J.
  Quantum Chem., 24 (1983), pp.~243--277.

\bibitem{LieLos-01}
{\sc E.~H. Lieb and M.~Loss}, {\em Analysis}, vol.~14 of Graduate Studies in
  Mathematics, American Mathematical Society, Providence, RI, second~ed., 2001.

\bibitem{LieNac-95b}
{\sc E.~H. Lieb and B.~Nachtergaele}, {\em Dimerization in ring-shaped
  molecules: the stability of the {P}eierls instability}, in X{I}th
  {I}nternational {C}ongress of {M}athematical {P}hysics ({P}aris, 1994), Int.
  Press, Cambridge, MA, 1995, pp.~423--431.

\bibitem{LieNac-95}
\leavevmode\vrule height 2pt depth -1.6pt width 23pt, {\em Stability of the
  {P}eierls instability for ring-shaped molecules}, Phys. Rev. B, 51 (1995),
  p.~4777.

\bibitem{LieNac-95c}
\leavevmode\vrule height 2pt depth -1.6pt width 23pt, {\em Bond alternation in
  ring-shaped molecules: The stability of the {P}eierls instability}, Int. J.
  Quantum Chemistry, 58 (1996), pp.~699--706.

\bibitem{LieSei-09}
{\sc E.~H. Lieb and R.~Seiringer}, {\em The {S}tability of {M}atter in
  {Q}uantum {M}echanics}, Cambridge Univ. Press, 2010.

\bibitem{LieThi-75}
{\sc E.~H. Lieb and W.~E. Thirring}, {\em Bound on kinetic energy of fermions
  which proves stability of matter}, Phys. Rev. Lett., 35 (1975), pp.~687--689.

\bibitem{LieThi-76}
\leavevmode\vrule height 2pt depth -1.6pt width 23pt, {\em Inequalities for the
  moments of the eigenvalues of the {S}chr{\"o}dinger hamiltonian and their
  relation to {S}obolev inequalities}, Studies in Mathematical Physics,
  Princeton University Press, 1976, pp.~269--303.

\bibitem{Migdal-58}
{\sc A.~Migdal}, {\em Interactions between electrons and lattice vibrations in
  a normal metal (russian)}, Zh. Eksp. Teor. Fiz., 34 (1958), pp.~1438--1446.
\newblock English translation: \textit{Sov. Phys. JETP}, \textbf{7}, p. 996
  (1958).

\bibitem{Peierls}
{\sc R.~E. Peierls}, {\em Quantum {T}heory of {S}olids}, Clarendon Press, 1955.

\bibitem{Pushnitski-08}
{\sc A.~Pushnitski}, {\em The scattering matrix and the differences of spectral
  projections}, Bulletin London Math. Soc., 40 (2008), pp.~227--238.

\bibitem{PutYaf-09}
{\sc A.~Pushnitski and D.~Yafaev}, {\em Spectral theory of discontinuous
  functions of self-adjoint operators and scattering theory}.
\newblock Preprint arXiv:0907.1518, 2009.

\bibitem{ReeSim1}
{\sc M.~Reed and B.~Simon}, {\em Methods of {M}odern {M}athematical {P}hysics.
  {I}. Functional analysis}, Academic Press, 1972.

\bibitem{ReeSim2}
\leavevmode\vrule height 2pt depth -1.6pt width 23pt, {\em Methods of {M}odern
  {M}athematical {P}hysics. {II}. {F}ourier analysis, self-adjointness},
  Academic Press, New York, 1975.

\bibitem{ReeSim4}
\leavevmode\vrule height 2pt depth -1.6pt width 23pt, {\em Methods of {M}odern
  {M}athematical {P}hysics. {IV}. {A}nalysis of operators}, Academic Press, New
  York, 1978.

\bibitem{Rumin-11}
{\sc M.~Rumin}, {\em Balanced distribution-energy inequalities and related
  entropy bounds}, Duke Math. J., 160 (2011), pp.~567--597.

\bibitem{Simon-79}
{\sc B.~Simon}, {\em Trace ideals and their applications}, vol.~35 of London
  Mathematical Society Lecture Note Series, Cambridge University Press,
  Cambridge, 1979.

\bibitem{Sobolev-91}
{\sc A.~V. Sobolev}, {\em Weyl asymptotics for the discrete spectrum of the
  perturbed {H}ill operator}, in Estimates and asymptotics for discrete spectra
  of integral and differential equations ({L}eningrad, 1989--90), vol.~7 of
  Adv. Soviet Math., Amer. Math. Soc., Providence, RI, 1991, pp.~159--178.

\bibitem{Voit-95}
{\sc J.~Voit}, {\em One-dimensional {F}ermi liquids}, Rep. Prog. Phys., 58
  (1995), p.~977.

\bibitem{NeuWig-29}
{\sc J.~{von Neumann} and E.~Wigner}, {\em {\"U}ber merkw{\"u}rdige diskrete
  {E}igenwerte}, Phys. Z, 30 (1929), pp.~465--467.

\bibitem{Yafaev-10}
{\sc D.~R. Yafaev}, {\em Mathematical scattering theory}, vol.~158 of
  Mathematical Surveys and Monographs, American Mathematical Society,
  Providence, RI, 2010.
\newblock Analytic theory.

\end{thebibliography}

\end{document}